\documentclass[a4paper,11pt]{article}
\usepackage{jcappub} 
\usepackage{lineno}
\usepackage{dsfont}
\usepackage{forest}
\usepackage{graphicx}   
\usepackage{changepage} 
\usepackage{makecell}
\title{\texttt{BROOM}: a python package for model-independent analysis of microwave astronomical data}

\def\reff@jnl#1{{\rm#1\/}}

\def\aj{\reff@jnl{AJ}}                  
\def\araa{\reff@jnl{ARA\&A}}            
\def\apj{\reff@jnl{ApJ}}                
\def\apjl{\reff@jnl{ApJ}}               
\def\apjs{\reff@jnl{ApJS}}              
\def\ao{\reff@jnl{Appl.Optics}}         
\def\apss{\reff@jnl{Ap\&SS}}            
\def\aap{\reff@jnl{A\&A}}               
\def\aapr{\reff@jnl{A\&A~Rev.}}         
\def\aaps{\reff@jnl{A\&AS}}             
\def\azh{\reff@jnl{AZh}}                        
\def\baas{\reff@jnl{BAAS}}              
\def\jcap{\reff@jnl{JCAP}}              
\def\jrasc{\reff@jnl{JRASC}}            
\def\memras{\reff@jnl{MmRAS}}           
\def\mnras{\reff@jnl{MNRAS}}            
\def\pra{\reff@jnl{Phys.Rev.A}}         
\def\prb{\reff@jnl{Phys.Rev.B}}         
\def\prc{\reff@jnl{Phys.Rev.C}}         
\def\prd{\reff@jnl{Phys.Rev.D}}         
\def\prl{\reff@jnl{Phys.Rev.Lett}}      
\def\physrep{\reff@jnl{Phys.Rep.}}      
\def\pasp{\reff@jnl{PASP}}              
\def\pasj{\reff@jnl{PASJ}}              
\def\qjras{\reff@jnl{QJRAS}}            
\def\skytel{\reff@jnl{S\&T}}            
\def\solphys{\reff@jnl{Solar~Phys.}}    
\def\sovast{\reff@jnl{Soviet~Ast.}}     
 \def\ssr{\reff@jnl{Space~Sci.Rev.}}    
\def\zap{\reff@jnl{ZAp}}                
\def\nat{\reff@jnl{Nature}}             

\author[*,1,2,3]{A. Carones,}
\author[1]{S. Jose,}
\author[1,4]{A. Mustafa,}
\author[1,2,3]{N. Krachmalnicoff,}
\author[1,2,3]{C. Baccigalupi}
\affiliation[1]{\it International School for Advanced Studies (SISSA), Via Bonomea 265, Trieste 34136, Italy}
\affiliation[2]{\it INFN Sezione di Trieste, Via Valerio 2, Trieste 34127, Italy}
\affiliation[3]{\it IFPU, Via Beirut 2, Trieste 34151, Italy}
\affiliation[4]{Università di Trento, Dipartimento di Fisica, Via Sommarive 14, 38123, Trento, Italy}

\emailAdd{acarones@sissa.it}

\abstract{We present \texttt{BROOM}, a new python package for the application of blind, minimum-variance component-separation techniques to microwave observations. The package enables the reconstruction of signals with known spectral energy distributions—such as the Cosmic Microwave Background (CMB), Sunyaev--Zeldovich distortions, or foreground moments—in both temperature and polarization through a suite of Internal Linear Combination (ILC) implementations, in the presence of astrophysical and instrumental contaminants. 
In addition, \texttt{BROOM} supports the blind reconstruction of coherent emission components with unknown covariance properties via a Generalized ILC (GILC) framework. Beyond component separation, the package provides tools to diagnose foreground complexity and to estimate residual contamination leaking into reconstructed maps across angular scales and sky regions. It also includes utilities to generate realistic microwave simulations for arbitrary CMB experiments and to compute angular power spectra of the resulting products.
We present a comprehensive description and validation of the implemented pipelines in two representative experimental configurations: a full-sky satellite mission and a ground-based experiment. \texttt{BROOM} is publicly available, fully documented, and easily installable at \url{https://github.com/alecarones/broom}.
}

\begin{document}
\maketitle
\flushbottom

\section{Introduction}
\label{sec:intro}

The study of the Cosmic Microwave Background (CMB) has been pivotal in shaping our understanding of the Universe’s origin, composition, and evolution. High-precision measurements of temperature and polarization anisotropies have established the standard cosmological model and placed tight constraints on fundamental parameters \cite{Planck_cosmopars, ACT2025params, SPT2025}. Nonetheless, measurements of CMB anisotropies with enhanced sensitivity may open new frontiers in fundamental physics and astrophysics, including — to name a few — cosmic inflation, the stochastic background of gravitational waves \cite{1997PhRvL..78.2058K}, the history of cosmic reionization \cite{Planck_reio, 2020MNRAS.499..550Q}, constraints on neutrino masses \cite{2015PhRvD..92l3535A}, and cosmic birefringence \cite{2019PhRvD.100b3507P,2022PhRvL.128i1302D}. This anticipated wealth of information hinges on the accurate reconstruction of the CMB polarization field, particularly its scalar patterns \cite{1997PhRvD..55.1830Z} (the $E$ and $B$ modes), of the CMB lensing potential \cite{Zaldarriaga_lensing, Okamoto_lensing, Lewis_lensing}, and other secondary anisotropies, which are or will be probed by ongoing and forthcoming experiments such as the Simons Observatory (SO) \cite{SO_2019},  \textit{LiteBIRD} \cite{PTEP} and \textit{BICEP/Keck} \cite{BK2021}. 

A central challenge in the analysis of future microwave data is the separation of the CMB signals from Galactic and extragalactic foreground emissions. As a result, considerable effort is being devoted to the development of robust and effective techniques for separating or extracting the different sky components with the precision demanded by upcoming experiments.

Component separation techniques can broadly be divided into: (i) \texttt{parametric} methods \cite{Commander, Commander3, FGBuster, Azzoni2021, Vacher2022, Bsecret}, which construct detailed models of the data (in the time, pixel or spectral domain) and fit them to observations; and (ii) \texttt{blind} techniques \cite{ILC, NILC, SMICA, SEVEM, MCNILC}, which exploit the distinct spectral properties of astrophysical components without relying on explicit modeling. Each class has specific strengths and limitations, particularly regarding robustness to astrophysical uncertainties, sensitivity to instrumental systematics, and computational efficiency. In light of the ambitious scientific objectives of future CMB missions, combining both approaches will be essential to achieve reliable component separation and to ensure the consistency and validation of cosmological inferences.

Among the \texttt{blind} techniques, the most widely used approach is the Internal Linear Combination (ILC), which reconstructs the signal of interest by linearly combining multifrequency observations so as to minimize the contribution of all components with different spectral energy distributions (SEDs) to a chosen summary statistic—most often the variance of the combined map \cite{ILC, Dela_compsep}. The performance of ILC can be enhanced by applying it separately to wavelet-filtered data, thereby allowing contaminants that dominate on different angular scales to be addressed within the same framework. This idea led to the development of the Needlet ILC (NILC), the state-of-the-art blind cleaning technique. ILC methods have been successfully applied to various data sets, including \textit{WMAP} \cite{kim2009, basak2012} and \textit{Planck} \cite{Planck2018_compsep}, and are expected to remain a benchmark for the analysis of upcoming polarization data. In this context, several extensions of ILC have been proposed, including blind extraction of CMB foregrounds \cite{GNILC_intro, GNILC}, adaptations to spin-2 polarization data \cite{PILC,Adak2021}, techniques for deprojecting foreground moments \cite{cMILC, ocMILC}, and optimizations of spatial domains based on foreground properties \cite{MCNILC}.

Given the benchmark role currently played by ILC techniques in the data analysis of major CMB collaborations, and the active development of the field, we introduce in this manuscript the most complete and advanced (to our knowledge) public python code for blind microwave data analysis: \texttt{BROOM}\footnote{\url{https://github.com/alecarones/broom}}. We provide full details on the implemented methodologies in \texttt{BROOM}, along with validation results, making this paper a reference for researchers interested in employing the package. For further inquiries, readers are invited to consult the corresponding GitHub repository, which includes comprehensive documentation and up-to-date tutorials covering all relevant use cases.

The paper is structured as follows. Section~\ref{sec:methods} outlines the main methodological aspects of the \texttt{BROOM} pipelines, followed by a detailed description of the package in Section~\ref{sec:broom}. Validation results are presented in Section~\ref{sec:results}, and concluding remarks are given in Section~\ref{sec:concl}. A comprehensive description of many mandatory and optional parameters for \texttt{BROOM} runs is provided in Appendix~\ref{app:broom_pars}.

\section{Methodological background}
\label{sec:methods}

In this section, we outline the key methodological elements of the \texttt{BROOM} package. The fundamental concepts of microwave sky maps are introduced in Section~\ref{sec:basics}. Building on this, \texttt{BROOM} can be employed to: (i) generate realistic map-based simulations of microwave observations as performed by modern CMB experiments (Section~\ref{sec:sims}); (ii) carry out model-independent component separation (Section~\ref{sec:compsep}); and (iii) compute angular power spectra from the component-separation outputs (Section~\ref{sec:spectra}).

\subsection{General framework}
\label{sec:basics}
The \texttt{BROOM} package can handle pixelized maps of fields defined on the sphere. At present, the only supported pixelization scheme is \texttt{HEALPix} \cite{healpix}, accessed through its python implementation \texttt{healpy} \cite{healpy}. Additional pixelization schemes (e.g. alternative equal-area or hierarchical tessellations) are planned to be supported in future releases.
The \texttt{HEALPix} scheme divides the sphere into equal-area pixels whose centers are arranged along iso-latitude rings, a property that allows efficient spherical harmonic transforms. The resolution of a map is set by the \texttt{HEALPix} parameter $N_{\text{side}}$with the total number of pixels given by $N_{\text{pix}}=12\cdot N_{\text{side}}^{2}$.

Key physical quantities typically reconstructed from microwave observations are the total measured intensity and the Stokes parameters for the linear polarization $Q$ and $U$ \cite{Rybicki}.

A \texttt{HEALPix} map of temperature (intensity) anisotropies $\Delta T(\hat{n})$ can be rephrased in the spherical-harmonic basis as:
\begin{equation}
    \Delta T(\hat{n}) = \sum_{\ell m} a_{\ell m}^{T}\, Y_{\ell m}(\hat{n}) \, ,
\label{eq:alm_T}
\end{equation}
where \(Y_{\ell m}(\hat{n})\) denotes the spin-0 spherical harmonics and $\hat{n}$ a line-of-sight, which, in most cases, is centered in a specific pixel of the \texttt{HEALPix} grid of the celestial sphere.

The linear polarization field, composed of the $Q$ and $U$ Stokes parameters, is not rotationally invariant on the sphere and therefore constitutes a spin-$2$ field, which can be expanded in a harmonic basis through the spin-\(2\) spherical harmonics:
\begin{equation}
    Q(\hat{n})\pm iU(\hat{n}) = \sum_{\ell m} a^{\pm2}_{\ell m}\, {}_{\pm 2}Y_{\ell m}(\hat{n}) .
\label{eq:P_lm}
\end{equation}
From the above decomposition, the polarization spin-$2$ field can be decomposed into the scalar $E$ modes:
\begin{equation}
     a_{\ell m}^{E} = -\tfrac{1}{2}\left(a^{+2}_{\ell m} + a^{-2}_{\ell m}\right) \quad \rightarrow \quad E(\hat{n}) = \sum_{\ell m} a_{\ell m}^{E}\, Y_{\ell m}(\hat{n})
\end{equation}
and pseudo-scalar $B$ modes
\begin{equation}
    a_{\ell m}^{B} = -\tfrac{1}{2i}\left(a^{+2}_{\ell m} - a^{-2}_{\ell m}\right)\quad \rightarrow \quad B(\hat{n}) = \sum_{\ell m} a_{\ell m}^{B}\, Y_{\ell m}(\hat{n}) .
\label{eq:EB_lm}
\end{equation}


To reconstruct the intensity and/or polarization signals of the microwave sky components, observations across multiple frequency channels are required. Multi-frequency coverage enables the characterization of the spectral scaling of the various components and is therefore essential for their separation within the component-separation framework.
In the following, we introduce the formalism used to describe the instrument response to sky signals, which ultimately determines the observed maps at each frequency channel. This formalism is implemented in \texttt{BROOM} both for generating simulations of CMB experiments and for preprocessing simulated or externally provided data (either real or simulated) prior to and during the component-separation stage.

Let us consider a specific frequency channel of a CMB experiment at a reference frequency $\nu$. The corresponding set of observation maps \(X_{\nu}(\hat{n})\), collecting $TQU$,
amounts to the sky emission integrated over the instrumental bandpass $W_{\nu}$ (normalized to have unit integral) and instrumental noise $n_{\nu}(\hat{n})$:
\begin{equation}
X_{\nu}(\hat{n}) = n_{\nu}(\hat{n}) + \int d\nu^{\prime}W_{\nu^{\prime}}\cdot X_{\nu^{\prime}}(\hat{n}).
\label{eq:bandpass}
\end{equation}
In practice, for any procedure requiring bandpass knowledge within \texttt{BROOM}, this integral is evaluated as a discrete sum, where the user may specify both the bandpass transmission values $W_{\nu}$ and the frequency grid on which the finite integration must be performed. Another key instrumental effect is the finite angular resolution of any telescope used to observe the microwave sky. To model this, the observed harmonic coefficients of the maps \(X_{\nu}(\hat{n})\) are obtained by multiplying the true sky coefficients at infinite resolution ($a^{X}_{\nu,\ell m}$) with a harmonic transfer function $b^{X}_{\nu,\ell m}$:
\begin{equation}
   \tilde{a}^{X}_{\nu,\ell m} = a^{X}_{\nu,\ell m} \cdot b^{X}_{\nu,\ell m}.
\label{eq:beams}
\end{equation}
If the instrumental beam is azimuthally symmetric, the window function reduces to a dependence on the multipole only, \textit{i.e}
$b^{X}_{\nu,\ell m}\rightarrow b^{X}_{\nu,\ell}$. We note that a fully consistent treatment of instrumental beams in sky simulations would require operating at the time-ordered data (TOD) level, which lies beyond the current scope of the \texttt{BROOM} package. Instead, the package adopts the approximate harmonic-space description of Equation~\ref{eq:beams} when generating simulations of microwave experiments.


During the simulation stage, the instrumental angular response (beam) and frequency transmission (bandpass) are applied exclusively to the simulated sky signals and not to the instrumental noise realizations.

Overall, the inputs to the \texttt{BROOM} component-separation modules can be provided either as multifrequency sets of maps ($T$, $QU$, $E$, $B$, $EB$, $TQU$, $TEB$) or as their corresponding harmonic coefficients ($T$, $E$, $B$, $EB$, $TEB$). Even when inputs are provided as \texttt{HEALPix}-pixelized maps, they are first converted into harmonic coefficients—restricted to the fields requested by the user—which then serve as the common input to the component-separation modules. Although this step may be suboptimal for certain specific applications—for instance, pixel-based ILC reconstructions of CMB temperature anisotropies \cite{ILC} or of $QU$ maps optimized for minimum-variance polarization intensity $P$ \cite{PILC}—it is required in the majority of use cases. Examples include component-separation approaches implemented in the needlet domain \cite{Marinucci2008}, commonly adopted to address localized contaminants, or methods that operate directly on $E$ and/or $B$ modes starting from input $QU$ maps. For the analysis of polarization fields, we indeed emphasize—this will become clearer in the following sections—that the user has the option to perform component separation either independently on the $E$- and $B$-mode maps or jointly on the $Q$ and $U$ maps, thereby treating the $E$ and $B$ polarization modes simultaneously.

When analyzing polarization data from partial-sky observations (as is the case for all ground-based CMB experiments), the transformation from maps to harmonic coefficients is non-trivial. It inevitably introduces leakage between $E$ and $B$ modes, arising from ambiguous modes that cannot be uniquely assigned to either polarization pattern \cite{lewis_EBleak}. This phenomenon is especially critical in CMB studies, since the $E$-mode signal is expected to exceed the $B$-mode counterpart by several orders of magnitude. $E$–$B$ leakage can be addressed either prior to component separation, via a map-based correction, and/or afterward during the computation of the angular power spectrum of the reconstructed CMB signal. The former strategy—embedding map-based correction methods within model-independent component-separation pipelines—has already been explored in \cite{NILC_cutsky}. In \texttt{BROOM}, several map-based options are available: (i) the standard $E$–$B$ purification as implemented in \texttt{NaMaster} (see \cite{Bunn2003,lewis_EBleak,pymaster} and references therein), (ii) three different techniques, which reconstruct a template of the leakage contamination and subtracts it from the data \cite{2019PhRvD.100b3538L, NILC_cutsky}. The handling of $E$-$B$ leakage at the power spectrum computation step is detailed in Section \ref{sec:spectra}.

\subsection{Simulating microwave sky components}
\label{sec:sims}
The \texttt{BROOM} package enables the generation of realistic simulated maps of the observed microwave sky, which can be tailored to any space-based, balloon-borne, or ground-based CMB experiment or microwave observatory. This enables both forecasting analyses and pipeline validation within the same package. Specifically, it allows the simulation of CMB signal, its spectral distortions- such as the Sunyaev-Zeldovich (SZ) effect \cite{SZ_theo,SZ_obs,1998ApJ...499....1C}, Galactic and extragalactic foreground emissions, and instrumental noise. Details on the parameters required for simulation generation are given in Appendix~\ref{app:broom_sims}. 

Primary CMB anisotropies are simulated as Gaussian, statistically isotropic realizations generated from a theoretical angular power spectrum, which is assumed to be provided in units of $\mu\mathrm{K}^2$. 
Galactic and extragalactic emissions are modeled using the \texttt{PySM} python package \cite{pysm, pysm3, pysm2025}. The non-primordial microwave components that can be simulated within \texttt{BROOM} (with their corresponding \texttt{PySM} indices in parentheses) include: synchrotron radiation (\textbf{s}) \citep{Rybicki,krach_2018,2022ApJ...936...24W}, thermal dust emission (\textbf{d}) \cite{1998ApJ...500..525S,Planck2018_compsep}, anomalous microwave emission (\textbf{a}) \cite{2011ame,2015ame,AME_commander}, CO line emission (\textbf{co}) \cite{2001ApJ...547..792D,2014A&A...571A..13P}, free–free emission (\textbf{f}) \cite{spitzer2008physical,Rybicki,2003MNRAS.341..369D}, the cosmic infrared background (\textbf{cib}) \cite{cib_2001,cib_planck}, and radio galaxies (\textbf{rg}) \cite{2016A&A...594A..26P,puglisi_radiosources}. In addition, maps of the thermal (\textbf{tsz}) and kinetic (\textbf{ksz}) Sunyaev–Zel’dovich effects can also be generated via \texttt{PySM}.
Details on the models and assumptions for each component are provided in the corresponding \texttt{PySM} papers \cite{pysm, pysm3, pysm2025} and documentation\footnote{\url{https://pysm3.readthedocs.io/en/latest/models.html}}. Both CMB and foreground components are smoothed with the appropriate beam transfer function (Equation~\ref{eq:beams}) and integrated over the bandpass window function (Equation~\ref{eq:bandpass}), if requested. 

Instrumental noise in intensity and polarization is simulated by drawing Gaussian realizations from a theoretical power spectrum of the form:
\begin{equation}
N_{\ell}^{T/P} = \left(\frac{\pi \cdot \sigma^{T/P}}{180 \cdot 60}\right)^{2}
\cdot \left[1 + \left(\frac{\ell}{\ell_{k}^{T/P}}\right)^{\alpha^{T/P}}\right],
\label{eq:cl_noise}
\end{equation}
where $T$ and $P$ refer to temperature and polarization, respectively. Here, $\sigma^{T/P}$ is the map depth in units of $\mu\mathrm{K}\cdot \text{arcmin}$, while $\ell_{k}^{T/P}$ and $\alpha^{T/P}$ denote the knee multipole and the spectral index associated with the $1/f$ noise component. The parameters $\sigma$, $\ell_{k}$, and $\alpha$ can be specified independently for temperature and polarization, and for each frequency channel.
As detailed in Appendix~\ref{app:broom_sims}, the user can simulate: (i) pure white noise (no $1/f$ contribution) in both temperature and polarization; (ii) white noise only in polarization, mimicking the presence of a perfect polarization modulator; or (iii) noise including a $1/f$ component in both fields, mimicking absent or imperfect modulation or a transfer function correction after data filtering.

Anisotropic scanning strategies can be incorporated at the stage of noise generation by providing either (i) a hit-count map (one per frequency channel or shared across channels), which is used to appropriately weight isotropic noise realizations generated from Equation~\ref{eq:cl_noise}, or (ii) maps of the pixel-domain standard deviation (in $\mu\mathrm{K}$ units) for each frequency channel. 
In the latter case, the parameter $\sigma^{T/P}$ in Equation~\ref{eq:cl_noise} is set to $4\pi/N_{\rm pix}$, and the resulting isotropic noise realization is subsequently rescaled according to the supplied standard-deviation maps.

All the aforementioned components, although initially generated in their intrinsic units, are consistently converted into the units specified by the user. For the generation of CMB and noise realizations, the package allows the user to specify a seed for each simulation. This ensures full control over the random generation process and facilitates reproducibility whenever needed.

If requested, data splits can also be generated. In this case, the \texttt{BROOM} routine simulates two independent noise realizations, each with an increased noise level (corresponding to a factor $\sqrt{2}$ higher map depth), which are then added to the sky signals to obtain two full data splits (\texttt{split1} and \texttt{split2}). The full-mission data can also be produced, assuming the noise map is given by
\begin{equation}
    \text{noise} = \frac{\text{noise}\_\text{split1} + \text{noise}\_\text{split2}}{2}.
\end{equation}

\subsection{Component separation}
\label{sec:compsep}
All component separation methods implemented within \texttt{BROOM} require input data to be brought to a common resolution:
\begin{equation}
   \bar{a}^{X}_{\nu,\ell m} = \tilde{a}^{X}_{\nu,\ell m} \cdot \frac{b^{X}_{\text{out}, \ell}}{b^{X}_{\nu,\ell m}},
\label{eq:beams_1}
\end{equation}
where $b^{X}_{\text{out}, \ell}$ is the output harmonic transfer function, corresponding to a symmetric Gaussian beam with pre-defined Full Width at Half Maximum (FWHM). Alternatively, the user may perform this resolution-matching step independently and provide an input data set that already shares a common angular resolution. In that case, the component-separation routines will skip the \emph{a priori} correction of Equation~\ref{eq:beams_1}.

Once the frequency maps have been brought to a common angular resolution, \texttt{BROOM} provides a suite of modules to perform model-independent analyses. These include the reconstruction of a signal with known SED (e.g. CMB, the thermal SZ effect), the recovery of foreground emission across observed frequency channels, the derivation of templates of foreground residuals (after component separation), and diagnostics of foreground complexity. 

All these techniques can be carried out either in real space—where all angular scales are treated simultaneously—or in the needlet domain \cite{Marinucci2008, NILC}, where input data are filtered into user-defined needlet bands and analyzed separately at different angular scales. The latter approach is particularly powerful for targeting contaminants that dominate in distinct multipole ranges.

Details on reconstruction of signals with known SED are presented in Section~\ref{sec:cmb_rec}, while foreground recovery, residual estimation, and complexity diagnostics are discussed in Section~\ref{sec:fgds_rec}. In both cases, the formalisms are described for a generic multifrequency input data set $\mathcal{X}_{\nu}$, which may correspond either to pixel-domain maps (i.e. $X_{\nu}$) or to needlet coefficients at a given scale. Section~\ref{sec:needlets} provides an overview of the needlet configurations implemented in \texttt{BROOM} and explains how to reconstruct a real-space field from solutions obtained across different needlet scales.

We note that some of the techniques outlined below can also be formulated in harmonic space, i.e., applied to the multifrequency set of harmonic coefficients. Although this case is not covered here and is not included in the current package release, extensions to the harmonic domain will be incorporated in future versions.

\subsubsection{Reconstruction of signals with known SED}
\label{sec:cmb_rec}
In this section, we outline the component-separation methodologies implemented within \texttt{BROOM} capable of reconstructing a signal with known a-priori SED from a linear combination of the input data-set $\mathcal{X}_{\nu}$. Within \texttt{BROOM}, the signals with fixed SED that can be reconstructed include CMB primary anisotropies, the thermal SZ effect, $\mu$-type spectral distortions, and arbitrary moments of foreground emission characterized by either power-law or MBB fundamental SEDs.

We begin by briefly introducing the formalism of moments-based modeling of foreground emission, as this framework is adopted both for foreground moment reconstruction and for deprojection within specific component-separation pipelines. This overview is intended to provide the interested reader with the necessary background to properly use several of the pipelines implemented in \texttt{BROOM}. For a more comprehensive treatment, we refer the reader to the cited literature and references therein.

\paragraph{Foreground moments.} Each foreground emission process contributing to microwave observations can be described by a fundamental SED, $f$, which characterizes the frequency scaling of its (polarized) intensity per unit emitting volume. However, the spectral parameters governing this scaling may vary both along a given line of sight and across different lines of sight on the sky.
As a result, the observed foreground signal in a specific direction corresponds to a superposition of emissions from multiple sources with potentially different effective SEDs. This arises from the projection of all emitters along the same line of sight onto the celestial sphere and from the beam-weighted integration over neighboring lines of sight within the instrumental resolution. These observational effects lead to distortions of the effective SED of each observed foreground component relative to its underlying fundamental form. 

As shown in \cite{moments}, the emission $X_{\nu}^{(i)}(\hat{n})$ of a given foreground component $i$ at frequency $\nu$ and in the sky direction $\hat{n}$ can be expressed as a Taylor expansion around its fundamental SED to capture the distortions induced by line-of-sight averaging and instrumental effects:
\begin{equation}
    X_{\nu}^{(i)}(\hat{n}) = X_{\nu_0}^{(i)}(\hat{n}) \cdot \left[\sum_{k}\sum_{\sum \alpha=k} \frac{(\beta^{(i)}_{1}(\hat{n})-\bar{\beta}^{(i)}_{1})^{\alpha_1}\cdot...\cdot(\beta^{(i)}_{n}(\hat{n})-\bar{\beta}^{(i)}_{n})^{\alpha_n}}{\alpha_1!\cdot...\cdot\alpha_n!}\frac{\partial^{k}f^{(i)}(\nu,\bar{\beta}^{(i)})}{\partial{\beta^{(i)}_{1}}^{\alpha_1}...\partial{\beta^{(i)}_{n}}^{\alpha_n}} \right],
\label{eq:mom_exp}
\end{equation}
where $\beta^{(i)}_{1},\ldots,\beta^{(i)}_{n}$ are the spectral parameters of the fundamental SED $f^{(i)}$, and $\bar{\beta}^{(i)}$ are their pivot values.
This expansion shows that the observed emission can be decomposed into a set of \emph{statistical moment components}, defined as
\begin{equation}
    m_{\alpha_1...\alpha_n}(\hat{n})=X_{\nu_0}^{(i)}(\hat{n})\cdot \frac{(\beta^{(i)}_{1}(\hat{n})-\bar{\beta}^{(i)}_{1})^{\alpha_1}\cdot...\cdot(\beta^{(i)}_{n}(\hat{n})-\bar{\beta}^{(i)}_{n})^{\alpha_n}}{\alpha_1!\cdot...\cdot\alpha_n!},
\label{eq:SED_exp_1}
\end{equation}
each associated with a predictable SED, uniform across the sky, given by:
\begin{equation}
    A^{(i)}_{\alpha_1...\alpha_n,k}=\frac{\partial^{k}f^{(i)}(\nu,\bar{\beta}^{(i)})}{\partial{\beta^{(i)}_{1}}^{\alpha_1}...\partial{\beta^{(i)}_{n}}^{\alpha_n}}.
\label{eq:SED_exp_2}
\end{equation}

As mentioned in the introduction to this section, the moment components $m_{\alpha_1 \ldots \alpha_n}(\hat{n})$ can be either reconstructed or deprojected within the ILC framework, once the pivot spectral parameters $\bar{\beta}^{(i)}$ have been specified by the user.

At the moment, within the \texttt{BROOM} package, moment expansions of components having a fundamental power-law or modified blackbody (MBB) SED are at disposal for component separation.
Multi-frequency observed emission from synchrotron, free-free or radio-galaxies \cite{radio_seds} can commonly be expanded around a baseline power-law SED up to second order in $\beta_{\textrm{s}}(\hat{n})$ as:
\begin{equation}
    \begin{aligned}
        X_{\nu}(\hat{n})=X_{\nu_{\textrm{s}}}(\hat{n})\, &\Bigg[f_{\textrm{pl}}(\nu,\bar{\beta}_{\textrm{s}})+  \left(\beta_{\textrm{s}}(\hat{n})-\bar{\beta}_{\textrm{s}}\right) \partial_{\beta_{\textrm{s}}}f_{\textrm{pl}}(\nu,\bar{\beta}_{\textrm{s}})\\
        &+\frac{1}{2} \left(\beta_{\textrm{s}}(\hat{n})-\bar{\beta}_{\textrm{s}}\right)^2 \partial^{2}_{\beta_{\textrm{s}}}f_{\textrm{pl}}(\nu,\bar{\beta}_{\textrm{s}})+\mathcal{O}\left( \left(\beta_{\textrm{s}}(\hat{n})-\bar{\beta}_{\textrm{s}}\right)^3\right) \Bigg]\,,
    \end{aligned}
\label{eq:sync_exp}
\end{equation}
where
 \begin{equation}
    f_{\textrm{pl}}(\nu,\bar{\beta}_{\textrm{s}})=\left(\frac{\nu}{\nu_{\textrm{s}}} \right)^{\bar{\beta}_{\textrm{s}}},
\end{equation}
is a power-law with fixed pivot spectral index $\bar{\beta}_{\textrm{s}}$. Equation~\ref{eq:sync_exp} highlights moments of the spectral index, $(\beta_{\textrm{s}}(\hat{n})-\bar{\beta}_{\textrm{s}})^k$, with uniform SEDs $\partial^{k}_{\beta_{\textrm{s}}}f_{\textrm{pl}}(\nu,\bar{\beta}_{\textrm{s}})$.
On the other hand, fundamental SED of thermal dust and Cosmic Infrared Background is expected to be very close to a MBB, therefore their moment expansion around a baseline MBB SED up to second order reads as:
\begin{equation}
    \begin{aligned}
        X_{\nu}^{(d)}(\hat{n})=X_{\nu_{\textrm{d}}}^{(d)}(\hat{n})\,  &\Bigg[f_{\textrm{mbb}}(\nu,\bar{\beta}_{\textrm{d}},\bar{T}_{\textrm{d}})+ \left(\beta_{\textrm{d}}(\hat{n}) - \bar{\beta}_{\textrm{d}}\right)  \partial_{\beta_{\textrm{d}}}f_{\textrm{mbb}}(\nu,\bar{\beta}_{\textrm{d}},\bar{T}_{\textrm{d}})\\
        &+\left(T_{\textrm{d}}(\hat{n}) - \bar{T}_{\textrm{d}}\right)  \partial_{T_{\textrm{d}}}f_{\textrm{mbb}}(\nu,\bar{\beta}_{\textrm{d}},\bar{T}_{\textrm{d}}) + \frac{1}{2}\left(\beta_{\textrm{d}}(\hat{n}) - \bar{\beta}_{\textrm{d}}\right)^2  \partial^{2}_{\beta_{\textrm{d}}}f_{\textrm{mbb}}(\nu,\bar{\beta}_{\textrm{d}},\bar{T}_{\textrm{d}})\\
        &+  \left(\beta_{\textrm{d}}(\hat{n}) - \bar{\beta}_{\textrm{d}}\right) \left(T_{\textrm{d}}(\hat{n}) - \bar{T}_{\textrm{d}}\right)  \partial_{\beta_{\textrm{d}}}\partial_{T_{\textrm{d}}}f_{\textrm{mbb}}(\nu,\bar{\beta}_{\textrm{d}},\bar{T}_{\textrm{d}})\\
        &+ \frac{1}{2} \left(T_{\textrm{d}}(\hat{n}) - \bar{T}_{\textrm{d}}\right)^2  \partial^{2}_{T_{\textrm{d}}}f_{\textrm{mbb}}(\nu,\bar{\beta}_{\textrm{d}},\bar{T}_{\textrm{d}})+ \mathcal{O}\left(\left(\beta_{\textrm{d}}(\hat{n}) - \bar{\beta}_{\textrm{d}}\right)^3,\left(T_{\textrm{d}}(\hat{n}) - \bar{T}_{\textrm{d}}\right)^3\right)\Bigg]\,,
    \end{aligned}
\label{eq:dust_exp}
\end{equation}
where 
\begin{equation}
\label{eq:dust_baseline_krj}
         f_{\textrm{mbb}}(\nu,\bar{\beta}_{\textrm{d}},\bar{T}_{\textrm{d}})=\left(\frac{\nu}{\nu_{\textrm{d}}} \right)^{\bar{\beta}_{\textrm{d}} +1}\frac{\exp{\left(h\nu_{\textrm{d}}/k\bar{T}_{\textrm{d}}(\hat{n})\right)}-1}{\exp{\left(h\nu/k\bar{T}_{\textrm{d}}(\hat{n})\right)}-1}
\end{equation}
is the MBB function for data in brightness temperature units, with fixed pivot values of the spectral index and temperature $\bar{\beta}_{\textrm{d}}$ and $\bar{T}_{\textrm{d}}$. \\

We now introduce and briefly describe the component-separation techniques currently implemented in \texttt{BROOM} for the reconstruction of signals with known SEDs. This overview summarizes the key elements and implementation steps of each method and serves as a concise reference for users interested in applying the component-separation pipelines available in \texttt{BROOM}. For a more detailed treatment, we refer the reader to the cited literature and references therein.

\paragraph{Internal Linear Combination (ILC).}
In the standard ILC technique, the reconstruction of a fixed-SED signal $\mathcal{X}^{\text{signal}}$ is performed by linearly combining the input multifrequency maps with frequency- and pixel-dependent weights \cite{ILC}:
\begin{equation}
 \mathcal{X}_{\text{ILC}}(\hat{n}) 
   = \sum_{\nu} w_{\nu}(\hat{n})\, \mathcal{X}_{\nu}(\hat{n}) 
   = \sum_{\nu} w_{\nu}(\hat{n}) \left( A^{\nu}_{\text{signal}}\,\mathcal{X}^{\text{signal}} 
     + \mathcal{X}^{\text{cont}}_{\nu} 
     + \mathcal{X}^{\text{noi}}_{\nu} \right)(\hat{n}),
\end{equation}
where $\mathcal{X}^{\text{signal}}$, $\mathcal{X}^{\text{cont}}$, and $\mathcal{X}^{\text{noi}}$ denote the signal of interest, any astrophysical and cosmological sky contamination, and instrumental noise, respectively, while $A^{\nu}_{\text{signal}}$ is the known SED of the component of interest. 
The ILC weights are calibrated to: (i) fully preserve the signal of interest,
\begin{equation}
   \sum_{\nu} w_{\nu}(\hat{n})\, A^{\nu}_{\text{signal}} = 1,
\end{equation}
and (ii) minimize the variance of the reconstructed map, thereby suppressing contamination from other components with different spectral behaviours:
\begin{equation}
   \min_{w}\, \text{Var}\!\left[\mathcal{X}_{\text{ILC}}(\hat{n})\right].
\end{equation}
The weights satisfying the above constraints are given by the generalized least-squares solution:
\begin{equation}
    \mathbf{w}(\hat{n}) = 
    \left[\mathbf{A}^{T} C^{-1}(\hat{n}) \mathbf{A}\right]^{-1} 
    \mathbf{A}^{T} C^{-1}(\hat{n}),
\label{eq:ilc_w}
\end{equation}
where $\mathbf{w}$ is the weight vector, $\mathbf{A}$ is the SED vector of the signal of interest in the chosen units, and $C(\hat{n})$ is the covariance matrix of the input data, whose elements are computed as
\begin{equation}
    C_{\nu\nu^{\prime}}(\hat{n}) 
    = \left\langle \mathcal{X}_{\nu}(\hat{n}^{\prime})\,
      \mathcal{X}_{\nu^{\prime}}(\hat{n}^{\prime}) \right\rangle_{\hat{n}^{\prime}\in \mathcal{D}_{\hat{n}}},
    \label{eq:data_cov}
\end{equation}
i.e. as the local average over a sky domain $\mathcal{D}_{\hat{n}}$ centered around the pixel $\hat{n}$.  
This localization allows the ILC solution to adapt to spatial variations of the contaminants across the sky. The smaller the domains over which the covariance is computed, the more effectively the contaminants are deprojected. However, this also increases the anti-correlation between the reconstructed signal of interest and the residuals, an effect known as the \textit{ILC bias} \cite{2009LNP...665..159D,NILC}, as the ILC weights start to be correlated with the input data themselves. The same limitation constrains the choice of needlet band widths on large angular scales: if a band includes too few modes, the covariance is poorly estimated, leading to a significant negative bias in the recovered power.

We note that, in general—and within the \texttt{BROOM} modules in particular—the computation of the SED vector $\mathbf{A}$ in Equation~\ref{eq:ilc_w} requires knowledge of the instrumental bandpasses, whose effect is modeled according to Equation~\ref{eq:bandpass}. An important exception arises when reconstructing the CMB signal and the input data are provided in thermodynamic units, in which case the bandpass integration does not need to be explicitly accounted for.
This consideration applies to all pipelines introduced below that aim to reconstruct signals with known SEDs.

The ILC technique can be applied to any scalar field, such as $T$, $E$, or $B$ modes. Its extension to the needlet domain \cite{Marinucci2008}, known as NILC \cite{NILC}, performs the minimum-variance combination independently at each needlet scale. The resulting scale-dependent CMB estimates are then recombined through an inverse needlet transform to produce the final real-space CMB reconstruction, as described in Section~\ref{sec:needlets}.

\paragraph{Polarization ILC (PILC).}
The ILC approach can be generalized to the spin-$2$ polarization field, giving rise to the Polarization ILC (PILC) \cite{PILC}:
\begin{equation}
	\left[\mathcal{Q}\pm \text{i}\ \mathcal{U}\right]_{\mathrm{PILC}} (\hat{n}) 
	= \sum_{\nu}\left[w_{\nu}^{(R)}\pm \text{i}\, w_{\nu}^{(I)}\right](\hat{n}) 
	\cdot \left[\mathcal{Q}\pm \text{i}\ \mathcal{U}\right]_{\nu}(\hat{n}),
	\label{eq:ILC_QU}
\end{equation}
where $\mathcal{Q}$ and $\mathcal{U}$ are the Stokes parameters in either pixel or needlet space. The complex weights are determined so as to minimize the total polarization power, $\mathcal{P}^2 = \mathcal{Q}^2 + \mathcal{U}^2$. This approach enables the user to deal with a covariant quantity related to the full polarization field. 

To preserve the input polarized component of interest, analogous conditions to the ILC case have to be imposed on the real and imaginary parts of the weights:
\begin{equation}
\sum_{\nu}w_{\nu}^{(R)}(\hat{n})\cdot A^{\nu}_{\text{signal}} = 1 
\qquad 
\text{and} 
\qquad 
\sum_{\nu}w_{\nu}^{(I)}(\hat{n})\cdot A^{\nu}_{\text{signal}} = 0
\label{eq:ILC_QU_w} 
\end{equation}
leading to \citep{PILC}:
\begin{equation}
\begin{aligned}
w_{\nu}^{(R)}(\hat{n})=& \frac{\lambda_{R}(\hat{n})}{2}\sum_{\nu^{\prime}=1}^{N_{\nu}}A^{\nu^{\prime}}_{\text{signal}}C^{-1}_{\nu\nu^{\prime}}(\hat{n})+\frac{\lambda_{I}(\hat{n})}{2}\sum_{\nu^{\prime}=N_{\nu}+1}^{2N_{\nu}}A^{\nu^{\prime}-N_{\nu}}_{\text{signal}}C^{-1}_{\nu\nu^{\prime}}(\hat{n}) \\
\omega_{\nu}^{(I)}(\hat{n})=& \frac{\lambda_{R}(\hat{n})}{2}\sum_{\nu^{\prime}=1}^{N_{\nu}}A^{\nu^{\prime}}_{\text{signal}}C^{-1}_{N_{\nu}+\nu,\nu^{\prime}}(\hat{n})+\frac{\lambda_{I}(\hat{n})}{2}\sum_{\nu^{\prime}=N_{\nu}+1}^{2N_{\nu}}A^{\nu^{\prime}-N_{\nu}}_{\text{signal}}C^{-1}_{N_{\nu}+\nu,\nu^{\prime}}(\hat{n}),
\end{aligned}
\label{eq:ILC_QU_w_1} 
\end{equation}
where $N_{\nu}$ is the number of observed frequency channels and $C$ is the polarization covariance matrix:
\begin{equation}
\textbf{C}(\hat{n})\equiv
\begin{pmatrix}
C^{+}(\hat{n}) & -C^{-}(\hat{n}) \\
C^{-}(\hat{n}) & C^{+}(\hat{n})
\end{pmatrix} 
\label{eq:cov_QU} 
 \end{equation}	
with explicit components: 
 \begin{align}
C^{+}_{\nu\nu^{\prime}}(\hat{n})=\langle \textrm{Q}_{\nu}(\hat{n}^{\prime})\textrm{Q}_{\nu^{\prime}}(\hat{n}^{\prime})+\textrm{U}_{\nu}(\hat{n}^{\prime})\textrm{U}_{\nu^{\prime}}(\hat{n}^{\prime})\rangle_{\hat{n}^{\prime}\in \mathcal{D}_{\hat{n}}} \\
C^{-}_{\nu\nu^{\prime}}(\hat{n})=\langle \textrm{Q}_{\nu}(\hat{n}^{\prime})\textrm{U}_{\nu^{\prime}}(\hat{n}^{\prime})-\textrm{U}_{\nu}(\hat{n}^{\prime})\textrm{Q}_{\nu^{\prime}}(\hat{n}^{\prime})\rangle_{\hat{n}^{\prime}\in \mathcal{D}_{\hat{n}}}
\end{align}	
and $\lambda_{R}$, $\lambda_{I}$ are the Lagrange multipliers:
\begin{equation}
\frac{\lambda_{R}(\hat{n})}{2}= \frac{S_{+}(\hat{n})}{S_{+}^{2}(\hat{n})-S_{-}^{2}(\hat{n})}  
\qquad 
\text{and} 
\qquad 
\frac{\lambda_{I}(\hat{n})}{2}= \frac{-S_{-}(\hat{n})}{S_{+}^{2}(\hat{n})-S_{-}^{2}(\hat{n})}
\label{eq:ILC_QU_lambda} 
\end{equation}
with:
\begin{equation}
S_{+}(\hat{n}) \equiv  \mathbf{A}^{T} (C^{+})^{-1}(\hat{n}) \mathbf{A} \qquad 
\text{and} 
\qquad 
S_{-}(\hat{n}) \equiv  \mathbf{A}^{T} (C^{-})^{-1}(\hat{n}) \mathbf{A}.
\label{eq:ILC_QU_lambda_1} 
\end{equation}
PILC can be applied, both in pixel and needlet domain, to spin-$2$ fields and therefore just to polarization maps. Once PILC has been performed, the \texttt{BROOM} package can return either cleaned $QU$ or $EB$ maps. It can also consider in the minimization just the $E$- or $B$-family of the Stokes parameters, i.e. $QU$ maps only sourced by $E$ or $B$ modes. A subcase of PILC, referred to as Polarization Real ILC (PRILC) and also implemented in the package, consists of assuming the component-separation weights to be purely real, thereby neglecting the second constraint in Equation~\ref{eq:ILC_QU_w}. In this case, the PRILC weight estimation reduces to the standard ILC formulation, with the data covariance given solely by the $C^{+}$ block of Equation~\ref{eq:cov_QU}.

\paragraph{Constrained ILC (cILC).}

ILC and PILC techniques allow for the reconstruction of a signal without making any assumption about the spectral properties of the contaminants, except that they are uncorrelated with the signal of interest. However, additional constraints can be imposed on the linear combination so that the weights explicitly deproject selected contaminating components, if their SEDs are known or well characterized. If one wants to partially or totally deproject $n$ components with SEDs $\mathbf{f}_{1},\ \mathbf{f}_{2},\dots,\mathbf{f}_{n}$, the overall constraints on the weights estimation are:
\begin{equation}
    \begin{cases}
    \min_{\textbf{w}}\, \text{Var}\!\left[\mathcal{X}_{\text{ILC}}(\hat{n})\right] \\
    \textbf{w}^{T}(\hat{n})\cdot \textbf{A}_{\text{signal}} = 1 \\
    \textbf{w}^{T}(\hat{n})\cdot \textbf{f}_{1}=\epsilon_{1} \\
    \textbf{w}^{T}(\hat{n})\cdot \textbf{f}_{2}=\epsilon_{2} \\
    \vdots \\
    \textbf{w}^{T}(\hat{n})\cdot \textbf{f}_{n}=\epsilon_{n}
    \end{cases}\,,
\label{eq:w_constraints}
\end{equation}
where $\epsilon_1,\epsilon_2,\ldots,\epsilon_{n}$ denote the desired residual amount of each component in the final solution, with $\epsilon=0$ corresponding to full deprojection. This prescription defines the constrained ILC (cILC) \cite{cILC}. The weights satisfying the constraints in Equation~\ref{eq:w_constraints} can be computed in closed form:
\begin{equation} \textbf{w}_{\textrm{cILC}}(\hat{n})=\textbf{e}^{T}\left[\textbf{A}^{T}C^{-1}(\hat{n})\,\textbf{A}\right]^{-1}\textbf{A}^{T}C^{-1}(\hat{n})\,. \label{eq:cMILC_weights} \end{equation}
where $C$ is the data covariance matrix introduced in Equation~\ref{eq:data_cov}, and $\mathbf{A}$ is the mixing matrix collecting the SEDs of the signal to be reconstructed and of the components to be (partially or fully) deprojected:
\begin{equation*} \textbf{A}=\left(\textbf{A}_{\text{CMB}}\quad \textbf{f}_{1} \quad \textbf{f}_{2} \quad \ldots \quad \textbf{f}_{n}\right).
\end{equation*}
The vector $\mathbf{e}$ encodes the deprojection coefficients corresponding to Equation~\ref{eq:w_constraints}, namely
\begin{equation} \textbf{e} = (1\quad \epsilon_{1} \quad \epsilon_{2} \quad \ldots \quad \epsilon_{n}). 
\end{equation}
In this framework, the standard ILC appears as a special case of cILC, where only the first element of $\boldsymbol{e}$ and the first column of $\mathbf{A}$ are retained in Equation~\ref{eq:cMILC_weights}, leading to Equation~\ref{eq:ilc_w}.

Another specific subcase of the cILC approach is the constrained Moments ILC (cMILC), introduced in \cite{cMILC}, where the CMB primary anisotropies are reconstructed and statistical moments of foreground emission are fully deprojected to reduce leakage of foregrounds into the CMB solution.
The general outcome of applying cMILC is a reduction in foreground contamination, achieved at the expense of increased reconstruction noise \cite{cMILC}.

The current implementation of cMILC within \texttt{BROOM} allows for the deprojection of moment expansions (up to second order in all parameters) for sky signals whose fundamental SED follows either a MBB (see Equation \ref{eq:dust_exp}) or a power law (see Equation \ref{eq:sync_exp}). Moment expansions for additional components with different basic SEDs may be included in future releases. 

The cMILC technique has been further extended in \cite{ocMILC} with the introduction of the optimized cMILC (ocMILC) pipeline. This method employs a fully data-driven strategy to automatically determine, as a function of angular scale, the optimal number of moments to be deprojected before the reconstruction becomes dominated by noise. In addition, ocMILC selects the optimal combination of moments, along with their associated deprojection coefficients in Equation~\ref{eq:w_constraints}. Although the current release of \texttt{BROOM} does not yet include an ocMILC module, its implementation is planned for future versions of the package.

In addition to cMILC, which improves the recovery of primary CMB anisotropies through an enhanced mitigation of foreground contamination, cILC can be employed, for instance, to improve the reconstruction of the tSZ signal by deprojecting Galactic and extragalactic components, or even the primary CMB anisotropies themselves \cite{2011MNRAS.410.2481R,2023MNRAS.526.5682C,tsz_PR4}.

The cILC approach can be applied both to spin-0 fields ($T$, $E$, and $B$ modes) and to the spin-2 polarization field ($Q,U$), in which case it is referred to as cPILC or cPRILC (depending on the nature of the weights) in \texttt{BROOM} \cite{Adak2021}.

\paragraph{Multi-Clustering ILC (MC-ILC).} 
As discussed earlier, the ILC reconstructs the signal of interest by minimizing the variance of the combined map, relying on local estimates of the data covariance (through the choice of $\mathcal{D}_{\hat{n}}$ in Equation~\ref{eq:data_cov}). However, this approach may become suboptimal in the presence of complex foregrounds with strong spatial variability across the sky. To address this challenge within the framework of model-independent ILC, the Multi-Clustering ILC (MC-ILC) method was introduced \cite{MCNILC}. MC-ILC optimizes the variance minimization principle by grouping together sky pixels that exhibit similar foreground spectral properties. The covariance is then estimated independently within each cluster, effectively replacing Equation~\ref{eq:data_cov} with:
\begin{equation}
    C_{\nu\nu^{\prime}}(i) 
    = \left\langle \mathcal{X}_{\nu}(\hat{n}^{\prime})\,
      \mathcal{X}_{\nu^{\prime}}(\hat{n}^{\prime}) \right\rangle_{\hat{n}^{\prime}\in \mathcal{D}_{i}},
    \label{eq:data_cov_mcilc}
\end{equation}
where $\mathcal{D}_{i}$ denotes a specific sky patch defined by homogeneous foreground properties.

In \cite{MCNILC}, it was demonstrated that an effective tracer of the spatial variability of foreground spectral properties is given by the ratio of two observed frequency channels: one at high frequency, which is dominated by dust emission, and one at a so-called “CMB channel,” thereby capturing the interplay between dust and synchrotron contributions in polarization. A detailed discussion of this tracer can be found in \cite{MCNILC}. Within \texttt{BROOM}, this tracer can be constructed in two ways: (i) in an idealized setting, where the user provides input maps of foreground-only emission, allowing validation of the method on controlled simulations; or (ii) in a more realistic scenario, where the tracer is derived directly from the data after applying a denoising procedure based on the Generalized ILC (GILC), which will be introduced in Section~\ref{sec:fgds_rec}.

Once the tracer has been constructed, sky partitioning for the computation of the data covariance (Equation~\ref{eq:data_cov_mcilc}) can be carried out using two approaches:
(i) \textit{Clusters with Equal Area} (CEA), in which the sky is divided into patches each containing the same number of pixels; and
(ii) \textit{Random Partitioning} (RP), in which multiple partitions are generated from the same tracer and for each partition a random number of pixels is assigned to each cluster. In the RP approach, variance minimization is performed independently for each partition, and the resulting solutions are subsequently averaged. This averaging naturally mitigates border effects that may arise in the CEA scheme.

At present, MC-ILC has been shown to be particularly effective in low signal-to-noise regimes, such as in the extraction of CMB polarization $B$ modes. For this reason, its implementation within the \texttt{BROOM} package, in the framework of realistic tracers, is currently restricted to this specific reconstruction case.

\subsubsection{Foreground reconstruction and diagnostic}
\label{sec:fgds_rec}
In this section, we describe the methodologies implemented in \texttt{BROOM} for handling Galactic foregrounds. These include: (i) blind reconstruction of foreground emission using the GILC technique, (ii) diagnostics of foreground complexity, and (iii) derivation of templates of foreground residuals contaminating the final solution after applying one of the component separation techniques introduced in Section~\ref{sec:cmb_rec}.

\paragraph{Generalized ILC (GILC).}
The Generalized ILC (GILC) is a multidimensional extension of the ILC method, designed to reconstruct signals across all observed frequency channels without requiring any assumptions about their spectral properties. The only necessary input is the knowledge of the two-point statistics of the contaminating components to be deprojected \cite{GNILC_intro}. Thanks to this property, GILC is widely used in CMB data analysis as a model-independent tool for recovering foreground emission across frequency channels. For example, it has been successfully applied in the past to reconstruct the intensity of thermal dust emission at \textit{Planck} frequencies \cite{GNILC} or to construct the MC-ILC foreground tracer for optimal sky partitioning \cite{MCNILC}.

GILC yields a set of multifrequency templates, $\mathcal{X}_{\text{GILC}}$, defined at the same frequencies as the input data set, which capture all independent modes in the data whose power dominates over that of the provided nuisance components. They are obtained by linearly combining the $N_{\nu}$ input maps $\mathcal{X}$:
\begin{equation}
    \mathcal{X}_{\text{GILC}}(\hat{n})=W_{\text{GILC}}(\hat{n}) \cdot \mathcal{X}(\hat{n}).
\label{eq:gnilc_f}
\end{equation}
where $W_{\text{GILC}}$ is an $N_{\nu}\times N_{\nu}$ weight matrix defined for each pixel. The weights are designed to provide unit response to all not nuisant modes while minimizing the variance of $\mathcal{X}_{\text{GILC}}$. As shown in \cite{GNILC}, the optimal weight matrix can be expressed as:
\begin{equation}
    W_{\text{GILC}}(\hat{n})=\left[F(F^TC^{-1}{F})^{-1}F^TC^{-1}\right](\hat{n}),
\label{eq:w_gnilc}
\end{equation}
where $C$ denotes the empirical data covariance matrix defined in Equation~\ref{eq:data_cov}, and $F$ is the mixing matrix encoding the spectral scaling of the foreground components to be reconstructed. Both $C$ and $F$ are, in general, spatially varying. The key objective of GILC is thus to obtain a blind estimate of $F$ directly from the data, without imposing parametric assumptions on the spectral properties of the signals to be preserved, e.g. of the foreground emission.

The empirical covariance matrix $C$ can be decomposed as the sum of the covariance of the signal of interest, namely the microwave foreground emission ($C_{f}$), and a \emph{nuisance} term ($C_{N}$) that accounts for all components to be deprojected (e.g.~CMB, instrumental noise):
\begin{equation}
    C(\hat{n}) = C_{f} (\hat{n}) + C_{N}(\hat{n}),
\label{eq:cov_gnilc}
\end{equation}
under the assumption that the two contributions are uncorrelated.
The foreground covariance can be expressed as:
\begin{equation}
    C_{f}(\hat{n})=F (\hat{n})C_{t}(\hat{n})F^T(\hat{n}),
\label{eq:R_f}
\end{equation}
where $C_{t}=\langle \mathcal{X}_{t}\mathcal{X}_{t}^T\rangle$ and $\mathcal{X}_{t}$ denotes a set of independent (i.e.\ uncorrelated, not necessarily physical) templates that serve as a basis for describing the correlated full foreground signal \cite{GNILC_intro, ocMILC}.  Within \texttt{BROOM}, the set of nuisance components considered in a GILC run can be defined by the user. By default, this includes the CMB and instrumental noise.

Once both $C$ (from the data) and $C_{N}$ (from simulations and prior knowledge of the nuisance statistical properties) are available, we can construct the whitened covariance matrix as:
\begin{equation}
\tilde{C}(\hat{n}) = \left( C_{N}^{-1/2} C C_{N}^{-1/2} \right)(\hat{n})\simeq \left(C_{N}^{-1/2} C_{f} C_{N}^{-1/2}\right)(\hat{n}) + \mathds{1}\,,
\label{eq:whitened_covar}
\end{equation}
where the contribution of the nuisance components is encoded in the identity matrix $\mathds{1}$.

The whitened empirical covariance can be diagonalized as:
\begin{equation}
    \tilde{C}(\hat{n})=\left[\textbf{U}_{s}D_{s}\textbf{U}_{s}^T+\textbf{U}_{N}\textbf{U}_{N}^T\right](\hat{n}),
\label{eq:gnilc_diag}
\end{equation}
where $\mathbf{U}_{s}$ denotes the set of eigenvectors associated with the modes of input components which are not included in the nuisance covariance. The corresponding eigenvalues are collected in the diagonal matrix $D_{s}$:
\begin{equation}
    D_s(\hat{n})=\textrm{diag}\left[\lambda_1^{(s)}(\hat{n}),...,\lambda_{m(\hat{n})}^{(s)}(\hat{n})\right]
\label{eq:D_s}
\end{equation}
and all of them are expected to be significantly greater than unity, corresponding to excess variance above the nuisance floor.  
The dimension of this subspace (i.e.\ the rank of $\mathbf{U}_{s}D_{s}\mathbf{U}_{s}^T$) is $m$, which gives the number of independent templates in $\mathcal{X}_{t}$ required to fully describe the non-nuisance signal.  
The matrix $\mathbf{U}_{N}$ in Equation~\ref{eq:gnilc_diag} instead refers to the eigenvectors spanning the nuisance subspace.

From Equation \ref{eq:gnilc_diag} and the orthonormality condition, $\textbf{U}_{s}\textbf{U}_{s}^T+\textbf{U}_{N}\textbf{U}_{N}^T=\mathds{1}$,
the non-nuisance covariance term in Equation~\ref{eq:cov_gnilc} can be written as:
\begin{equation}
    C_{f}(\hat{n})=C_{N}^{1/2}(\hat{n})\left(\textbf{U}_{s}(\hat{n})\left(D_s(\hat{n})-\mathds{1}\right)\textbf{U}_{s}(\hat{n})^T\right)C_{N}^{1/2}(\hat{n}),
\end{equation}
which, when compared to Equation~\ref{eq:R_f}, allows one to obtain an estimate of the non-nuisance signal mixing matrix $\hat{F}$:
\begin{equation}
    \hat{F}(\hat{n})=C_{N}^{1/2}(\hat{n})\textbf{U}_{s}(\hat{n}).
\end{equation}
This estimated mixing matrix $\hat{F}$ can then be used to compute the GILC weights $W_{\text{GILC}}$ from Equation~\ref{eq:w_gnilc}, and thereby perform a blind reconstruction of the sky emission of interest across the observed frequency channels with reduced nuisance contamination via Equation~\ref{eq:gnilc_f}. We recall that the nuisance term in Equation~\ref{eq:cov_gnilc} can also include components beyond CMB and noise, such as, for instance, the CIB covariance. This allows, in particular at high frequencies, for a better separation of Galactic non-thermal dust emission from that originating in extragalactic sources \cite{GNILC}.

Foreground reconstruction with the GILC technique can be applied both to spin-0 scalar fields, such as $T$, $E$, or $B$ modes, using in Equation \ref{eq:cov_gnilc} the data covariance defined in Equation~\ref{eq:data_cov}, and to spin-2 fields (the Stokes parameters $Q$ and $U$), using the polarization covariance of Equation~\ref{eq:cov_QU}. The latter case corresponds to the GPILC implementation available within the \texttt{BROOM} package. Moreover, it can be performed either in pixel- or in needlet-space.

\paragraph{Diagnostic of foreground complexity.} 

The implementation of the GILC technique naturally provides a diagnostic tool for assessing the complexity of foreground emission across the sky, and—if applied in the needlet domain—also as a function of angular scales. Indeed, as shown in Equation~\ref{eq:D_s}, the diagonalization of the whitened data covariance matrix yields an estimate of the number $m$ of independent templates required to fully describe the non-nuisance signal at each sky location (see Equation \ref{eq:D_s}). Such a value, for each pixel (i.e. line of sight), is derived by minimizing the Akaike Information Criterion (AIC) \cite{GNILC}:
\begin{equation}
    \mathrm{AIC}(m,\hat{n}) = 2m +  \sum_{k=m+1}^{N_{\nu}} \left[ \lambda_k(\hat{n}) - \log \lambda_k(\hat{n}) - 1 \right],
\label{eq:AIC}
\end{equation}
with $N_{\nu}$ the number of observed frequency channels and $[\lambda_1,\dots,\lambda_{N_{\nu}}]$ the set of eigenvalues of the whitened data covariance $\tilde{C}$ introduced in Equation \ref{eq:whitened_covar}.
The obatined maps $m(\hat{n})$ directly provide a robust and model-independent estimate of the complexity and structure of the non-nuisance emission across the sky. Consequently, within the \texttt{BROOM} package it is possible not only to reconstruct the foreground emission at the observed frequency channels using GILC, but also to return diagnostic maps of the foreground subspace rank.  

Such diagnostic information can in principle provide valuable feedback to map-based component separation pipelines, guiding them on the level of complexity that must be accounted for across the sky. For instance, this strategy is adopted in the ocMILC pipeline \cite{ocMILC} to determine the optimal number of foreground moments to deproject across different sky regions and needlet scales. Analogously to the GILC approach, this diagnostic can be implemented in either pixel space or needlet space and applied to any scalar field ($T$, $E$, and/or $B$), or to the polarization intensity.

Complementary diagnostics of foreground complexity have also been explored in other works \cite{2025ApJS..276...45L,2026ApJS..283...80L}.

\paragraph{Template of component-separation foreground residuals}

In addition to delivering a valuable set of cleaned foreground maps (useful for Galactic and extragalactic astrophysics studies) \cite{GNILC_intro, pysm2025}, GILC provides a cleaned tracer of foreground spectral properties for MC-ILC \cite{MCNILC}, as well as a diagnostic of foreground complexity that can be exploited within multiple component separation pipelines. Moreover, GILC forms the foundation of the procedure presented in \cite{Carones_marg}, which derives a spectral template of foreground residuals after component separation. 

After applying GILC to the same data set used for CMB reconstruction, and adopting a nuisance covariance that includes both instrumental noise and the CMB signal, the resulting output consists of foreground emission templates, $\mathcal{X}^{f}_{\text{GILC}}$, which represent the actual contaminants present in the observations targeted by CMB-reconstruction methodologies. These cleaned templates, with reduced noise and CMB leakage, are then combined with the component separation weights employed to recover the cleaned CMB signal:
\begin{equation}
    \tilde{\mathcal{X}}^{f}(\hat{n}) 
   = \sum_{\nu} w_{\nu}(\hat{n})\, \mathcal{X}^{f}_{\nu,\text{GILC}}(\hat{n}).
\label{eq:fgd_res}
\end{equation}
The resulting combined map, $\tilde{\mathcal{X}}^{f}$, provides a template of the residual foregrounds that remain in the cosmological solution. As demonstrated in \cite{Carones_marg}, once properly debiased from noise reconstruction, the angular power spectrum of this template closely matches that of the true foreground residuals estimated from realistic simulations. It can therefore be included in the cosmological likelihood to account for Galactic contamination and to debias parameter estimation. This procedure is particularly relevant in analysis cases where significant residual foreground contamination is expected, such as constraints on the lensing amplitude and the tensor-to-scalar ratio from $B$ modes \cite{PTEP, Wolz2024}, as well as large-scale inference of the optical depth to reionization $\tau$ from $E$ modes.

This procedure for deriving a spectral template of foreground residuals from component separation has been shown to be effective across different methods and masking strategies, marking an important step forward for cosmological inference from CMB maps with significant foreground contamination. It is implemented in \texttt{BROOM} and can be applied to any of its component separation techniques---including ILC, PILC, MC-ILC, cILC, and cPILC---making it suitable for both spin-$0$ and spin-$2$ fields on the sphere.

\subsubsection{Formalism of needlet filtering}
\label{sec:needlets}
All component separation routines within the \texttt{BROOM} package can be implemented either in pixel space (\textit{i.e.}, accounting for all angular scales simultaneously) or in needlet space, which allows provided data to be processed separately in different ranges of angular scales. This last option, for instance, enhances component separation, as different contaminants typically dominate in different multipole ranges and can therefore be addressed separately. In this section, we provide a concise overview of needlet decomposition, serving as the theoretical background for the needlet filtering implemented in the \texttt{BROOM} package.

Needlets were first introduced in the statistical literature by \cite{Baldi2006} and subsequently applied to CMB data analysis in \cite{Pietrobon2006}. They possess several useful properties \citep{Marinucci2008}: (i) they do not rely on the tangent-plane (\textit{i.e.}, local flat-sky) approximation; (ii) they admit a simple reconstruction formula, with the same needlet functions appearing in both the forward and inverse transforms; and (iii) their pixel-space window function decays faster than any polynomial.  

The spherical needlet kernel $\psi_{jk}$ is defined as:
\begin{equation}
	\psi_{jk}(\hat{n})=\sqrt{\lambda_{jk}}\sum_{\ell}b_{j}(\ell)\left(\sum_{m=-\ell}^{\ell}\textrm{Y}_{\ell m}^{*}(\hat{n})\cdot \textrm{Y}_{\ell m}(\xi_{jk})\right),
\label{eq:Psi}
\end{equation}
where $\lambda_{jk}$ and $\xi_{jk}$ denote, respectively, the cubature weights, which can be identified in practice with the typical pixel area, $4\pi/N_{\textrm{pix}}$, and the pixel centers of the map.  
The key ingredient in constructing needlet wavelets in Equation~\ref{eq:Psi} is the choice of the harmonic window function $b_j(\ell)$. The width of $b_j(\ell)$ in harmonic space determines the localization properties of $\psi_{jk}$ in real space: the broader $b_j(\ell)$ is in multipole space, the more compact the corresponding kernel $\psi_{jk}$. The index $j$ labels a specific needlet scale and thus corresponds to a well-defined multipole range, with lower values of $j$ probing large angular scales and higher values of $j$ progressively probing smaller angular scales.

Three main needlet constructions are commonly adopted in the literature:  
\begin{enumerate}
    \item \emph{Standard needlets} \cite{doi:10.1137/040614359}, whose harmonic filter is defined as:
    \begin{equation}
    b_{j}(\ell)
    = \sqrt{\;\phi\!\left(\frac{\ell}{B^{j+1}}, B\right)
    - \phi\!\left(\frac{\ell}{B^{j}}, B\right)}\,,
\end{equation}
where the tapering function $\phi:[0,\infty)\to[0,1]$ is given by
\begin{equation}
\phi(l, B) =
\begin{cases}
1, & 0\le l \le \dfrac{1}{B}, \\[8pt]
\psi\!\left(1-\dfrac{2B}{B-1}\Big(l-\dfrac{1}{B}\Big)\right), & \dfrac{1}{B} < l < 1, \\[12pt]
0, & l \ge 1,
\end{cases}
\end{equation}
with
\begin{equation}
    \psi(u) = \dfrac{\displaystyle\int_{-1}^{\,u} \exp{\left[1/(t^2-1)\right]}\,dt}
    {\displaystyle\int_{-1}^{\,1} \exp{\left[1/(t^2-1)\right]}\,dt}\,.
\end{equation}
    Standard filters have compact support in $\ell \in [B^{j-1}, B^{j+1}]$.
    \item \emph{Mexican needlets}, where $b_j(\ell)$ is defined through Gaussian-related weights \cite{mexican}:
    \begin{equation}
    b_{j}(\ell)\propto\left(\frac{\ell}{B^{j}}\right)^{p}\cdot \exp\left(-\frac{1}{2}\cdot\Bigg(\frac{\ell}{B^{j}}\Bigg)^{2}\right),
    \label{eq:b_mex}
    \end{equation}
    with $p$ an additional free parameter (typically $p=1$). Unlike standard needlets, mexican needlets do not have bounded harmonic support, but in practice their Gaussian tails ensure excellent localization both in multipole and pixel space.
    
    \item \emph{Cosine needlets}, with bandpass filters defined as:
\begin{equation}
   b_{j}(\ell) = \begin{cases}
       \cos\left( \frac{\ell_{\rm peak}^j\,  -\, \ell}{\ell_{\rm peak}^j\,  -\, \ell_{\rm peak}^{j-1}}\frac{\pi}{2}\right)   & \text{if}\ \ell_{\rm peak}^{j-1}\leq \ell < \ell_{\rm peak}^j \\
       \\
	\cos\left( \frac{\ell\, -\, \ell_{\rm peak}^j}{\ell_{\rm peak}^{j+1}\, -\, \ell_{\rm peak}^j}\frac{\pi}{2}\right)   & \text{if}\ \ell_{\rm peak}^j\leq \ell < \ell_{\rm peak}^{j+1}
   \end{cases}\,,
\end{equation}
where $\ell^{j}_{\rm peak}$ denotes the multipole at which the $j$-th needlet band is peaked. Like standard needlets, cosine filters have compact support in multipole space, but their pixel-space localization is somewhat less sharp compared to Mexican needlets. Nevertheless, cosine needlets have been widely used in the analysis of CMB data and simulations \cite{Planck2018_compsep, SO_2019}.
\end{enumerate}

Needlets form a \emph{tight frame}. A tight frame is a countable set of functions $\eta_{j}$, defined on the sphere, such that for any square-integrable field $f\in L^{2}(\mathbb{S}^{2})$ one has:
\begin{equation*}
\sum_{j}\langle f,\eta_{j}\rangle^{2}\equiv \int_{\mathbb{S}^{2}}f(\hat{n})^{2}d\Omega,
\end{equation*}
where $\langle f,\eta_{j}\rangle$ denotes the scalar product.  
Thanks to this property, any scalar field $X(\hat{n})$ defined on the sphere can be decomposed as:
\begin{equation}
X(\hat{n})=\sum_{j,k}\beta_{jk}\cdot \psi_{jk}(\hat{n}),
\label{eq:needletsdT}
\end{equation}
with $\beta_{jk}$ the needlet coefficients, defined by:
\begin{equation}
\begin{aligned}
\beta_{jk}=&\int_{S^{2}}X(\hat{n})\cdot\psi_{jk}(\hat{n})d\Omega= \\
=&\sqrt{\lambda_{jk}}\sum_{\ell}b_{j}(\ell)\left(\sum_{m=-\ell}^{\ell}\left\{\int_{S^{2}}X(\hat{n})\cdot \textrm{Y}_{\ell m}^{*}(\hat{n})d\Omega\right\}\textrm{Y}_{\ell m}(\xi_{jk})\right)= \\
=&\sqrt{\lambda_{jk}}\sum_{\ell}\sum_{m=-\ell}^{\ell}\left(b_j(\ell)\cdot a^{X}_{\ell m}\right)\textrm{Y}_{\ell m}(\xi_{jk}),
\end{aligned}
\end{equation}
where in the second step Equation~\ref{eq:Psi} has been used.  

In practice, needlet filtering corresponds to applying the weighting function $b_j(\ell)$ to the harmonic coefficients $a_{\ell m}$ of a map for each needlet scale $j$. The needlet coefficients $\beta_{jk}$ then correspond to a map reconstructed from these filtered harmonic terms. Equivalently, in real space, this procedure amounts to convolving the map with the needlet function $\psi_{jk}$ associated with the filter $b_{j}(\ell)$.

To reconstruct a field in real space from a set of needlet coefficients, an inverse needlet transform must be applied. This procedure consists of convolving, for each needlet scale $j$, the corresponding needlet coefficients $\beta_{jk}$ with the kernel $\psi_{jk}$ and summing the resulting contributions over all needlet scales \cite{Marinucci2008}:
\begin{equation}
    X(\hat{n}) = \sum_{jk}\beta_{jk}\ast \psi_{jk}(\hat{n}).
\label{eq:inv_needlet}
\end{equation}
Equation \ref{eq:inv_needlet} leads to the main requirement for the construction of $b_j(\ell)$ (satisfied by all introduced needlet functions):
\begin{equation}
    \sum_{j=1}^{\infty}b_{j}^{2}(\ell) = 1\qquad \forall \ell.
\end{equation}

Owing to their localization properties, needlets are particularly well suited for analyzing data sets with incomplete sky coverage, as demonstrated in \cite{NILC_cutsky}. In such cases, the needlet coefficients are only mildly affected by the presence of masks or missing data.

In needlet-based component separation, the input maps are first filtered with a user-defined set of needlet bands, yielding for each scale $j$ a corresponding set of multifrequency needlet coefficient maps $\beta^{\nu}_{jk}$. At each scale, a minimum-variance linear combination is then applied independently. Finally, the different scale-dependent solutions are recombined through Equation~\ref{eq:inv_needlet}, producing the final real-space map that consistently incorporates all angular scales. Within the framework of standard and Mexican needlets, adjacent needlet bands—constructed using specific choices of the hyperparameters ($B$ and $p$)—can be combined into broader bands through the following equation:
\begin{equation}
b^{\textit{new}}(\ell) = \sqrt{\sum_{j=j_{\min}}^{j_{\max}} b_j^{2}(\ell)} .
\label{eq:b_merge}
\end{equation}
This procedure effectively enlarges the harmonic windows, thereby ensuring that each resulting needlet band contains a sufficient number of harmonic modes for robust sampling of the data covariance during the component-separation stage.
\subsection{Computation of angular power spectrum}
\label{sec:spectra}

Since CMB anisotropies are predicted to be Gaussian (to leading order) and statistically isotropic \cite{Planck2020_I&S,2024MNRAS.527..756C}, the angular power spectrum is in most cases a sufficient statistical probe to extract the full cosmological information from CMB data. The spectrum quantifies the variance of the observed CMB spin-0 harmonic coefficients at each multipole $\ell$. Within \texttt{BROOM}, the angular power spectrum can be estimated either using the \texttt{healpy} function \texttt{anafast} \cite{healpy} or through the \texttt{MASTER} formalism \cite{MASTER,MASTER2}, as implemented in the \texttt{NaMaster} package \cite{pymaster}.

The observed angular power spectrum is empirically derived as:
\begin{equation}
 \tilde{C}_{\ell}^{XY} = \frac{1}{2\ell + 1} \sum_{m=-\ell}^{\ell} a^{X\ast}_{\ell m}\, a^{Y}_{\ell m},
 \label{eq:pseudo_cls}
\end{equation}
where $a_{\ell m}$ are estimated harmonic coefficients from a map and $X$ and $Y$ are two spin-$0$ fields with $X=Y$ for auto power spectrum. In the current \texttt{BROOM} implementation $X$ and $Y$ can be $T$, $E$ or $B$.

The computation of angular power spectra with \texttt{anafast}, as implemented in \texttt{BROOM}, automatically accounts for common observational effects, including beam smoothing, the pixel window function (when applicable), and the overall bias introduced by masking.

Some masking-induced effects are not corrected when using \texttt{anafast}, such as correlations between multipoles and $EB$-leakage (already introduced in Section~\ref{sec:basics}), with the latter arising from the estimation of angular power spectra from Stokes $Q$ and $U$ maps. These effects are properly treated within the \texttt{MASTER} power spectrum estimation. For this reason, we recommend the use of the \texttt{anafast} approach only in cases of full-sky analyses or when the applied mask is minimal.

In the \texttt{MASTER} implementation, all distortions affecting the angular power spectrum---including those induced by smoothing and sky masking---are properly taken into account. The corrected power spectrum is obtained by applying the inverse of the so-called \textit{mode-coupling} matrix to the observed angular power spectrum of Equation \ref{eq:pseudo_cls}:
\begin{equation}
    C_{\ell}^{XY} = \left[\left(M_{\ell\ell^{\prime}}^{s_{X}s_{Y}}\right)^{-1}\right]_{XY}\tilde{C}_{\ell}^{XY}.
\label{eq:cls_master}
\end{equation}
Here, the explicit form of the matrix $M_{\ell\ell'}^{s_X s_Y}$ depends on the spins of the fields from which the harmonic coefficients $a^{X}_{\ell m}$ and $a^{Y}_{\ell m}$ are derived and of course on the employed masks. The different forms of $M_{\ell\ell'}^{s_X s_Y}$ can be found in \cite{pymaster} and references therein. 

When the angular power spectra of $E$ and/or $B$ modes are computed from Stokes $Q$ and $U$ maps, the formalism of Equation~\ref{eq:cls_master} corrects the mean bias associated with $EB$-leakage. However, such leakage still propagates into the estimator’s uncertainty and therefore ambiguous $E$ modes leak into the variance of the estimated $C_{\ell}^{BB}$. To mitigate this, a correction at the level of the harmonic coefficients is required. In \texttt{BROOM}, this is implemented through the \texttt{NaMaster} routines, which apply the purification technique \cite{Bunn2003,lewis_EBleak,Smith2006} prior to computing the pseudo-$C_{\ell}$s. We note that, when purification is applied, the form of the mode-coupling matrix must be slightly modified, as described in Equation~35 in \cite{pymaster}.

In several cases, the mode-coupling matrix cannot be inverted multipole by multipole, and the estimated angular power spectrum must therefore be binned into bandpowers:
\begin{equation}
    C_{\ell_b} = \frac{1}{\ell^{(b)}_{\rm max}-\ell^{(b)}_{\rm min}}\sum_{\ell=\ell^{(b)}_{\rm min}}^{\ell^{(b)}_{\rm max}} C_{\ell},
\label{eq:binning}
\end{equation}
where $\ell^{(b)}_{\rm min}$ and $\ell^{(b)}_{\rm max}$ denote the minimum and maximum multipoles defining the band.
In the current \texttt{BROOM} implementation, two binning schemes are available: either averaging the spectrum over bins of constant width $\Delta\ell$ (default), or specifying an array of $\ell^{(b)}_{\rm min}$ values to define custom bandpower intervals.

Binning can be applied to power spectra computed either with \texttt{anafast} or with \texttt{NaMaster}. In the former case, since no mode-coupling matrix is involved, binning primarily serves to reduce the statistical scatter of the bandpowers. In the latter case, binning is generally required to enable a stable inversion of the mode-coupling matrix.

\section{The \texttt{BROOM} package}
\label{sec:broom}

In Section~\ref{sec:methods}, we introduced the theoretical framework on which the data-analysis and simulation tools implemented in the package are based. We now outline the structure of \texttt{BROOM} and its main routines. In particular, the package can be used to:

\begin{itemize}
\item[(i)] simulate or load microwave observations;
\item[(ii)] perform component separation;
\item[(iii)] compute angular power spectra;
\item[(iv)] estimate foreground residuals from data;
\item[(v)] propagate arbitrary sets of maps through a component-separation pipeline by combining them with the corresponding weights.
\end{itemize}



On the \texttt{BROOM} GitHub page\footnote{\url{https://github.com/alecarones/broom}}, users can find the full package documentation, including installation procedure and complete tutorials designed to guide them through the different applications of the implemented routines.

\subsection{Code structure and main routines}
\label{sec:broom_func}
The \texttt{BROOM} package is currently implemented entirely in \textit{python}. However, future performance enhancements may warrant the integration of a \textit{C++} or \textit{Fortran} backend to accelerate computationally intensive routines.

The package is available on \texttt{PyPI} and can be installed with the following command:
\begin{verbatim}
pip install cmbroom
\end{verbatim}
It builds upon several external libraries, which are automatically installed if not already available:
\begin{itemize}
\item \texttt{HEALPix}, introduced in Section~\ref{sec:basics};
\item \texttt{astropy}\footnote{\url{https://www.astropy.org/}}, used to handle astronomical and cosmological data;
\item \texttt{scipy}\footnote{\url{https://scipy.org/}}, employed for numerical routines and access to physical constants;
\item \texttt{numpy}\footnote{\url{https://numpy.org/}}, for numerical implementations and data manipulation;
\item \texttt{pysm3}\footnote{\url{https://pysm3.readthedocs.io/en/latest/}}~\cite{pysm,pysm3,pysm2025}, for simulating Galactic and extragalactic microwave-sky components (see Section~\ref{sec:sims} for details);
\item \texttt{mtneedlet}\footnote{\url{https://javicarron.github.io/mtneedlet/}}~\cite{mtneedlet}, used for generating some of the available needlet configurations.
\end{itemize}
In addition to the packages mentioned above, users who wish to perform map-based purification or compute angular power spectra using the \texttt{MASTER} approach must also install the \texttt{NaMaster} package. By default, the \texttt{BROOM} installation does not automatically install \texttt{NaMaster}; it must therefore be installed separately by the user.
Optional automatic installation may be possible, but it requires additional system libraries to be present. For detailed installation instructions and requirements, users are encouraged to consult the GitHub repository.

\begin{figure}
\centering
\includegraphics[width=.98\textwidth]{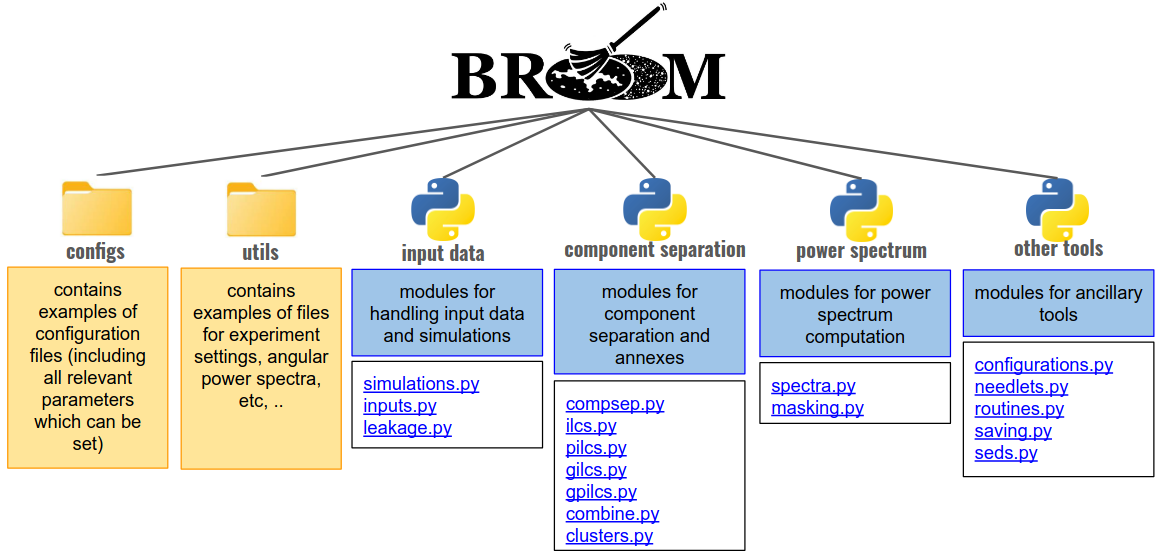}
\caption{Folders and main modules of the \texttt{BROOM} package.}
\label{fig:broom_chart}
\end{figure}

Once installed, \texttt{BROOM} can be imported into any python script as follows:
\begin{verbatim}
import broom
\end{verbatim}
The overall structure of the package is shown in Figure~\ref{fig:broom_chart}. The package is organized in different modules, which are grouped according to their different functionalities in the figure. Moreover, \texttt{BROOM} includes two additional folders (\texttt{configs} and \texttt{utils}) containing useful examples of needed files and ancillary data.

\begin{figure}
\centering
\includegraphics[width=.98\textwidth]{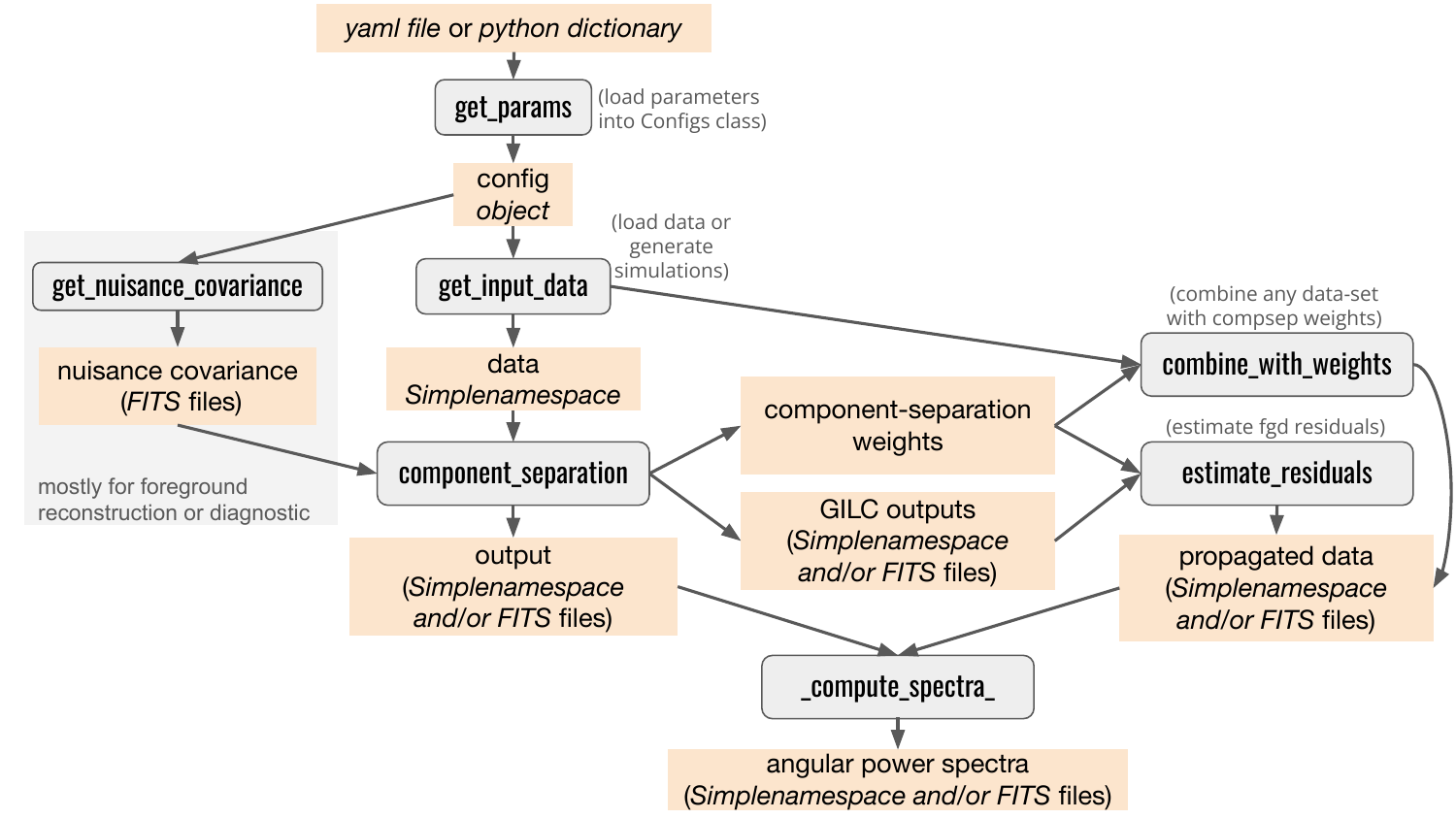}
\caption{Flowchart of the main \texttt{BROOM} functions and their corresponding outputs.}
\label{fig:broom_functions}
\end{figure}

Figure~\ref{fig:broom_functions} instead reports a summary flowchart illustrating the main \texttt{BROOM} functions and their corresponding required inputs and provided outputs. In the remaining part of this section we will provide further details of these key routines of the package.

\subsubsection{Loading or setting parameters: \texttt{get\_params}}

To execute most of the root-level functions in \texttt{BROOM}, a python dataclass containing all required configuration parameters must first be defined. This can be achieved in two ways.
The first one is to load the parameters from a \texttt{YAML} configuration file:
\begin{verbatim}
from broom import Configs, get_params

config: Configs = get_params(config_path=config_path)
\end{verbatim}
where \texttt{config\_path} is the full path to the aforementioned file.
Alternatively, users may directly initialize the dataclass within their python script by providing a dictionary of configuration values:
\begin{verbatim}
from broom import Configs

config_values = {
    'lmax': 256, 
    'nside': 128,
    .
    .
    .}
config = Configs(config=config_values)
\end{verbatim}
Example \texttt{YAML} configuration files are available in the \texttt{configs} folder of the \texttt{BROOM} GitHub repository. A complete list and detailed description of all user-configurable parameters is provided in Appendix~\ref{app:broom_pars}, as well as in the \texttt{config\_demo.yaml} file and in provided tutorials.

The configuration dataclass serves as the central input to all top-level functions in \texttt{BROOM}. The needed parameters can be broadly grouped into the following categories:
\begin{enumerate}
\item \textbf{General parameters}: used by most \texttt{BROOM} functionalities, such as \texttt{lmax}, \texttt{nside}, \texttt{coordinates}, and \texttt{instrument}.

\item \textbf{Input-data parameters}: used to generate simulations of CMB experiments or to load simulated or real data from disk. Examples include \texttt{generate\_input\_foregrounds}, \texttt{generate\_input\_cmb}, \texttt{generate\_input\_noise}, \texttt{fgds\_path}, and \texttt{noise\_path}.

\item \textbf{Component-separation parameters}: used to define the settings related to component separation and associated functionalities. Examples include \texttt{fwhm\_out}, \texttt{field\_out}, \texttt{compsep}, \texttt{compsep\_residuals}, and \texttt{compsep\_propagate}.

\item \textbf{Power-spectrum parameters}: used for angular power-spectrum computation, such as \texttt{field\_cls\_out}, \texttt{delta\_ell}, and \texttt{spectra\_comp}.

\end{enumerate}

To properly simulate microwave observations for a given CMB experiment or to perform specific pre-processing steps before component separation, information about the instrument under consideration must be provided. This information should be included as a dictionary within a separate \texttt{YAML} file. In the configuration file (or python dataclass), the full path to this file is specified via the \texttt{experiments\_file} parameter, while the label identifying the relevant instrument dictionary is provided through the \texttt{experiment} parameter. Specifying these two keywords automatically initializes an \texttt{instrument} dictionary within the configuration dataclass, containing all relevant (and potentially required) information about the selected experiment. Alternatively, the user can directly initialize the \texttt{instrument} dictionary within the python script as follows:
\begin{verbatim}
config.config["instrument"] =
    {"frequency": [...],
    "depth_I": [...],    
    "depth_P": [...],
    "fwhm": [...],
    "alpha_knee": [...],
    "ell_knee": [[...],[...]],
    "beams": "gaussian",
    "bandwidth": [...]}
config._store_passed_settings().
\end{verbatim}

A detailed description of the sub-parameters defining the instrumental configuration is provided in Appendix~\ref{app:broom_sims} and in the \texttt{config\_demo.yaml} file. Examples of instrument dictionaries can be found in the \texttt{experiments.yaml} file, located in the \texttt{utils} subdirectory of the \texttt{BROOM} package on GitHub. The same folder also includes default fiducial CMB power spectra, which will be used in case of need of generating simulations of CMB anisotropies.

\subsubsection{Loading or simulating input data: \texttt{get\_input\_data}}

The input data (and, optionally, the corresponding individual microwave components) to be provided to the component-separation routines can be either directly simulated (and optionally saved on disk) or loaded from disk (both for simulations and real observations) with the \texttt{get\_input\_data} function (implemented in the \texttt{simulations.py} module) as follows:
\begin{verbatim}
from broom import get_input_data
data = get_input_data(config, nsim = nsim)
\end{verbatim}
Here, \texttt{config} refers to the configuration dataclass, while \texttt{nsim} is an optional argument that specifies the label of the simulation to be generated or retrieved, when simulations are being used. In the case of loading actual data, there is no need to define this keyword.

The \texttt{get\_input\_data} function returns a \textsc{SimpleNamespace} object, which serves as the central data container for most \texttt{BROOM} functionalities. The \textsc{SimpleNamespace} class in python (from the \texttt{types} module) is a lightweight structure for dynamically storing attributes. It behaves similarly to a dictionary but allows access to entries using attribute notation rather than key indexing.

If simulation generation is requested, the returned output consists of a set of multifrequency maps for the three Stokes parameters ($T$, $Q$, and $U$), or alternatively of harmonic coefficients corresponding to the three scalar fields ($T$, $E$, and $B$), depending on the value of the \texttt{data\_type} parameter in the configuration dictionary (either \texttt{"maps"} or \texttt{"alms"}, respectively).

When data are instead loaded from disk, they may consist of any set or subset of maps (including directly $E$- and/or $B$-mode maps) or harmonic coefficients, without particular restrictions. However, the provided data must be consistent with the value specified in the \texttt{field\_in} parameter. All generated simulations can optionally be saved to disk as \texttt{.npy} files, which is also the expected format when loading simulated or real data from disk.

The choice between generating simulations and loading data is controlled by the parameters \texttt{generate\_input\_foregrounds}, \texttt{generate\_input\_cmb}, \texttt{generate\_input\_noise}, and \texttt{generate\_input\_data}. When any of these parameters is set to \texttt{False}, the corresponding component is loaded from disk using the paths specified by \texttt{fgds\_path}, \texttt{cmb\_path}, \texttt{noise\_path}, and \texttt{data\_path} parameters. When set to \texttt{True}, the component is generated through simulations. The different components can be handled independently, so that some may be simulated while others are loaded from disk. If \texttt{generate\_input\_data} is set to \texttt{True}, the code constructs the total data by summing the individual components, regardless of whether they were simulated or loaded. If any of the \texttt{generate\_input\_x} parameters is set to \texttt{False} and the corresponding \texttt{x\_path} is not provided, the associated component will not be returned by \texttt{get\_input\_data} and will not be included in the co-addition to generate the total data.

The user can also initialize the data \textsc{SimpleNamespace} object directly, given a predefined multifrequency set of maps \texttt{your\_data} and, for instance, the associated noise simulations \texttt{your\_noise}, as follows:
\begin{verbatim}
from types import SimpleNamespace
data = SimpleNamespace(total=your_data, noise=your_noise)
\end{verbatim}

To successfully run the component-separation pipeline, the input \textsc{SimpleNamespace} object must include either a \texttt{total} attribute, containing the full-mission co-added data, or the attributes \texttt{total\_split1} and \texttt{total\_split2}, which correspond to a division of the observations into two data sets with uncorrelated noise. The component-separation weights are derived from the covariance estimated from one of these inputs, with priority given to the full-mission data when available.
When the covariance is computed from the full-mission data, the associated noise bias is propagated into the variance minimization. The \texttt{total} attribute is therefore strictly required for runs involving foreground reconstruction and diagnostics, where the noise bias in the data covariance must always be accounted for.

Optionally, the \textsc{SimpleNamespace} object may also contain additional attributes such as, for example, \texttt{fgds} or \texttt{cmb}, representing the individual microwave components of the simulated data or their simulated counterparts for real observations. If present when the data object is passed to component-separation routines, these components are then propagated through the component-separation pipeline, contributing to the final output products. For example, propagating the \texttt{fgds} attribute allows the evaluation of foreground residuals in the reconstructed CMB maps for all ILC and PILC techniques, or the recovery of foreground-only emission in the case of GILC and GPILC methods.

Additional attributes of the \textsc{SimpleNamespace} object—for example \texttt{dust}, \texttt{synch}, \texttt{ame}, \texttt{co}, \texttt{freefree}, \texttt{cib}, \texttt{tsz}, \texttt{ksz}, and \texttt{radio\_galaxies}—can be used to store other specific microwave components. There are no restrictions for attributes in the \textsc{SimpleNamespace} object.

The shape of each of this attribute is expected to be $N_{\nu} \times N_{\text{fields}} \times N_{\text{pix/lm}}$, where $N_{\nu}$, $N_{\text{fields}}$, and $N_{\text{pix/lm}}$ denote, respectively, the number of observed frequency channels, the number of simulated or stored fields, and the number of pixels or harmonic coefficients. 

\subsubsection{Component separation: \texttt{component\_separation}}

The loaded or simulated data can then be passed to the main component-separation routine, which executes all the methods listed in the \texttt{compsep} parameter of the configuration dataclass \texttt{config}:
\begin{verbatim}
from broom import component_separation
outputs = component_separation(config, data, nsim=nsim)
\end{verbatim}
The \texttt{nsim} parameter specifies the simulation index when running multiple realizations; it is used, for example, to tag and save (if requested) the component-separation weights and/or output products for that particular simulation.

The \texttt{component\_separation} function relies on several additional modules—\texttt{ilc.py}, \texttt{pilcs.py}, \texttt{gilcs.py}, \texttt{gpilcs.py}, and \texttt{clusters.py}—to perform the actual component-separation procedures.

The input \texttt{data} provided to the \texttt{component\_separation} function can consist either of \texttt{HEALPix} maps, with acceptable fields in \texttt{["T", "E", "B", "QU", "TQU", "EB", "TEB"]}, or of harmonic coefficients, with allowed fields in \texttt{["T", "E", "B", "EB", "TEB"]}. 

By setting the corresponding keywords in the configuration, this function executes all selected component-separation techniques and subsequently saves the resulting products to disk (with file paths automatically generated from the configuration and component-separation parameters), returns them in \texttt{outputs}, or performs both actions. The \texttt{outputs} variable will be a \textsc{SimpleNamespace} object with the same attribute structure as the \texttt{data} object, where each attribute contains the result of propagating the corresponding multifrequency set through the applied component-separation pipelines. Each methodology to be applied to \texttt{data} should correspond to a different entry in the \texttt{compsep} dictionary.
A variety of hyperparameters can be specified for each component-separation technique within \texttt{BROOM}. All of them can be defined under the \texttt{compsep} entry, either in the configuration \texttt{YAML} file or directly within the python dataclass object. For example, within a python script this can be done as follows:
\begin{verbatim}
config_run = {
    'compsep': [
    {'method': method_1,
    'domain': domain_1,
    .
    .
    .},
    {'method': method_2,
    'domain': domain_2,
    .
    .
    .},
    ...]
config.config["compsep"] = config_run["compsep"]
config._store_passed_settings().
\end{verbatim}
Here the \texttt{domain} keyword specifies the domain in which the method is applied (either \texttt{'pixel'} or \texttt{'needlet'}), while \texttt{method} identifies the particular technique to be used. We emphasize that future \texttt{BROOM} releases will incorporate component-separation extensions in the harmonic domain.
The \texttt{component\_separation} function can execute several methodologies (corresponding labels to be provided as \texttt{method} are reported in parenthesis) for reconstruction of a signal of interest with pre-defined SED (see Section~\ref{sec:cmb_rec}), including ILC (\texttt{ilc}), cILC (\texttt{cilc}), MC-ILC (\texttt{mcilc}), PILC (\texttt{pilc}), PRILC (\texttt{prilc}), cPILC (\texttt{cpilc}), and cPRILC (\texttt{cprilc}). It can also perform techniques for foreground reconstruction (see Section~\ref{sec:fgds_rec}), such as GILC (\texttt{gilc}) and GPILC (\texttt{gpilc}). All the different hyperparameters that can be specified for each method are detailed in the configuration file example \texttt{config\_demo.yaml}, and in the GitHub tutorial notebooks \texttt{tutorial\_satellite.ipynb} and \texttt{tutorial\_satellite\_part2.ipynb}.

When the selected domain is \texttt{needlet}, hybrid approaches can also be implemented, with different component-separation techniques applied to distinct sets of needlet bands. This strategy enables one to exploit the relative strengths of each method over specific ranges of angular scales.
The currently implemented hybrid configurations—where the techniques listed in parentheses are applied to complementary subsets of needlet bands—include: \texttt{c\_ilc} (cILC and ILC), \texttt{mc\_ilc} (MC-ILC and ILC), \texttt{mc\_cilc} (MC-ILC and cILC), \texttt{c\_pilc} (cPILC and PILC), and \texttt{c\_prilc} (cPRILC and PRILC). In each case, the technique specified before the underscore is applied to the needlet bands whose indices are provided through the \texttt{special\_nls} parameter (a list). 

When dealing with polarization data, the input and output representations may differ. For example, even if the component-separation routines are applied to scalar fields (i.e., $E$ and $B$ modes), the corresponding outputs can be returned as Stokes parameter maps (i.e., $Q$ and $U$), and vice versa. The map representation to be returned by all component-separation runs specified in the \texttt{compsep} dictionary is controlled by the global parameter \texttt{field\_out}.

The \texttt{component\_separation} function can also be used to perform diagnostics of foreground complexity, either across the full range of angular scales (i.e., in pixel space) or within separate multipole intervals (i.e., in needlet space), as described in Section~\ref{sec:fgds_rec}. This functionality is enabled by setting the \texttt{method} keyword to \texttt{fgd\_diagnostic} for scalar fields, or to \texttt{fgd\_P\_diagnostic} for spin-2 fields.

To run the MC-ILC pipeline using sky patches derived from realistic tracers constructed as described in Section~\ref{sec:cmb_rec}, the user should execute the following commands:
\begin{verbatim}
from broom import get_mcilc_tracers
get_mcilc_tracers(config, foregrounds=input_fgds, systematics=input_syst)
\end{verbatim}
Here, \texttt{input\_fgds} and \texttt{input\_syst} are optional arguments corresponding to the sets of input foreground and systematic-effect maps to be used for tracer generation. These inputs must consist of multifrequency maps or harmonic coefficients (depending on the \texttt{data\_type} parameter) with dimensions $N_{\nu} \times 3 \times N_{\text{pix/lm}}$, thus including both temperature and polarization components.
At present, the pipeline has been validated only for the reconstruction of CMB $B$ modes. Accordingly, it currently processes polarization inputs exclusively and stores only $B$-mode tracers. The same restriction applies when running MC-ILC itself: the only supported output field is the $B$-mode map or its transformation to $QU$ (i.e., the so-called $B$-mode family of Stokes parameters).
The configuration of the sky patches depends on the instrumental specifications and the assumed foreground model, but not on the specific realizations of the CMB signal or instrumental noise. Consequently, this function needs to be executed only once, thereby reducing the overall computational cost.

\subsubsection{Estimating nuisance covariance: \texttt{get\_nuisance\_covariance}}
In several component-separation pipelines, an estimate of the noise (or more generally nuisance) covariance is required. For example, in foreground-oriented routines such as G(P)ILC and in the foreground-complexity diagnostic (see Section~\ref{sec:fgds_rec}), it corresponds to the matrix $C_{N}$ introduced in Equation~\ref{eq:cov_gnilc} and is used to identify covariance modes dominated by nuisance contributions. In pipelines designed for the reconstruction of signals with fixed SEDs, this covariance can optionally be used to (partially or fully) debias the estimated data covariance from noise, thereby improving foreground suppression at the expense of increased reconstruction noise.

To generate and save an estimate of the nuisance covariance to disk, the user can execute:
\begin{verbatim}
from broom import get_nuisance_covariance
get_nuisance_covariance(config, nuisance_path=path_nuis)
\end{verbatim}
where \texttt{path\_nuis} is an optional directory containing precomputed nuisance maps. If not provided, nuisance simulations are generated on the fly. The function will search for a \texttt{nuisance\_covariance} dictionary within the configuration dataclass, which specifies the settings for the desired component-separation run (e.g., the needlet bands). As for any component-separation run, this dictionary may contain multiple subdictionaries with different sets of hyperparameters, thus enabling multiple different nuisance covariance estimations. 

A key hyperparameter for each run is \texttt{nuisance} (a string or list of strings), which specifies the components to be treated as nuisance. It may include \texttt{"cmb"}, \texttt{"noise"}, and any identifier corresponding to a \texttt{PySM} foreground model. For G(P)ILC runs and foreground diagnostics, the same string or list can be provided through the same keyword in the relevant \texttt{compsep} subdictionary.
If the user does not explicitly request loading precomputed nuisance covariances during component-separation runs by setting \texttt{load\_nuisance\_covariance} to \texttt{True}, the code extracts them directly from the inputs supplied to the routine. In this case, the input \textsc{SimpleNamespace} object must include attributes matching the entries specified in the \texttt{nuisance} keyword (e.g., \texttt{noise}, \texttt{cmb}, or other component names). The string \texttt{"nuisance"} is accepted as a valid value of the \texttt{nuisance} parameter only within the component-separation dictionaries and only when loading a precomputed nuisance covariance is not requested. In this case, it serves as a shortcut indicating that the covariance should be computed using the \texttt{nuisance} attribute provided in the input data.
For ILC-based component-separation runs, where a signal with known SED is reconstructed, the \texttt{nuisance} keyword will not be used, since noise is the only term relevant for data-covariance debiasing in this context. Therefore, if the user intends to debias the data covariance and load a nuisance-covariance estimate from disk, a corresponding entry must be defined in the \texttt{nuisance\_covariance} dictionary with \texttt{"noise"} as the single element in the \texttt{nuisance} keyword.

\subsubsection{Combine with weights: \texttt{estimate\_residuals} and \texttt{combine\_with\_weights}}
As outlined in Section~\ref{sec:fgds_rec}, it is possible to estimate the spatial distribution (i.e., a \texttt{HEALPix} map) of the foreground residuals present in a reconstructed signal with a pre-defined SED after a given component-separation run. Within the \texttt{BROOM} package, this can be achieved using the \texttt{estimate\_residuals} function:
\begin{verbatim}
from broom import estimate_residuals
residuals = estimate_residuals(config, nsim=nsim, gilc_outputs=outputs)
\end{verbatim}
Here, \texttt{gilc\_outputs} is an optional argument that may contain the outputs from G(P)ILC runs defined in the \texttt{compsep} dictionary. If multiple component-separation runs have been performed through \texttt{component\_separation}, the function automatically selects the relevant G(P)ILC runs. The parameters governing the residual estimation must be specified in the \texttt{compsep\_residuals} dictionary within the configuration file or dataclass. Multiple cases, corresponding to different G(P)ILC or component-separation runs, can be executed sequentially.
If \texttt{gilc\_outputs} is not provided, the function requires that the outputs of a previous GILC or GPILC run are saved to disk; the corresponding path must then be specified as a required keyword (\texttt{gilc\_path}) in each \texttt{compsep\_residuals} subdictionary. This routine propagates the G(P)ILC multifrequency foreground templates through the component-separation weights associated with the reconstruction of a signal with fixed SED (i.e., one of the pipelines described in Section~\ref{sec:cmb_rec}). For each run, the corresponding set of weights used for the reconstruction must therefore be loaded from disk, implying that they must have been previously saved during the associated component-separation run.
If \texttt{gilc\_outputs} is provided and contains multiple G(P)ILC runs, the user may either specify a single hyperparameter dictionary within \texttt{compsep\_residuals}, which will be applied to all G(P)ILC runs, or define as many dictionaries as G(P)ILC runs, in which case each dictionary will be associated sequentially with the corresponding set of outputs.

The estimation of residuals in any ILC solution is part of a more general procedure for propagating individual components through the component-separation pipelines by combining them with the previously computed corresponding multi-frequency weights. This can be performed using:
\begin{verbatim}
from broom import combine_with_weights
combined_outputs = combine_with_weights(config, inputs, nsim=nsim)
\end{verbatim}
where \texttt{inputs} is a \textsc{SimpleNamespace} object with potentially multiple attributes containing the multi-frequency data to be propagated. This function requires a \texttt{compsep\_propagate} dictionary, which must specify at least the root directory where the component-separation weights to be used are stored.

\subsubsection{Angular power spectrum: \texttt{\_compute\_spectra\_}}

Finally, the angular power spectra of the component-separation products — including foreground-residual estimates and propagated data — can be computed either using the \texttt{HEALPix} \textit{anafast} routine or a \texttt{MASTER} estimator (see Section~\ref{sec:spectra} for details):
\begin{verbatim}
from broom import _compute_spectra_
cls_out = _compute_spectra_(config, nsim=nsim, outputs=outputs)
\end{verbatim}

As usual, the \texttt{nsim} parameter is used when multiple simulations are processed, allowing the code to tag a specific realization and its corresponding power spectra for loading and saving purposes. The \texttt{outputs} argument is optional and may be provided if the user wishes to compute the angular power spectra of products returned by the \texttt{component\_separation}, \texttt{estimate\_residuals}, or \texttt{combine\_with\_weights} functions. The code will then sequentially compute the angular power spectra for all associated runs. Products generated by foreground-diagnostic routines are automatically excluded, as their angular power spectra are not assigned physical meaning.

Specific hyperparameters controlling spectrum computation are defined in the configuration file or dataclass through the \texttt{compute\_spectra} dictionary, which may contain multiple subdictionaries. If \texttt{outputs} is provided and contains products from multiple runs, \texttt{compute\_spectra} may either include a single dictionary (applied to all runs) or as many subdictionaries as runs.

In the absence of \texttt{outputs}, each subdictionary of \texttt{compute\_spectra} must include the path to the component-separation products of interest via the \texttt{path\_method} keyword. In this case, spectra from different component-separation runs and/or masking configurations can be computed sequentially by defining multiple subdictionaries in \texttt{compute\_spectra}. 

When processing simulations, the number of output products can be large (e.g., total reconstructed maps, foreground residuals, noise residuals, propagated CMB signal, etc.). If the user is interested in computing angular power spectra for only a subset of products, this can be specified via the \texttt{components\_for\_cls} list. This list may include any valid output tag, such as \texttt{output\_total}, \texttt{fgds\_residuals}, or \texttt{noise\_residuals}.
The \texttt{components\_for\_cls} list is mandatory when \texttt{outputs} is not provided. If \texttt{outputs} is provided and \texttt{components\_for\_cls} is not specified, angular power spectra are computed for all attributes associated with the run. In this case, for G(P)ILC runs, all auto-spectra (excluding cross-spectra) are evaluated for every attribute and available frequency channel.
If the user specifies the \texttt{components\_for\_cls} parameter, then for G(P)ILC runs the attribute tags must include the labels of the frequency channels of interest, as described in Appendix~\ref{app:broom_spectra}.

The \texttt{\_compute\_spectra\_} function can also be executed serially over multiple simulations:
\begin{verbatim}
from broom import _compute_spectra
cls_out = _compute_spectra(config)
\end{verbatim}
In this case, the output paths must be specified in the subdictionaries of \texttt{compute\_spectra}. The function will compute spectra for simulations ranging from \texttt{nsim\_start} to \texttt{nsim\_start} $+$ \texttt{nsims}, which are optional parameters defined in the configuration dataclass.

When using either \texttt{\_compute\_spectra\_} or \texttt{\_compute\_spectra}, the estimated angular power spectra can then be returned, saved to disk, or both. The spectra computation also relies on masking routines implemented in the \texttt{masking.py} module. \\

We now proceed to present the results of a validation run of the \texttt{BROOM} package in the following section.

\section{Validation and results}
\label{sec:results}
In this section, we present the results of a validation run covering most of the functionalities of the \texttt{BROOM} package. The analysis is divided into two parts: the first one focuses on full-sky observations, representative of satellite CMB experiments (e.g.\ WMAP~\cite{ILC,sync_temp}, \textit{Planck}~\cite{Planck2018_overview, Planck2018_compsep}, LiteBIRD~\cite{Litebird_Hazumi}, PICO~\cite{PICO_inst,PICO}), while the second targets partial-sky measurements characteristic of ground-based experiments (e.g.\ POLARBEAR~\cite{POLARBEAR,POLARBEAR_Bmodes}, ACT~\cite{2016ApJS..227...21T,2025JCAP...11..061N}, SPT~\cite{SPT,SPT2025}, BICEP/Keck~\cite{BK2021,BK_compsep}, SO~\cite{SO_2019,SO-LATS_en}). For both cases, we run the complete analysis pipeline provided by the package, including the generation of tailored simulations, the application of component-separation routines, inspection of the resulting products, and the final computation of angular power spectra. Results for the satellite-like scenario are presented in Section~\ref{sec:results_satellite}, whereas those for partial-sky observations are shown in Section~\ref{sec:results_ground}. All input simulations and output products are generated at $N_{\rm side}=512$.

The total co-added maps are composed of CMB signal, astrophysical foregrounds and instrumental noise.
In both validation cases, the simulated foreground sky includes the following components (with the corresponding \texttt{PySM} models in parentheses): thermal dust (\texttt{d1}), synchrotron (\texttt{s1}), anomalous microwave emission (\texttt{a1}), free–free emission (\texttt{f1}), CO line emission (\texttt{co3}) and CIB (\texttt{cib1}). Among these, only thermal dust, synchrotron, and CO line emission possess non-zero polarization fractions. In addition to these Galactic and extragalactic components, also tSZ and kSZ effect are included according to the \texttt{tsz1} and \texttt{ksz1} \texttt{PySM} models. 

CMB realizations are generated as Gaussian random fields drawn from theoretical angular power spectra computed with the \texttt{CAMB} package \cite{camb}, assuming the best-fit cosmological parameters from the Planck 2018 results \cite{Planck_cosmopars}. No primordial tensor modes are included (i.e. $r=0$), so the CMB $B$-mode signal arises solely from gravitational lensing. We note that, within the current \texttt{BROOM} framework, it is not yet possible to generate a physically lensed realization of the primary CMB anisotropies. Instead, the simulated CMB sky corresponds to a Gaussian realization whose power spectrum matches that of the lensed CMB.

For both the satellite and ground-based configurations, we generate ten independent realizations of the CMB and instrumental noise, and process each of them through the full component-separation pipeline and subsequent power-spectrum estimation. Both total (with co-addition of all components) and noise maps are generated for two independent data splits and for the full mission. 
 
In both analysis cases, for simplicity, we assume that: (i) the instrumental bandpasses are $\delta$-functions centered at the nominal central frequency of each channel; and (ii) the beams have Gaussian profiles and are therefore uniquely characterized by their corresponding FWHM values. More realistic configurations—including finite-width bandpasses and non-Gaussian, asymmetric beams—can be simulated by specifying the appropriate instrument parameters, as detailed in Section~\ref{app:broom_sims}. For all cases considered, the angular resolution of the maps produced by each component-separation run corresponds to a Gaussian beam with $\mathrm{FWHM}=30'$. This resolution also defines the common target beam to which all input maps are smoothed, following Equation~\ref{eq:beams_1}.

\subsection{Satellite experiment case}
\label{sec:results_satellite}
For the validation run on full-sky satellite observations, we primarily rely on the script \texttt{script\_paper\_satellite.py} available in the \texttt{tutorials} directory of the \texttt{BROOM} GitHub repository. A corresponding example configuration file for a satellite CMB experiment is provided as \texttt{config\_satellite.yaml}\footnote{\url{https://github.com/alecarones/broom/blob/main/broom/configs/config_satellite.yaml}}, located in the \texttt{configs} subfolder. We assume a LiteBIRD-like instrumental setup, with frequency coverage, angular resolution, and map depth consistent with the specifications reported in Table~13 of \cite{PTEP}. The associated instrument dictionary, \texttt{LiteBIRD\_PTEP}, is defined in the \texttt{experiments.yaml} file within the \texttt{utils} directory. Additional examples of similar validation runs can be found in \texttt{tutorial\_satellite.} \texttt{ipynb} and \texttt{tutorial\_satellite\_part2.ipynb}, also located in the \texttt{tutorials} directory.

LiteBIRD is a satellite mission approved by the Japan Aerospace Exploration Agency (JAXA) as a large-class space mission in 2019 \cite{Litebird_Hazumi, PTEP}. It is designed to observe the microwave sky in a broad frequency range to facilitate component separation over a large portion of the sky. Its primary scientific objective is to achieve a high-precision reconstruction and characterization of the large-scale CMB polarization signal, thereby shedding new light on several open questions in cosmology \cite{PTEP}.

As already anticipated, $10$ different realization of simulated LiteBIRD-like data sets are generated with \texttt{BROOM}. For each simulation, the output of the \texttt{get\_input\_data} function is a \textsc{SimpleNamespace} object including different attributes each containing the multifrequency CMB, foreground or noise (both for full mission and splits) components and an attribute storing the  full co-added maps (both for full mission and splits). Within the foreground attribute, the tSZ and kSZ distortions are included. Each of this attribute contains a \emph{numpy} array of dimension 
$N_{\nu} \times 3 \times N_{\text{pix}}$ with $N_{\nu}=22$, $N_{\text{pix}}=12\cdot 512^{2}$ and $3$ stands for $TQU$ maps. Noise maps are generated as isotropic Gaussian realizations from a white-noise power spectrum whose amplitude is set according to Equation \ref{eq:cl_noise} from the map depths of the different frequency channels. 

The white-noise spectrum in polarization is provided under the assumption of an ideal HWP as the first optical element of the satellite telescope. In general, this assumption does not hold for total intensity measurements, for which a $1/f$ noise component should be included. However, given the purely validation-oriented scope of the present analysis, we neglect this contribution here. A dedicated validation case including non-white noise is nonetheless considered for the ground-based experiment discussed in Section~\ref{sec:results_ground}.

\subsubsection{Reconstruction of the CMB signal}
\label{subsec:cmb_LB}
We first apply several \texttt{BROOM} routines for the reconstruction of the CMB signal:
\begin{enumerate}
    \item \texttt{ILC}: standard ILC performed in pixel space;
    \item \texttt{NILC}: ILC performed in the needlet domain;
    \item \texttt{cNILC0}: cILC in the needlet domain with full deprojection of the zeroth-order moment of a power-law emission, and of the zeroth- and first-order (with respect to $\beta_{d}$) moments of a MBB SED at all needlet scales;
    \item \texttt{cNILC01}: cILC in the needlet domain with full deprojection of all zeroth- and first-order moments for both power-law and MBB emissions;
    \item \texttt{PNILC}: PILC performed in the needlet domain on polarization maps only; 
    \item \texttt{iMCNILC}: MC-ILC in the needlet domain using ideal clusters (see Section~\ref{sec:cmb_rec} for details);
    \item \texttt{(rMC)NILC}: MC-NILC applied in the first needlet band and standard NILC in all subsequent bands, where MC-NILC uses the realistic clusters reconstructed directly from the input data (see Section~\ref{sec:cmb_rec}). In this case, MC-NILC is restricted to the largest angular scales, for which the sky-patch reconstruction is sufficiently accurate.
\end{enumerate}
\begin{figure}
\centering
\includegraphics[width=.6\textwidth]{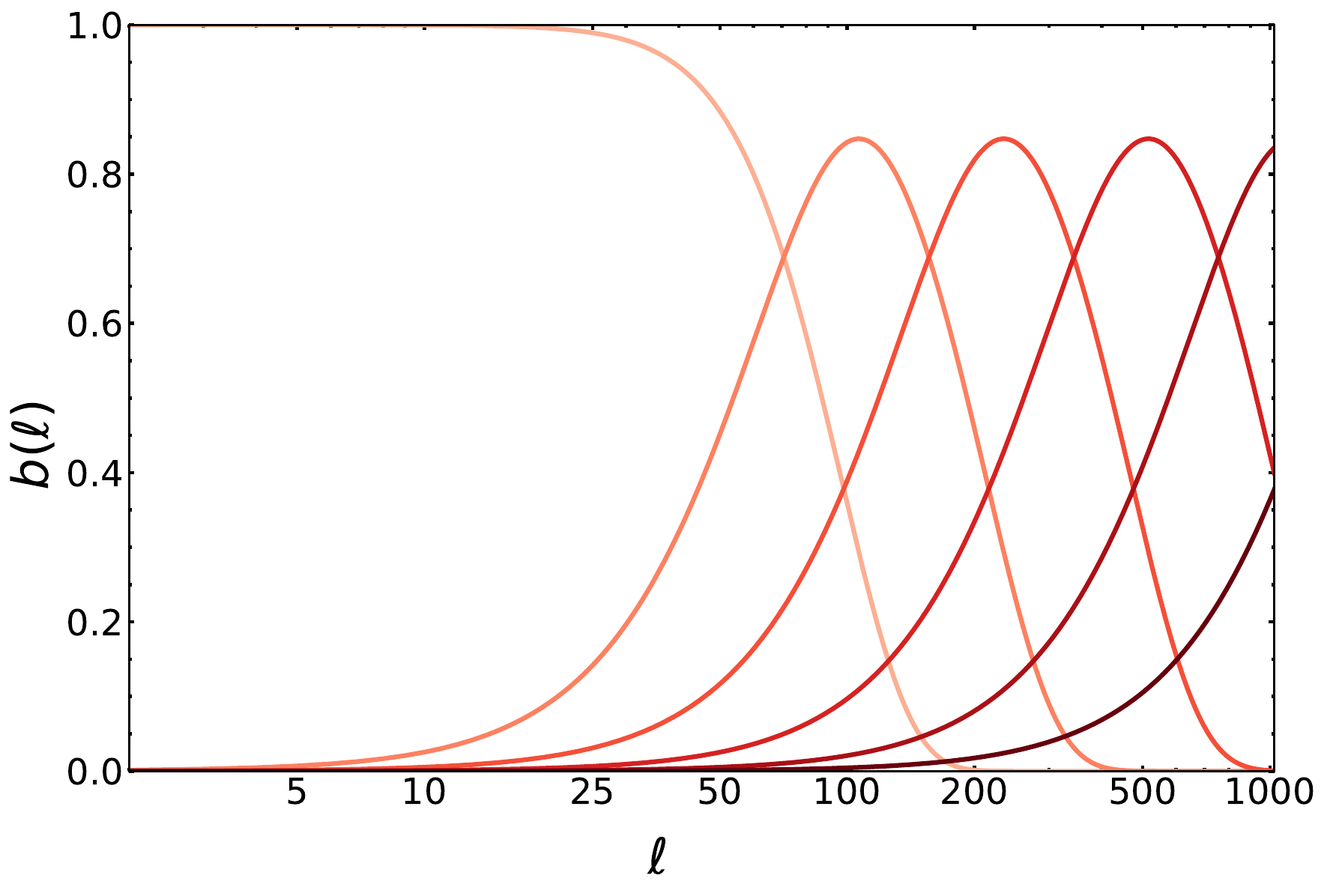}
\caption{Adopted configuration of needlet harmonic bands $b_{j}(\ell)$ used in Equation \ref{eq:Psi}, where the index $j$ identifies different needlet scales. In this specific construction, Mexican wavelets with $B=1.3$ and $p=1$ have been adopted. Original needlet bands are merged into larger hamonic windows according to Equation~\ref{eq:b_merge}. The resulting bands are formed by combining the original needlet scales within the following intervals: $[0, 16, 19, 22, 25, 28]$.}
\label{fig:needlets}
\end{figure}

All the aforementioned techniques have been introduced in Section \ref{sec:cmb_rec}.
All methodologies implemented in the needlet domain employ the needlet basis shown in Figure~\ref{fig:needlets}, constructed from Mexican needlets with $B=1.3$ and $p=1$ (see Section~\ref{sec:needlets} for details). To obtain the adopted needlet configuration, several consecutive original Mexican needlet bands are merged according to Equation~\ref{eq:b_merge}, thereby ensuring that a sufficient number of modes is sampled within each band. Specifically, the resulting bands are constructed by combining the original needlet scales corresponding to the following boundaries: $[0, 16, 19, 22, 25, 28]$. Details on how to configure these needlet parameters are provided in Appendix~\ref{app:compsep_broom}.
 
As anticipated in the previous section, the MC-NILC routines are currently effective only for $B$-mode reconstruction; therefore, results for \texttt{iMCNILC} and \texttt{(rMC)NILC} are presented exclusively for this spin-0 field.
On the other hand, ILC pipelines that minimize the output polarization intensity $P^2$ (such as \texttt{PNILC}) operate only on polarization. Consequently, although the inputs consist of $TQU$ maps, the outputs are returned as $EB$ maps for \texttt{PNILC}, as $B$-mode maps for \texttt{iMCNILC} and \texttt{(rMC)NILC}, and as $TEB$ maps for all other component-separation approaches.
We note that this choice may be sub-optimal when applying masking strategies, since foreground residuals can be redistributed across the sky due to the non-local nature of the $Q/U \rightarrow E/B$ transformation. However, given that our primary goal is to validate the pipeline, we do not explicitly account for this effect and proceed by computing angular power spectra from masked $E$- and $B$-mode maps.
For completeness, we also note that the \texttt{BROOM} routines support returning outputs in terms of $Q$ and $U$ Stokes parameter maps.

We now present the results obtained at both the map and power-spectrum levels for the three spin-$0$ fields ($T$, $E$, and $B$). For each field, we show representative outputs from a selection of the implemented methodologies.

At the map level, we display the total ILC solutions together with the corresponding foreground-only and noise-only residual maps, obtained by applying the component-separation weights to the input foreground-only and noise-only multifrequency data sets, respectively. For reference, we also show the target CMB-only map smoothed to the same angular resolution, along with representative input foreground and instrumental-noise maps at the simulated $119$\,GHz channel, which provides significant constraining power for CMB reconstruction in the LiteBIRD-like setup.
In all cases, just for visualization purposes, maps are reported without any masking. 
\begin{figure}
\centering
\includegraphics[width=.95\textwidth]{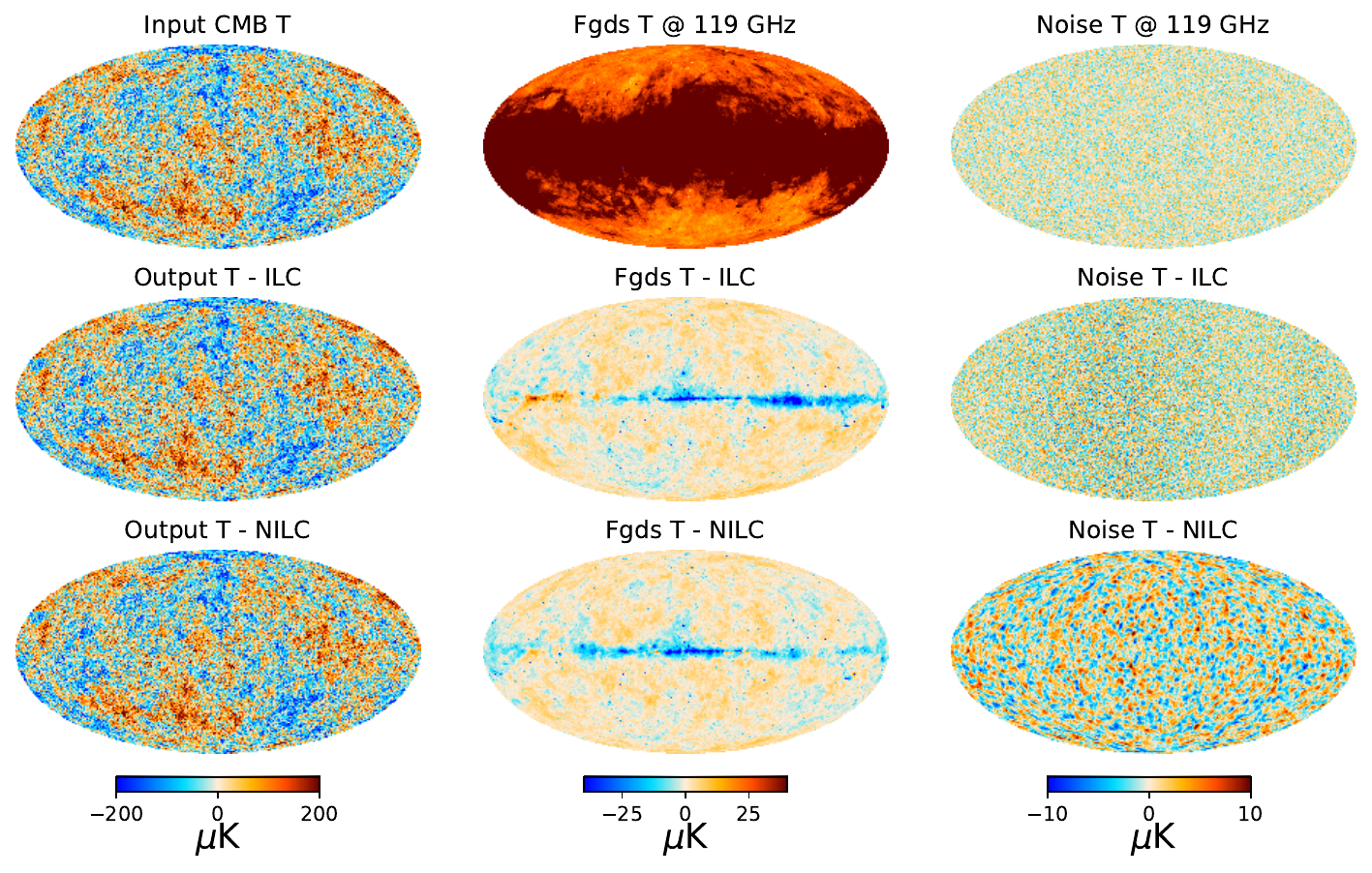}
\caption{Temperature maps. The first row illustrates the individual input components for a representative LiteBIRD-like simulation at $119\,\mathrm{GHz}$: the CMB signal (left), foreground emission (middle), and instrumental noise (right). The second and third rows show the corresponding outputs from the ILC (middle row) and NILC (bottom row) component-separation methods. The three columns display, respectively, the total reconstructed signal, the foreground residuals, and the noise residuals. The maps in the last two columns are obtained by applying the component-separation weights to the input foreground-only and noise-only maps. All maps are shown in Galactic coordinates and share a common angular resolution corresponding to a Gaussian beam with $\mathrm{FWHM}=30'$.
} 
\label{fig:lb_ouputs_T}
\end{figure}
\begin{figure}
\centering
\includegraphics[width=.9\textwidth]{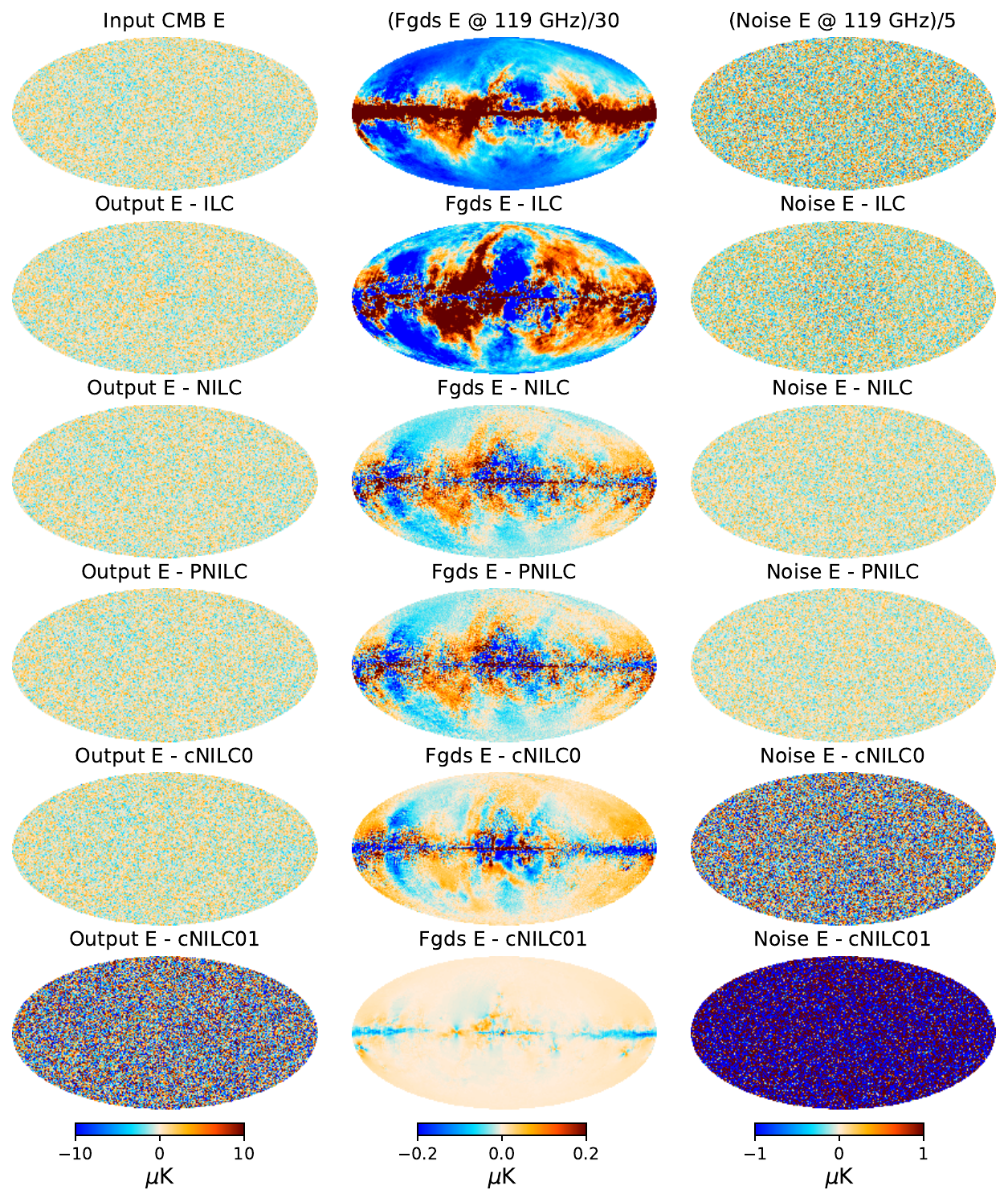}
\caption{Polarization $E$-mode maps. The first row illustrates the individual input components for a representative LiteBIRD-like simulation at $119\,\mathrm{GHz}$: the CMB signal (left), foreground emission (middle, reduced for visualization purposes by a factor $30$), and instrumental noise (right, reduced for visualization purposes by a factor $5$). The following rows show the corresponding outputs from the ILC (second), NILC (third), PNILC (fourth), cNILC0 (fifth) and cNILC01 (last) component separation. The three columns display, respectively, the total reconstructed signal, the foreground residuals, and the noise residuals. The maps in the last two columns are obtained by applying the component-separation weights to the input foreground and noise maps. All maps are shown in Galactic coordinates and share a common angular resolution corresponding to a Gaussian beam with $\mathrm{FWHM}=30'$.} 
\label{fig:lb_ouputs_E}
\end{figure}

Figure~\ref{fig:lb_ouputs_T} presents the temperature results for a representative simulation. In this example, outputs from both the ILC and NILC methods are shown to provide an illustrative comparison between pixel-domain and needlet-domain component-separation techniques. Both approaches achieve an accurate reconstruction of the CMB temperature anisotropies over the full sky, with no major differences in the large-scale spatial distribution of the residuals. Notably, NILC generally exhibits slightly lower levels of foreground residuals than ILC, particularly on small angular scales, whereas ILC shows reduced reconstruction noise. This behavior is expected: the ILC minimization is most sensitive to small-scale power, where instrumental noise is more significant, and therefore more efficiently suppresses noise on those scales. Compared to the input foreground emission at $119$ GHz, the residual contamination in both the ILC and NILC reconstructed maps is substantially reduced, highlighting the effectiveness of the component-separation methods. 
We also note that both ILC solutions display a reconstruction noise level that is enhanced relative to the amplitude of the input instrumental noise. This effect arises because only a limited number of modes of the multifrequency covariance matrix are noise dominated in intensity due to the complexity of foreground emission; as a result, the corresponding linear combination preferentially suppresses foreground emission at the expense of amplifying reconstruction noise.

Map-based results for the $E$-mode signal (again coming from a single simulation) are shown in Figure~\ref{fig:lb_ouputs_E} for several component-separation approaches. A visual inspection of these maps allows us to draw several conclusions. As in the temperature case, the CMB $E$-mode signal is generally well recovered over the full sky, with foreground residuals typically more than an order of magnitude smaller than the signal.
Overall, NILC and PNILC exhibit comparable levels of residual contamination, both lower than those obtained with the pixel-based ILC implementation. In contrast, the foreground-moment deprojection employed in \texttt{cNILC0} and \texttt{cNILC01} further reduces the level of foreground residuals, at the expected cost of increased reconstruction noise. The trade-off between residual suppression and noise amplification depends on the number of deprojected moments in the linear combination.
When comparing the overall distribution of foreground residuals across the different ILC-based solutions, we find that in all cases the foreground contamination is reduced by more than a factor of $30$ relative to the input emission in the $119$ GHz channel. At the same time, the reconstruction noise is also significantly reduced (by almost a factor $5$) with respect to the input noise level. This simultaneous reduction of both foreground residuals and reconstruction noise is attributed to the lower complexity of polarized foreground emission compared to that in intensity.

\begin{figure}
\centering
\includegraphics[width=.9\textwidth]{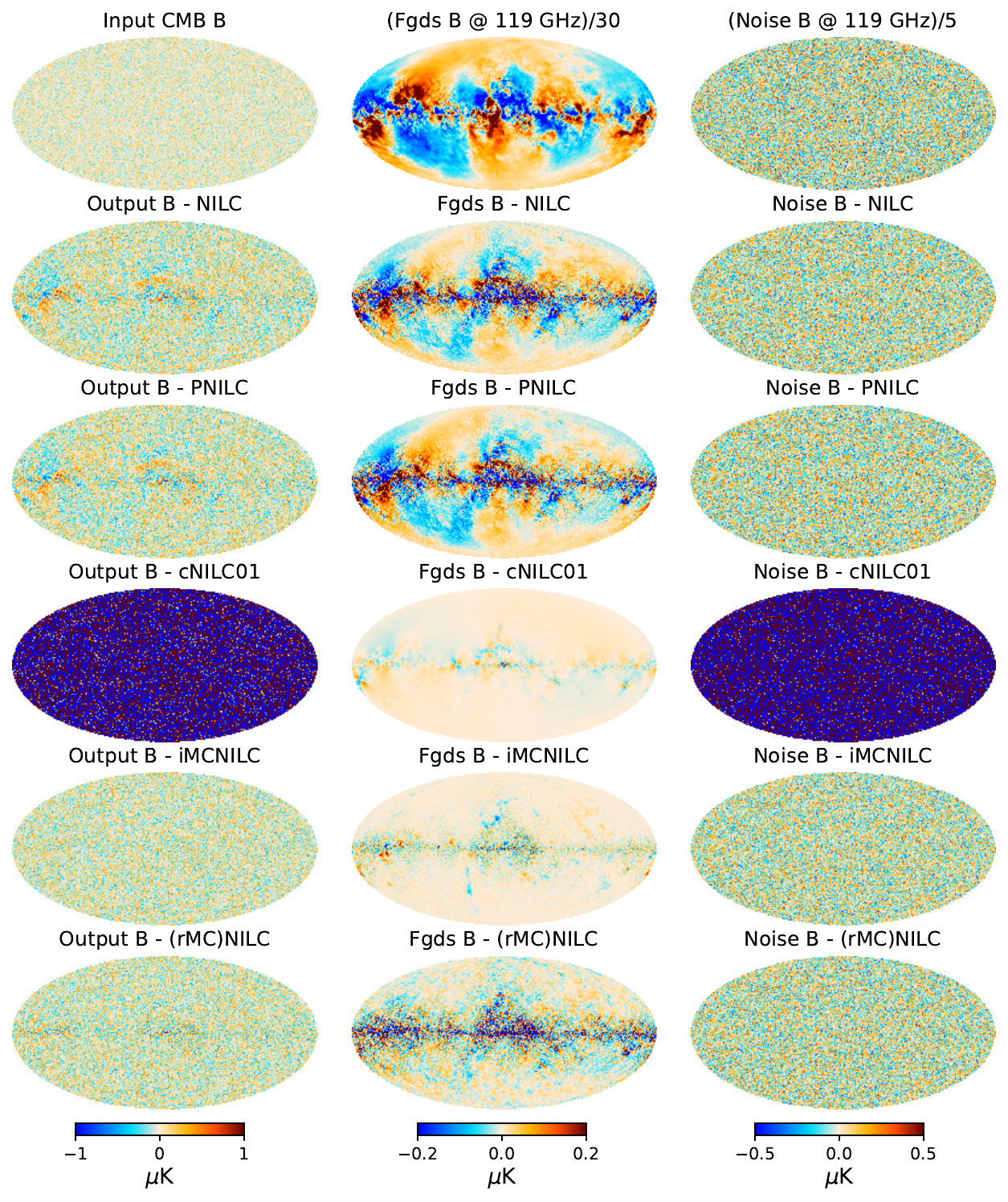}
\caption{Polarization $B$-mode maps. The first row illustrates the individual input components for a representative LiteBIRD-like simulation at $119\,\mathrm{GHz}$: the CMB signal (left), foreground emission (middle, reduced for visualization purposes by a factor $30$), and instrumental noise (right, reduced for visualization purposes by a factor $5$). The other rows show the corresponding outputs from the NILC (second), PNILC (third), cNILC01 (fourth), iMCNILC (fifth) and (rMC)NILC (last) component separation (we refer to main text for labelling of the adopted pipelines). The three columns display, respectively, the total reconstructed signal, the foreground residuals, and the noise residuals. The maps in the last two columns are obtained by applying the component-separation weights to the input foreground and noise maps. All maps are shown in Galactic coordinates and share a common angular resolution corresponding to a Gaussian beam with $\mathrm{FWHM}=30'$.} 
\label{fig:lb_ouputs_B}
\end{figure}
Finally, output maps for the CMB $B$-mode signal are compared with the targeted signal and the input foreground and noise contaminants in Figure~\ref{fig:lb_ouputs_B}. In this case, we present results from the following component-separation pipelines: \texttt{NILC}, \texttt{PNILC}, \texttt{cNILC01}, \texttt{iMCNILC}, and \texttt{(rMC)NILC}.
As expected for the $B$-mode case, the signal-to-noise ratio of the ILC-based solutions is significantly lower than for intensity and $E$ modes, owing to the much smaller amplitude of the targeted signal. Consequently, map-level biases become clearly visible, primarily driven by reconstruction noise and, in regions close to the Galactic plane, by residual Galactic emission (with extragalactic sources simulated as unpolarized).
For \texttt{NILC}, \texttt{PNILC}, and \texttt{cNILC01}, the comparative trends are analogous to those observed in the $E$-mode case. In contrast, the application of MC-NILC with ideal clusters—derived under the assumption of a perfect reconstruction of the foreground tracer (\texttt{iMCNILC})—yields foreground residual levels comparable to those obtained with \texttt{cNILC01} (which deprojects all zeroth- and first-order moments for power-law and MBB SEDs), while achieving lower reconstruction noise than the standard \texttt{NILC} case. This result highlights the improved effectiveness of an ILC pipeline which exploits variance minimization implemented in tailored sky partitions.
On the other hand, the application of MC-NILC on large angular scales using realistic clusters obtained from a tracer reconstructed directly from the data (\texttt{(rMC)NILC}; see Section~\ref{sec:cmb_rec}) produces output solutions with significantly reduced foreground and noise contamination at large scales compared to \texttt{NILC}. As in the $E$-mode analysis, we find that for all methods the foreground contamination in the reconstructed $B$-mode maps is reduced by more than a factor of $30$ relative to the input contamination at $119$ GHz. For instrumental noise, the suppression achieved reaches a factor of approximately $5$ in the most competitive cases.

\begin{figure}
\centering
\includegraphics[width=.325\textwidth]{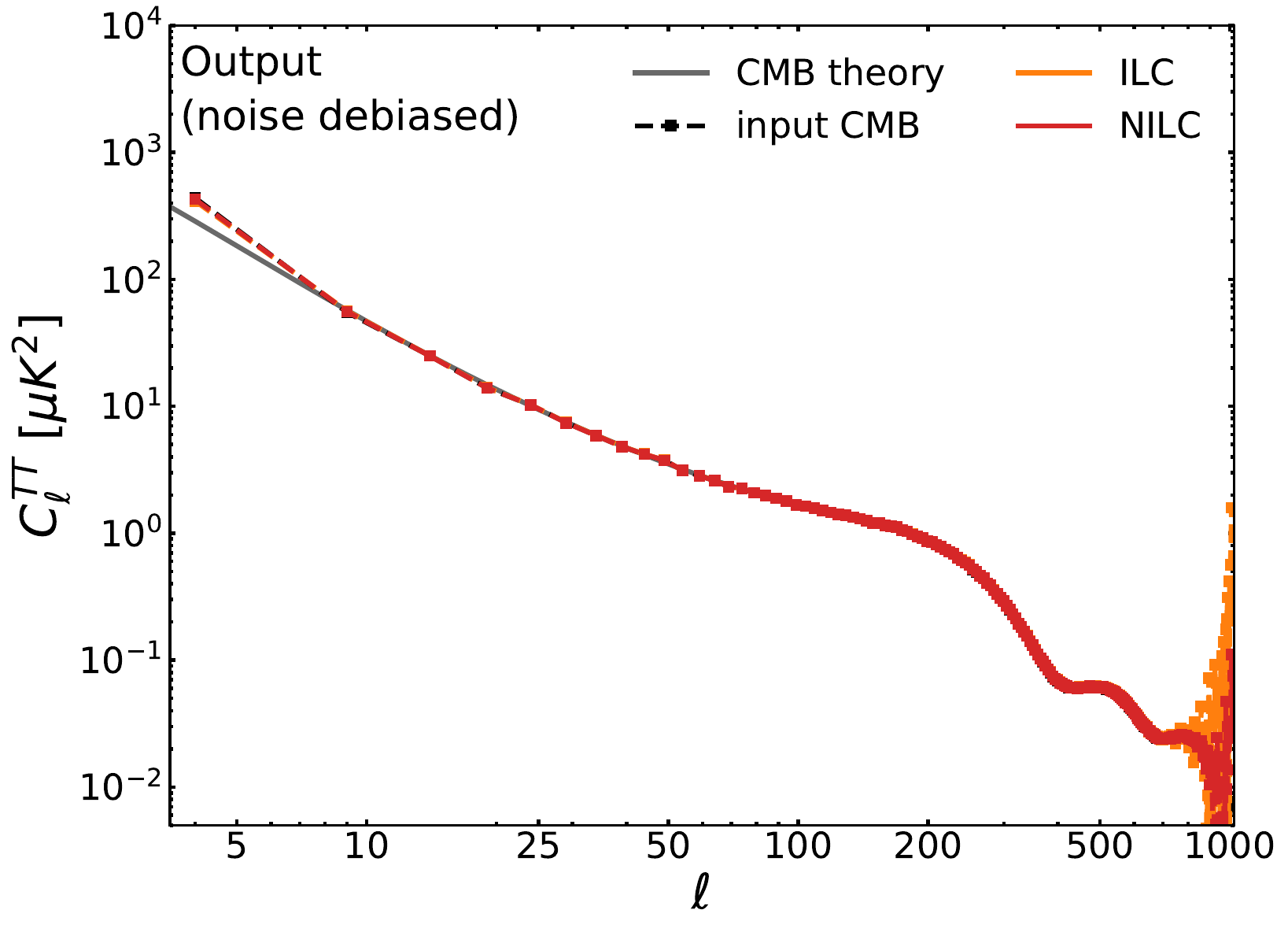}
\includegraphics[width=.325\textwidth]{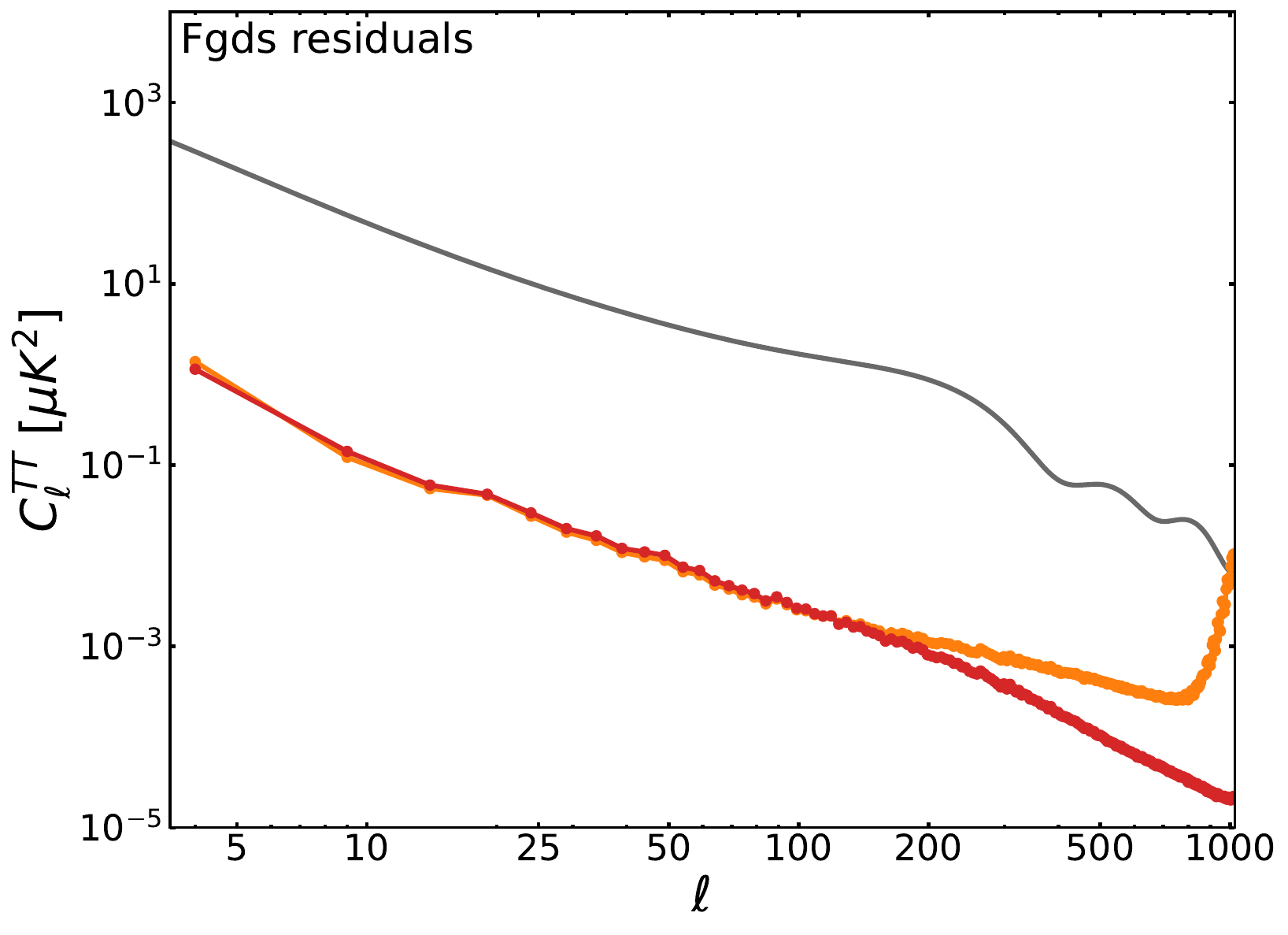}
\includegraphics[width=.325\textwidth]{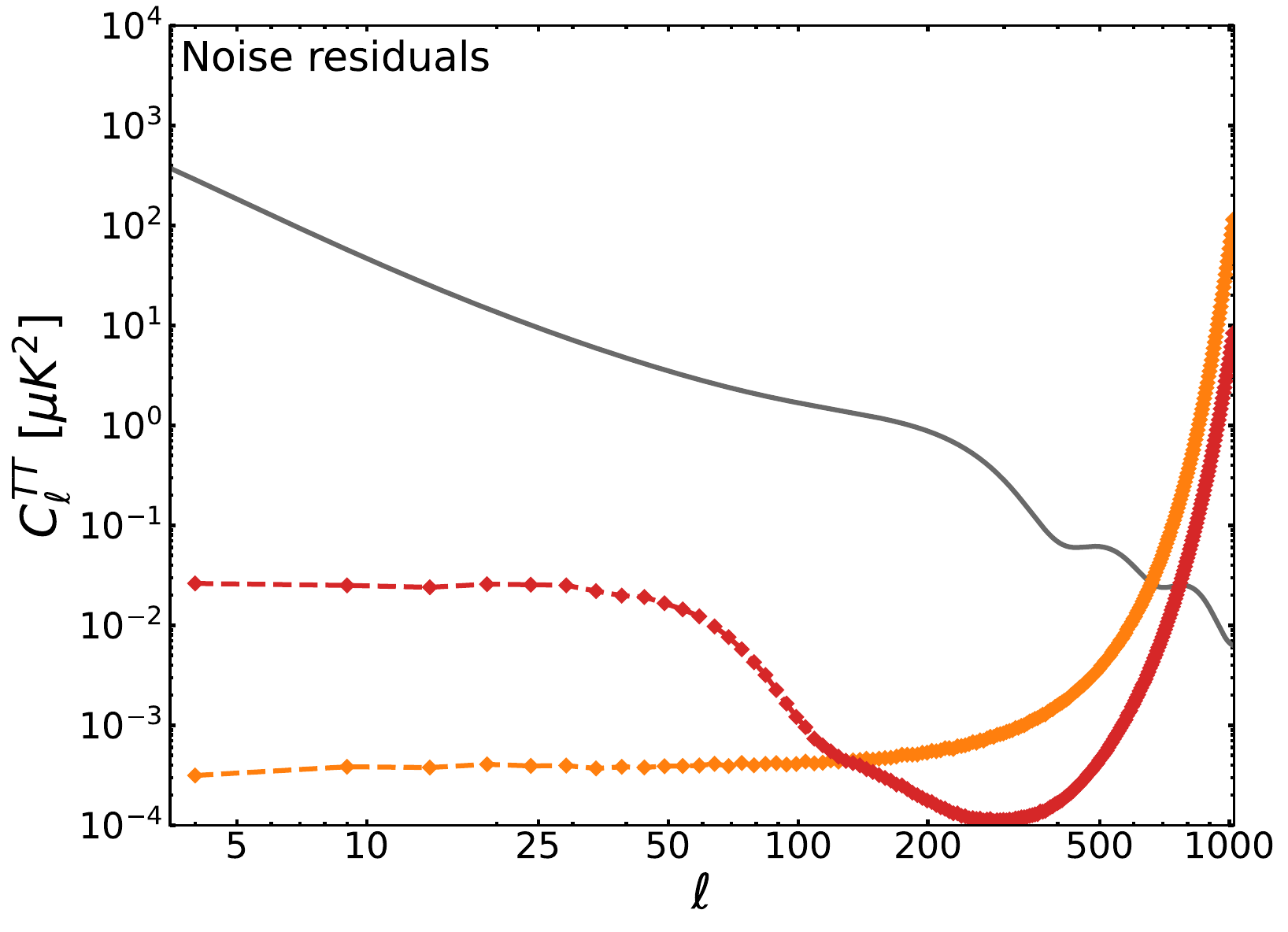}
\includegraphics[width=.325\textwidth]{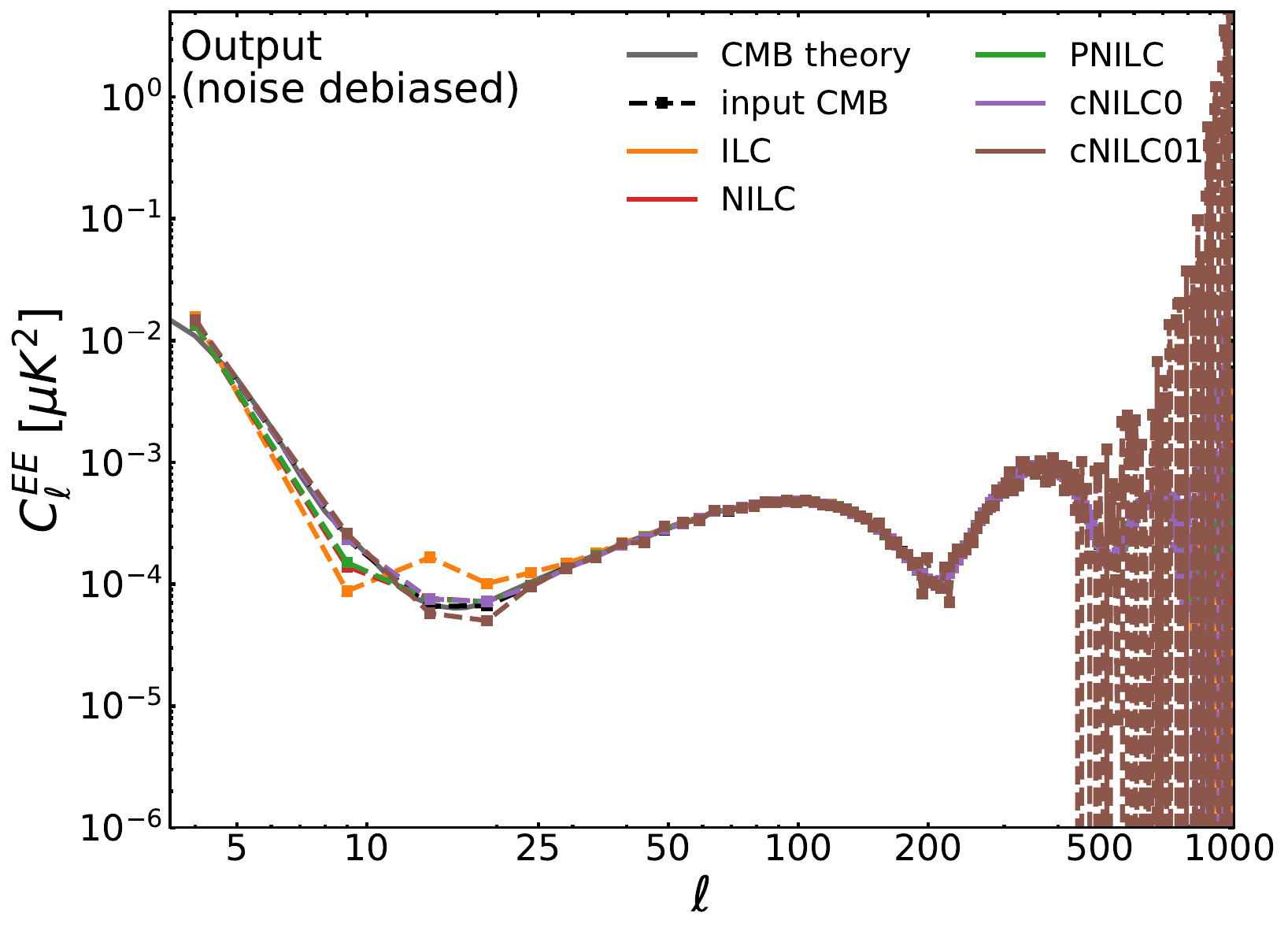}
\includegraphics[width=.325\textwidth]{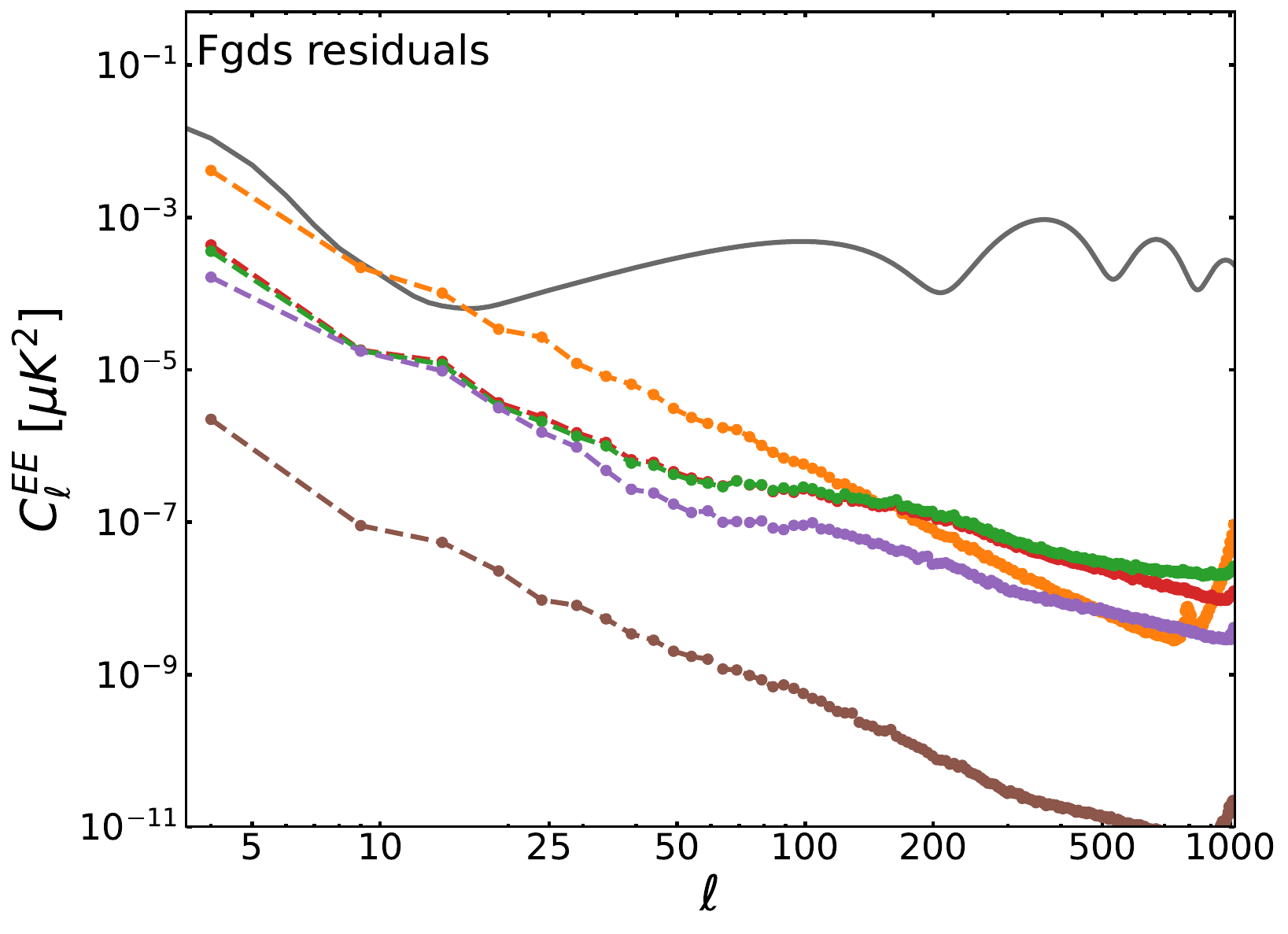}
\includegraphics[width=.325\textwidth]{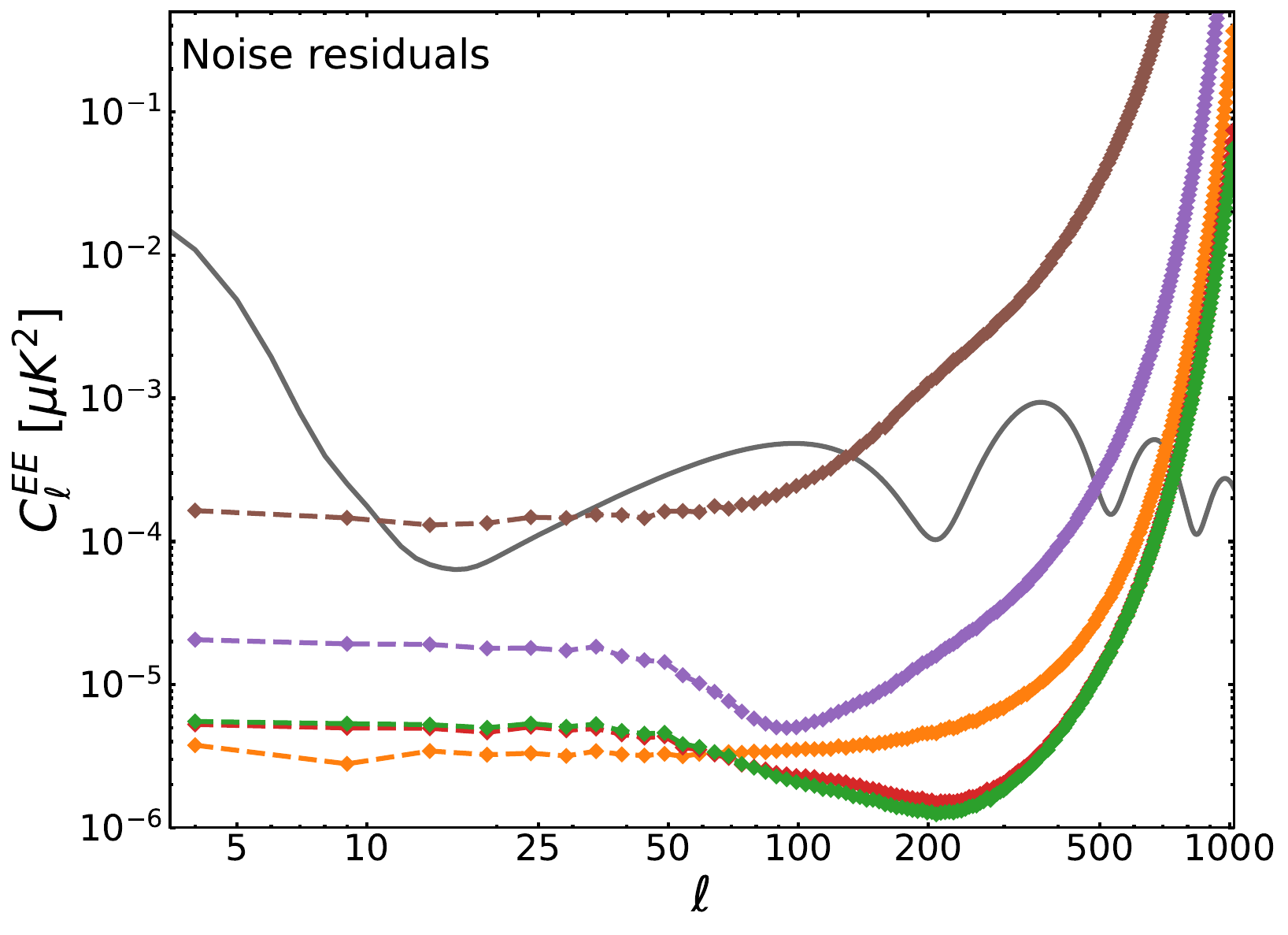}
\includegraphics[width=.325\textwidth]{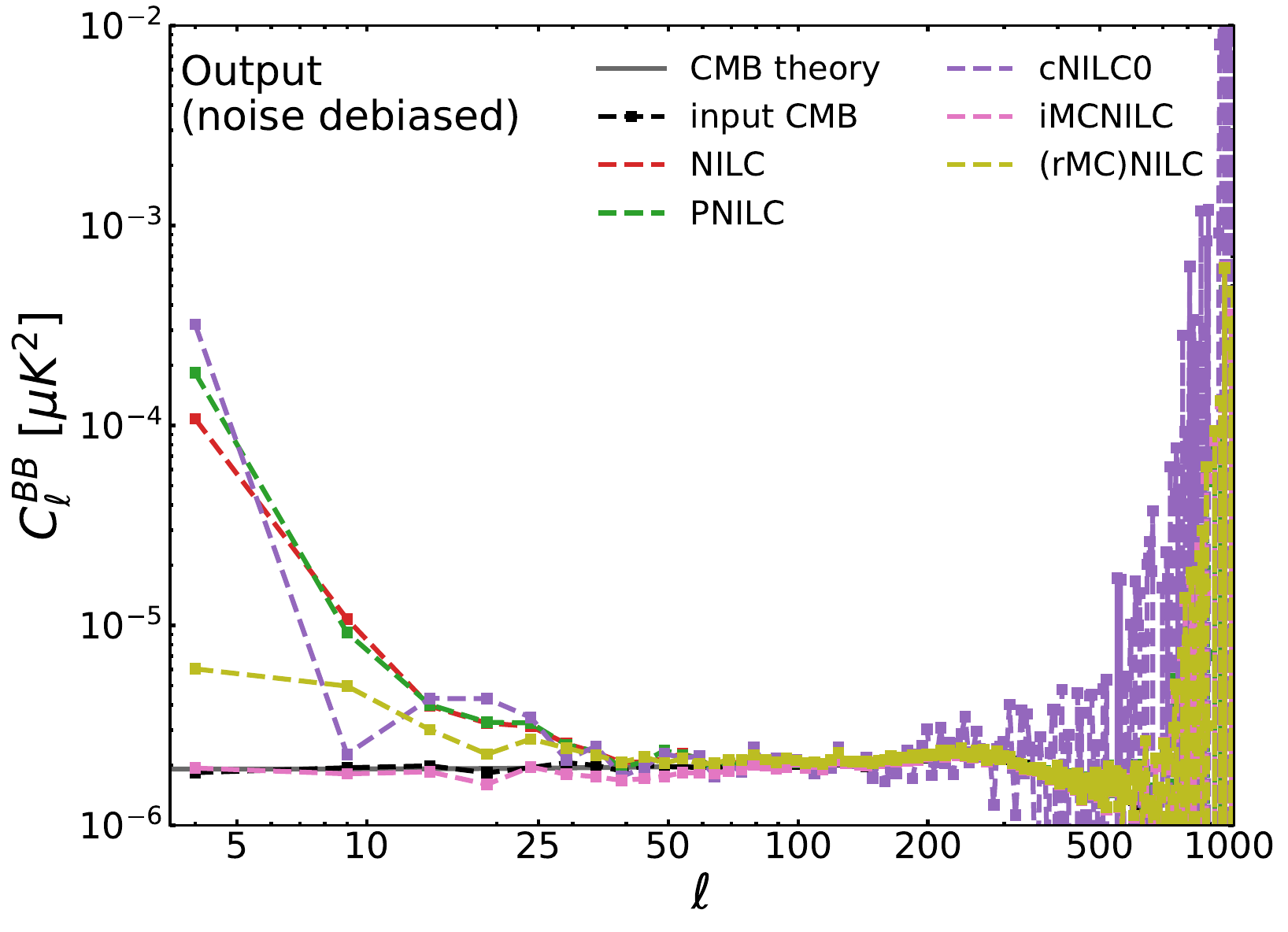}
\includegraphics[width=.325\textwidth]{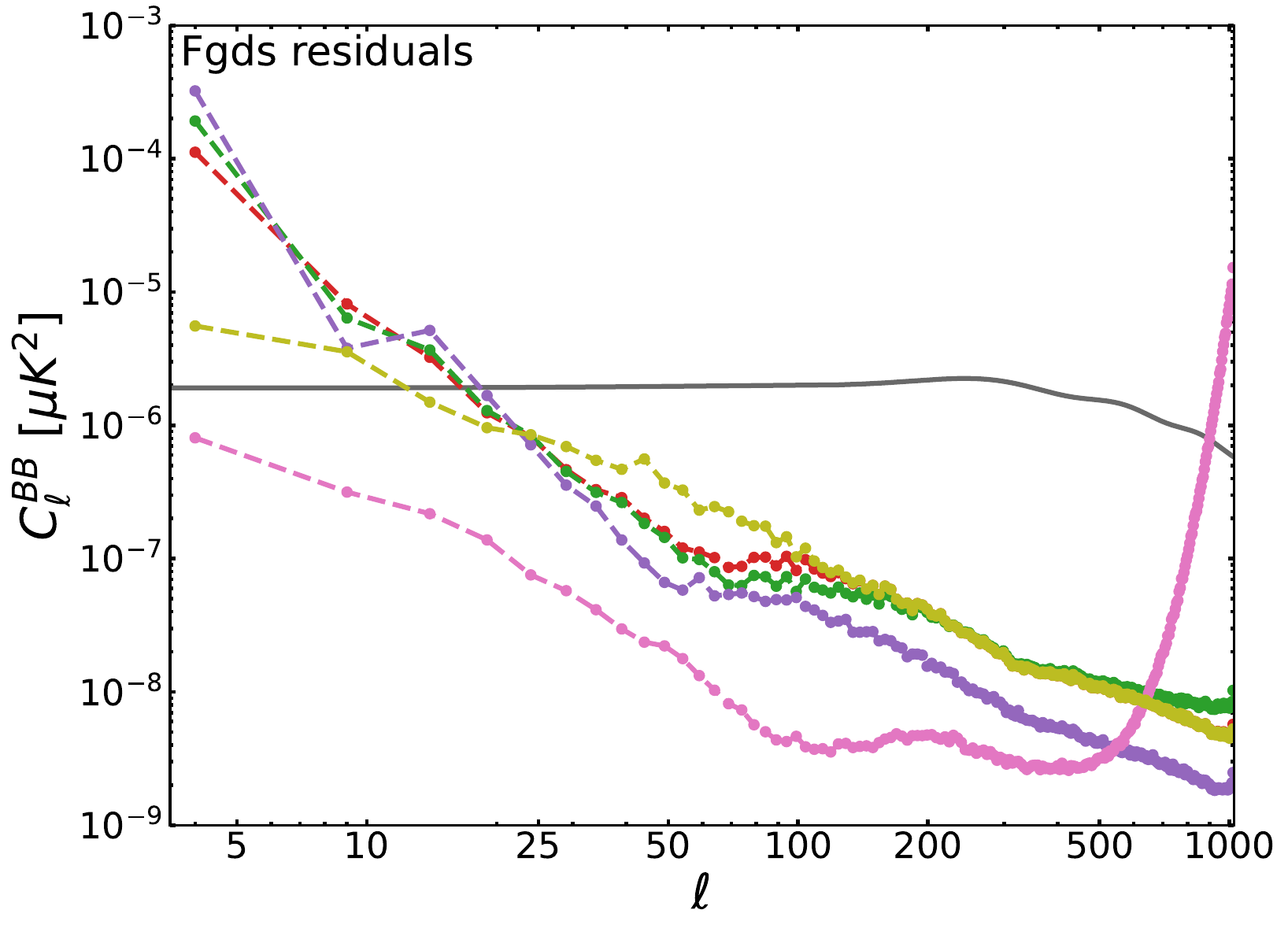}
\includegraphics[width=.325\textwidth]{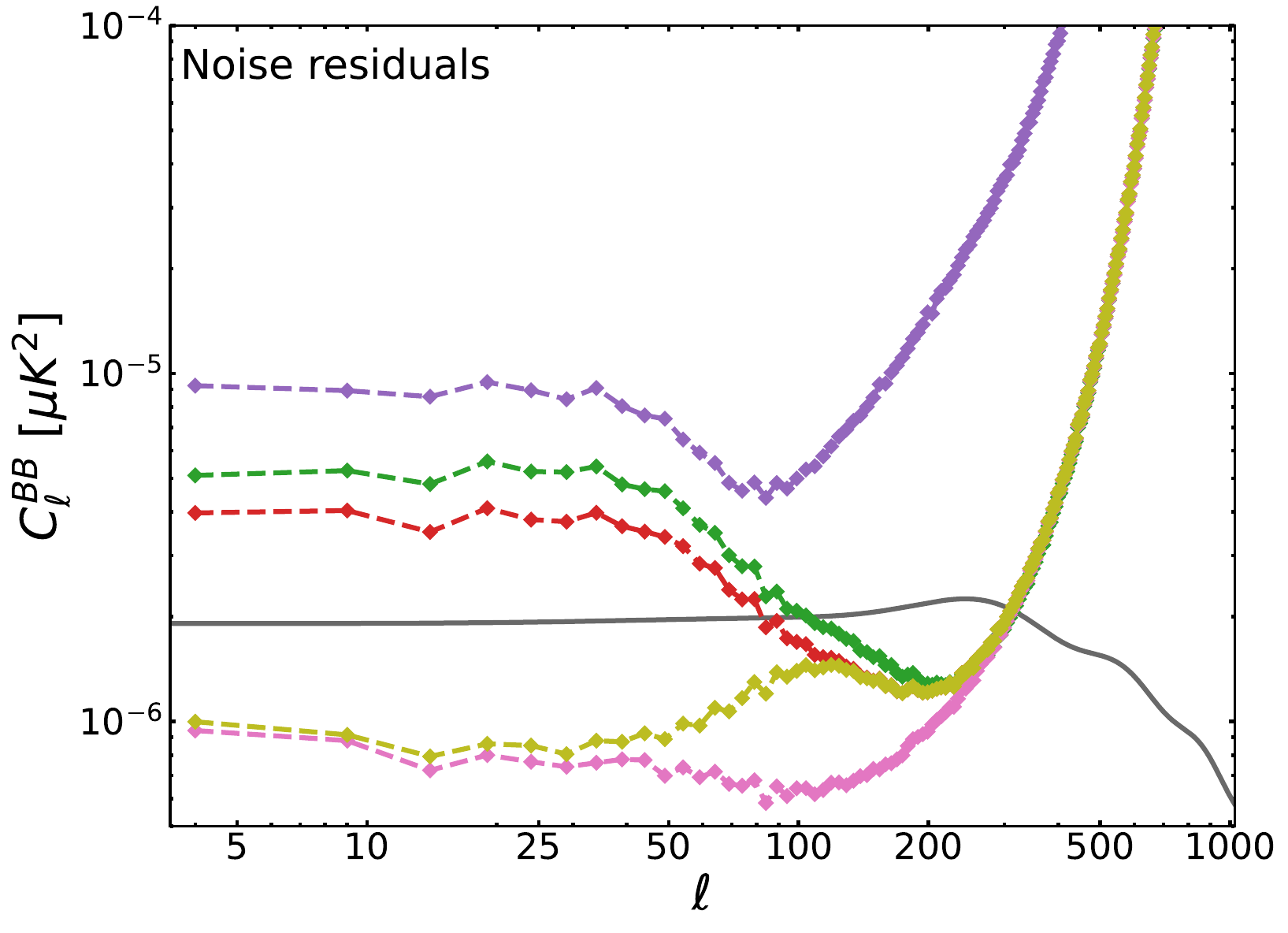}
\caption{Angular power spectra, averaged over $10$ distinct realizations. Left, middle and right panels report, respectively, power spectra of the output solution (denoised), foreground and noise residuals. The three rows report in order $C_{\ell}^{TT}$, $C_{\ell}^{EE}$ and $C_{\ell}^{BB}$. The denoised output is obtained by computing cross spectra between two simulated solution splits with uncorrelated noise residuals. Different colors reflect results for different component separation pipelines considered for this comparison. Angular power spectra are computed adopting different masking strategies for the different scalar fields, reatining sky fractions of $80$, $70$ and $60\ \%$, respectively, for $C_{\ell}^{TT}$, $C_{\ell}^{EE}$ and $C_{\ell}^{BB}$. Average angular power spectrum of input CMB signal is shown in black in the left panels (in some cases it fully or partially overlaps with other curves). The theoretical input spectrum used to generate CMB realization is also shown in all panels with a grey solid line.} 
\label{fig:cls_LB}
\end{figure}

Corresponding results for the angular power spectra of the three CMB scalar fields are shown in Figure~\ref{fig:cls_LB}. The spectra are computed using the \texttt{BROOM} \texttt{\_compute\_spectra} function and the \texttt{MASTER} formalism~\cite{MASTER,MASTER2}, as implemented in the \texttt{NaMaster} package~\cite{pymaster}. We report the angular power spectra averaged over the $10$ realizations considered in this analysis.
All power spectra are computed from masked scalar maps, using masks inherited from the Planck 2018 data analysis, specifically the Galactic masks derived from intensity data~\cite{Planck2018_compsep}. The retained sky fraction differs among the scalar fields, being $80\%$, $70\%$, and $60\%$ for $T$, $E$, and $B$, respectively.
For each field, we present separate plots showing the power spectra of the reconstructed CMB signal, as well as those of the residual foregrounds and reconstruction noise. The power spectra of the reconstructed solutions are properly debiased for noise by computing cross-power spectra between two independent splits. These splits are obtained by combining pairs of input data splits—characterized by uncorrelated but higher noise levels—using component-separation weights derived from the full-mission dataset. The same set of component-separation methods presented in the map-based analysis is also included in this power-spectrum-based comparison. For visualization purposes, all computed angular power spectra have been binned with constant binning of $\Delta\ell=5$.

In temperature, we observe an excellent reconstruction of the input power spectrum, characterized by very low scatter and negligible bias. At small angular scales, ILC exhibits higher levels of both foreground and noise residual power, whereas the pixel-based variance minimization yields lower reconstruction noise at large scales, where foreground intensity is more significant.
The relative behavior of reconstruction noise between ILC and NILC is expected: variance minimization in needlet space typically leads, respectively, to lower and higher reconstruction noise on smaller and larger angular scales, while the opposite trend is generally expected for foreground residuals—namely, reduced residuals at large scales for NILC and at small scales for ILC. This behaviour is indeed observed in polarization.
In temperature, however, this trend is not fully reproduced. We speculate that this difference arises from the additional complexity of small-scale temperature foregrounds, dominated by unpolarized extragalactic emission. Such contamination is more effectively mitigated by a variance minimization tailored to small angular scales (NILC), rather than by a global minimization that simultaneously accounts for all scales (ILC). We expect these results to hold for high-sensitivity experiments, such as the LiteBIRD-like configuration considered in this validation study. 

In the $E$-mode case, where we compare several different ILC-based approaches, we recover the expected main trends. The reconstructed CMB power spectrum is generally unbiased for all techniques, although with a lower signal-to-noise ratio compared to the intensity case. NILC and PILC, both implemented in the needlet domain, exhibit very similar behavior in terms of residual power. Finally, the inclusion of moment deprojection (\texttt{cNILC}) leads to a progressive reduction of foreground contamination, at the expected cost of increased reconstruction noise.

For $B$ modes, given the pessimistic masking strategy, we observe a residual large-scale bias for $\ell \lesssim 30$ in all component-separation techniques, with the exception of MC-NILC employing an ideal sky partition. The same pipeline, when applied using realistic clusters, still yields a significantly reduced bias compared to more traditional ILC approaches, while also achieving a reduction in reconstruction noise on large angular scales.
At higher multipoles, the residual power in the \texttt{(rMC)NILC} case converges to that of the standard \texttt{NILC}, since at those needlet scales the component separation effectively reduces to a standard local ILC minimization. In the $B$-mode case, progressively higher reconstruction noise relative to \texttt{NILC} is observed for \texttt{PNILC} and \texttt{cNILC0}.

We conclude by highlighting two important points. First, all component-separation pipelines presented here that enable the reconstruction of a signal with a fixed SED, described in Section~\ref{sec:cmb_rec} and implemented in the \texttt{BROOM} package, can also be applied to recover other signals with known SEDs beyond the cosmological background. Notable examples include the SZ effect and moments of foreground emission. Although results for these cases are not shown here for shortness, the corresponding pipelines have been internally tested and validated within the \texttt{BROOM} framework.
Second, a crucial aspect of any ILC-based reconstruction is the assessment of the so-called ILC bias. As already mentioned in Section \ref{sec:cmb_rec}, this bias arises from imperfect statistical estimates of the input covariance matrix derived from finite data samples and manifests itself as the artificial creation of chance correlations between output components—for example, between the reconstructed CMB signal and foreground or noise residuals. In the case of CMB reconstruction, this effect can be estimated in simulations by computing
\begin{equation}
C_{\ell}^{\text{bias}} = \langle C_{\ell}^{\text{out}} - C_{\ell}^{\text{CMB}} - C_{\ell}^{\text{fres}} - C_{\ell}^{\text{nres}} \rangle ,
\end{equation}
where $C_{\ell}^{\text{out}}$ is the angular power spectrum of the reconstructed output, $C_{\ell}^{\text{CMB}}$ corresponds to the input preserved CMB signal, and $C_{\ell}^{\text{fres}}$ and $C_{\ell}^{\text{nres}}$ denote the foreground and noise residual contributions, respectively. If this quantity is consistent with zero at all multipoles or bandpowers within the expected statistical scatter, one can robustly conclude that the signal recovery is not affected by ILC bias. This condition has been verified for all pipelines considered in this validation study.
To keep the ILC bias under control in practical applications, users must adopt an appropriate configuration of needlet bands (for needlet-domain implementations) and carefully select the size of the domains over which the input covariance matrix is estimated from the data. This is controlled by the component-separation parameter \texttt{ilc\_bias}, which sets the targeted level of residual ILC bias in the reconstruction. For example, setting \texttt{ilc\_bias} to $0.001$ corresponds to a theoretical residual bias at the level of one per mille.

\subsubsection{Foreground reconstruction and diagnostic}
\label{sec:fgds_rec_LB}

As a second step, we proceed with the validation and analysis of the pipelines outlined in Section~\ref{sec:fgds_rec}, which are designed for the blind reconstruction of multifrequency foreground emission and for diagnosing foreground complexity. The results presented here refer to pipelines applied in the needlet domain, adopting the needlet configuration shown in Figure~\ref{fig:needlets}. The methodologies are applied to the same $10$ LiteBIRD-like simulated datasets introduced above.

The blind reconstruction of foreground emission is performed using the GILC technique (see Section~\ref{sec:fgds_rec} for details), which provides estimates of the foreground signal with reduced contamination from nuisance components—specifically the CMB and instrumental noise in this case—at each observed frequency channel. We note that, while the \texttt{BROOM} user may choose to derive cleaned foreground templates only at selected frequency channels of interest, the default implementation produces such templates for all channels in the provided dataset. If one is interested in polarization-only reconstruction, GPILC can also be applied, in which case the variance of the reconstructed polarized intensity is minimized. The performance of GPILC is comparable to that of GILC applied separately to the $E$- and $B$-mode components.

\begin{figure}
\centering
\includegraphics[width=.9\textwidth]{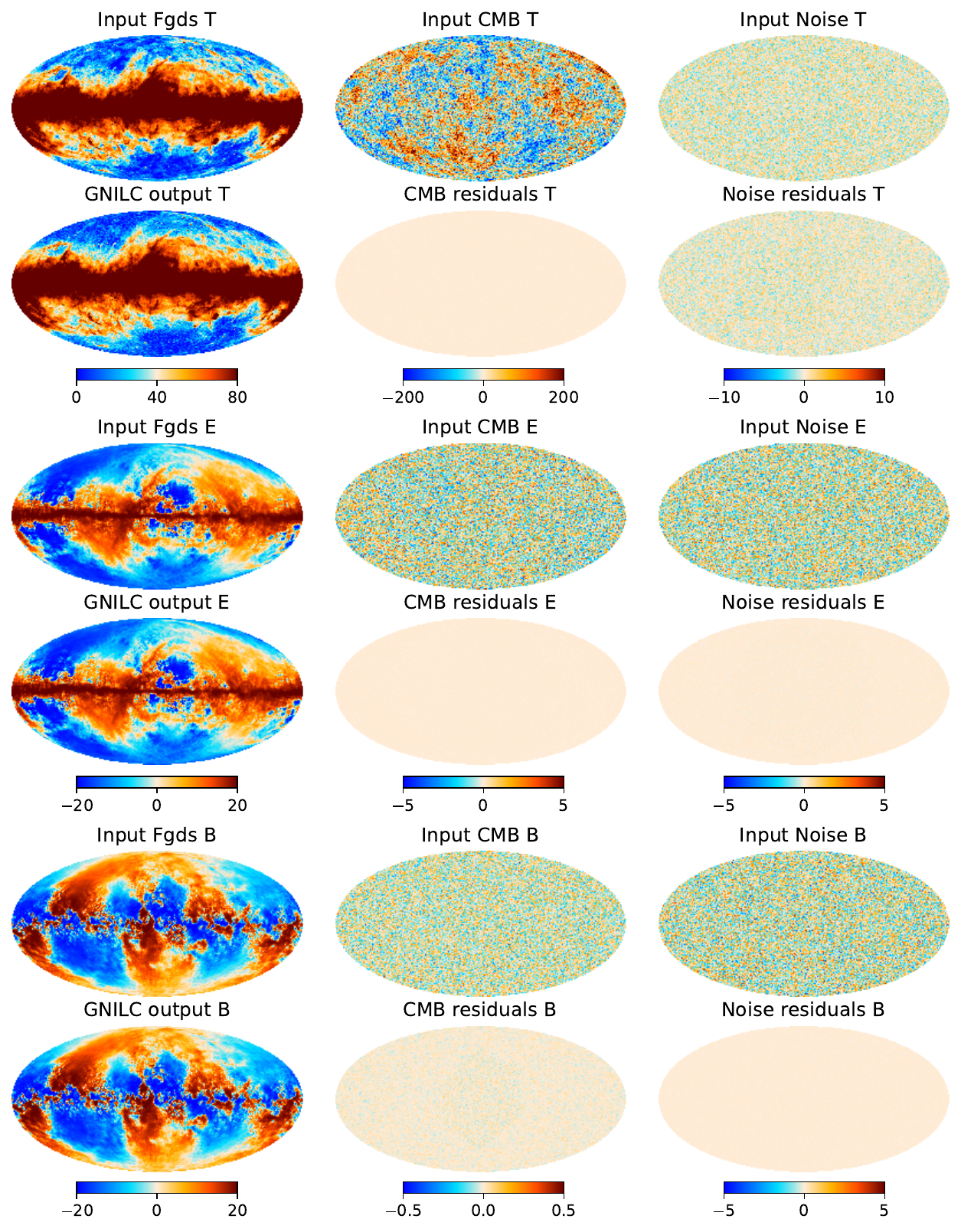}
\caption{Input and GILC-reconstructed maps for the $119$ GHz frequency channel of a simulated realization of the LiteBIRD-like experiment. Each pair of rows corresponds, from top to bottom, to a different scalar field: $T$, polarization $E$, and $B$ modes. In the left column, odd rows show the input total foreground emission, while the corresponding even rows display the total foreground reconstructed by GILC. The middle and right columns present, respectively, the input and output CMB and noise contributions. All maps are smoothed with a Gaussian beam of FWHM$=30'$.} 
\label{fig:maps_gnilc_lb}
\end{figure}
Figure~\ref{fig:maps_gnilc_lb} compares the GILC-reconstructed foreground maps with the corresponding input foreground emission for the LiteBIRD-like representative $119$ GHz frequency channel and for all scalar fields ($T$, $E$, and $B$). The figure also shows the residual CMB and noise maps obtained from GILC, alongside the corresponding input nuisance-component signals.
A visual inspection immediately reveals that, in temperature, the foreground emission is accurately recovered primarily close to the Galactic plane, with a minor loss of information at high Galactic latitudes and small angular scales. This behavior is attributable to the dominance of the CMB signal in temperature outside the Galactic plane over a substantial fraction of the observed frequency range. As a result, CMB modes dominate several entries of the input data covariance matrix, thereby limiting the effectiveness of the foreground reconstruction in regions where the CMB signal prevails.
In contrast, in polarization the Galactic foreground emission dominates over the CMB signal, leading to a visually robust reconstruction of the foregrounds across the full sky in both $E$ and $B$ modes. In all cases, the GILC-reconstructed maps exhibit substantially reduced contamination from CMB and instrumental noise compared to the corresponding input nuisance components.

\begin{figure}
\centering
\includegraphics[width=.32\textwidth]{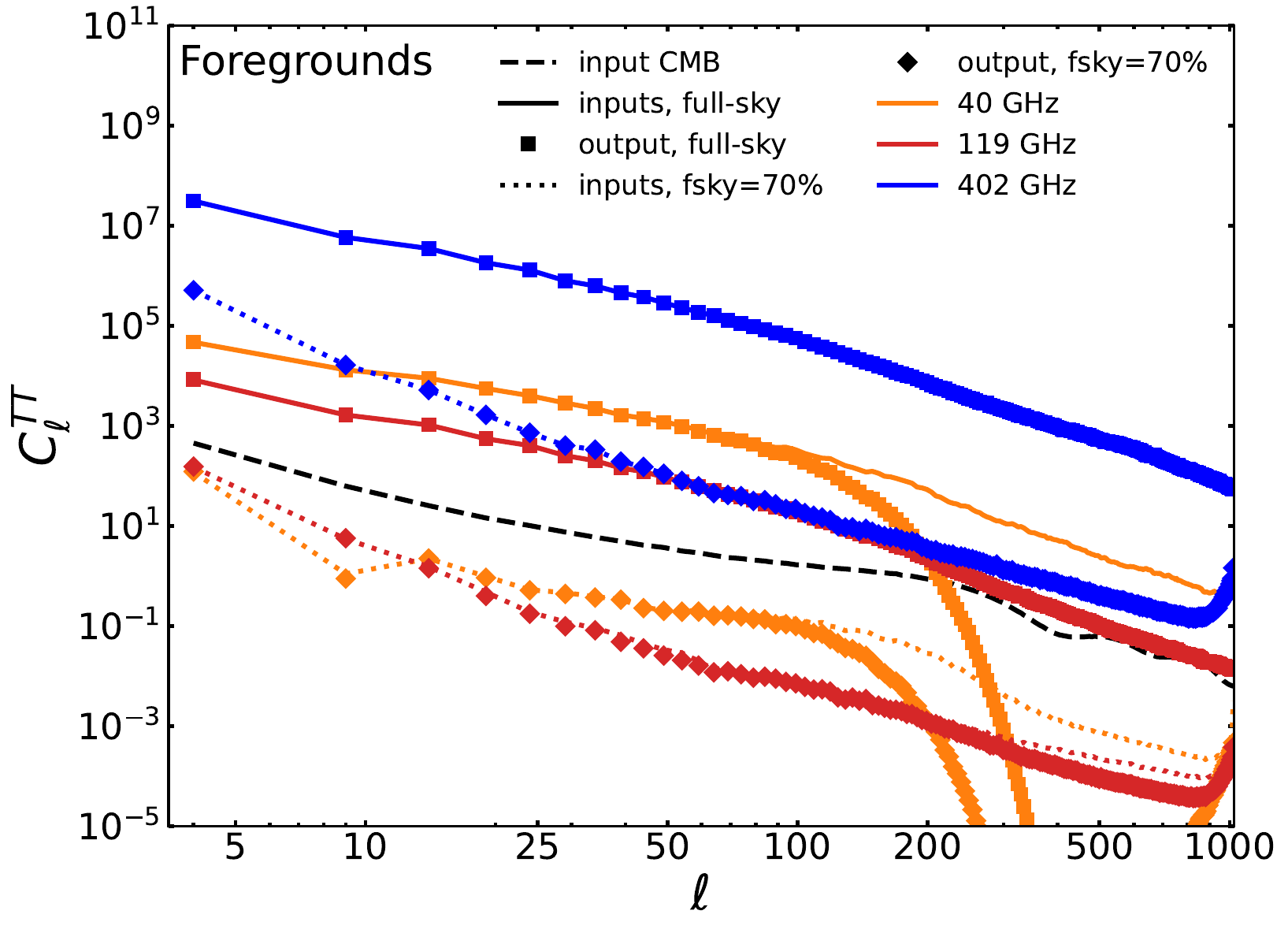}
\includegraphics[width=.32\textwidth]{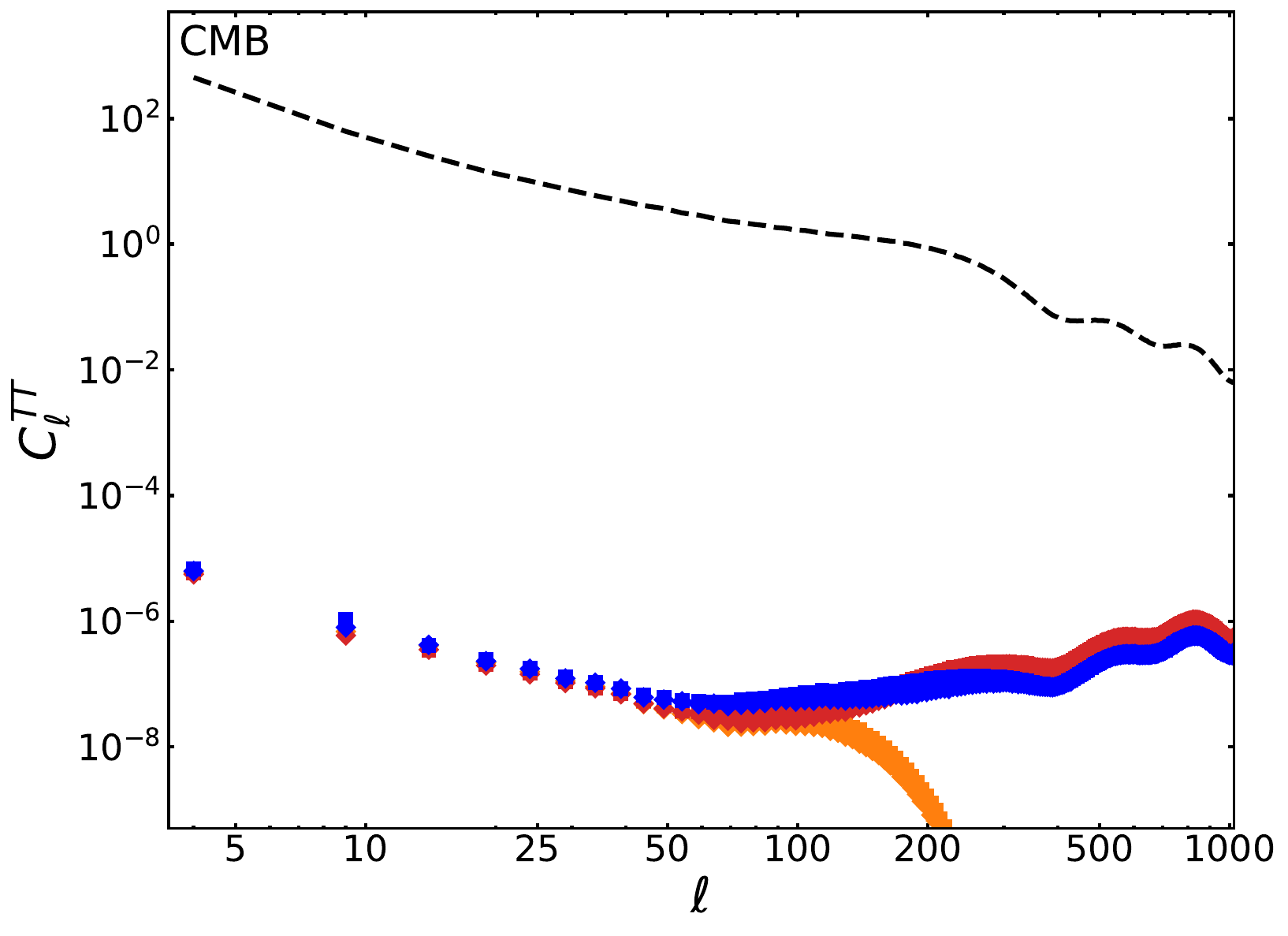}
\includegraphics[width=.32\textwidth]{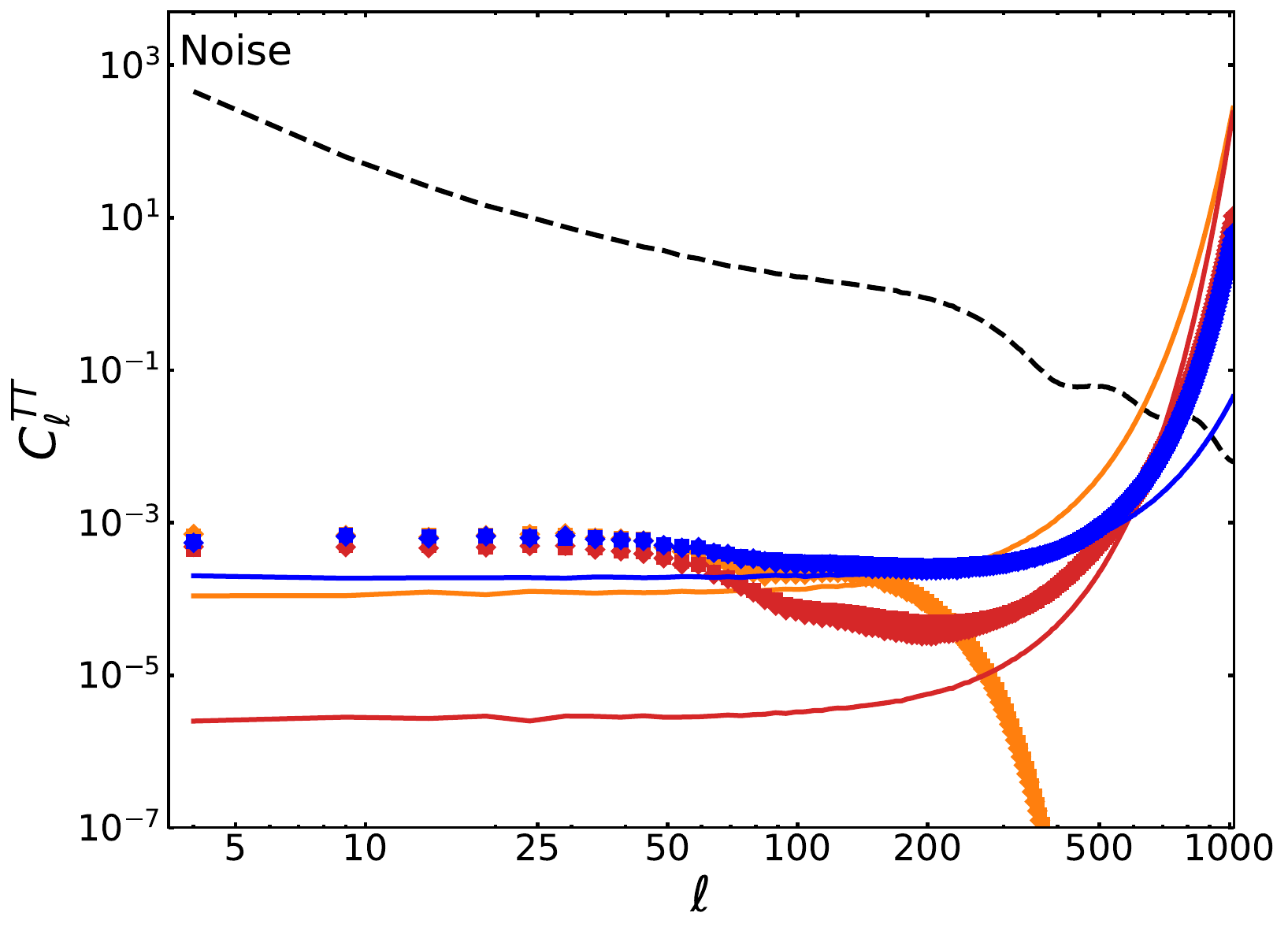}\\
\includegraphics[width=.32\textwidth]{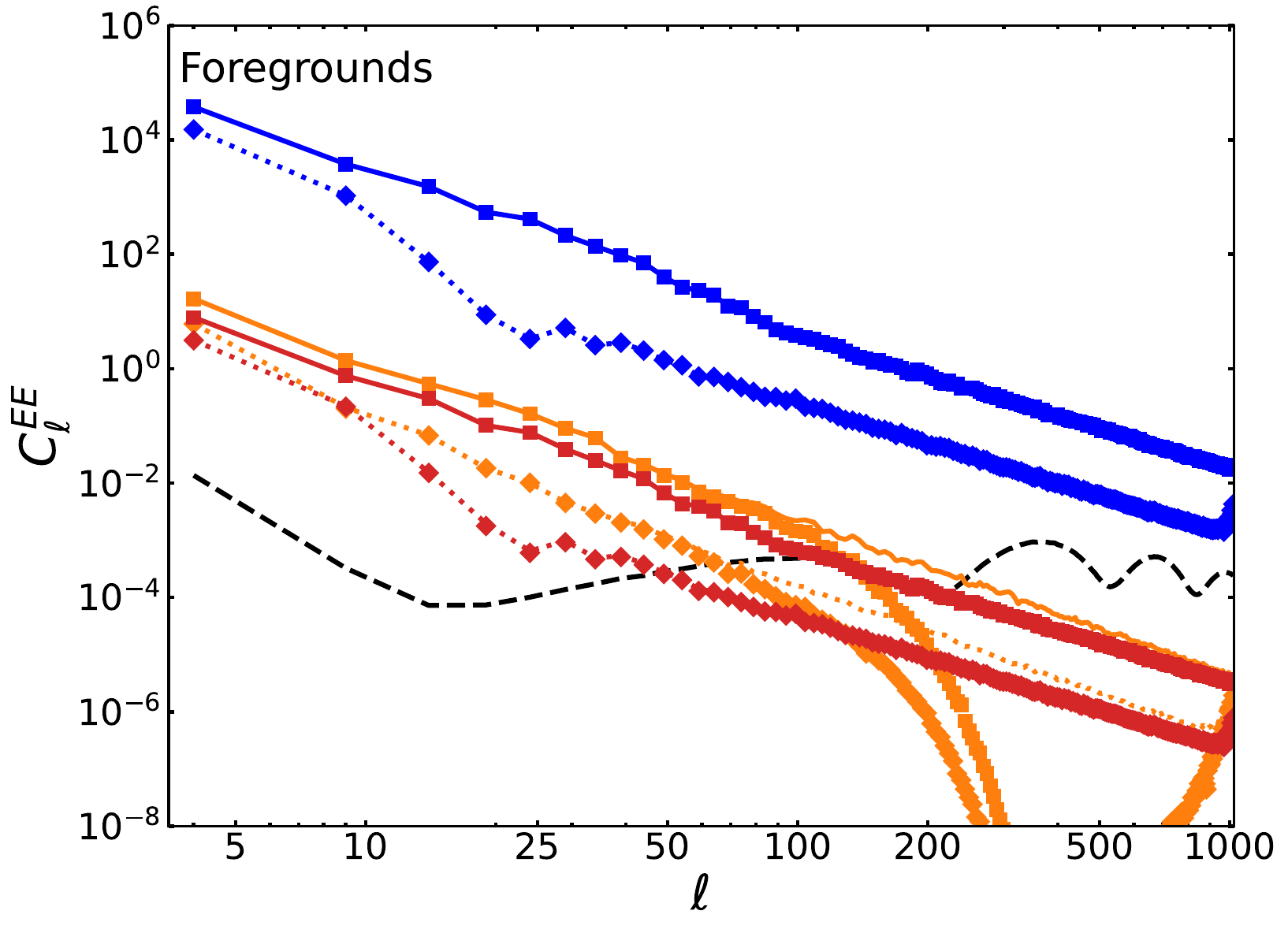} \includegraphics[width=.32\textwidth]{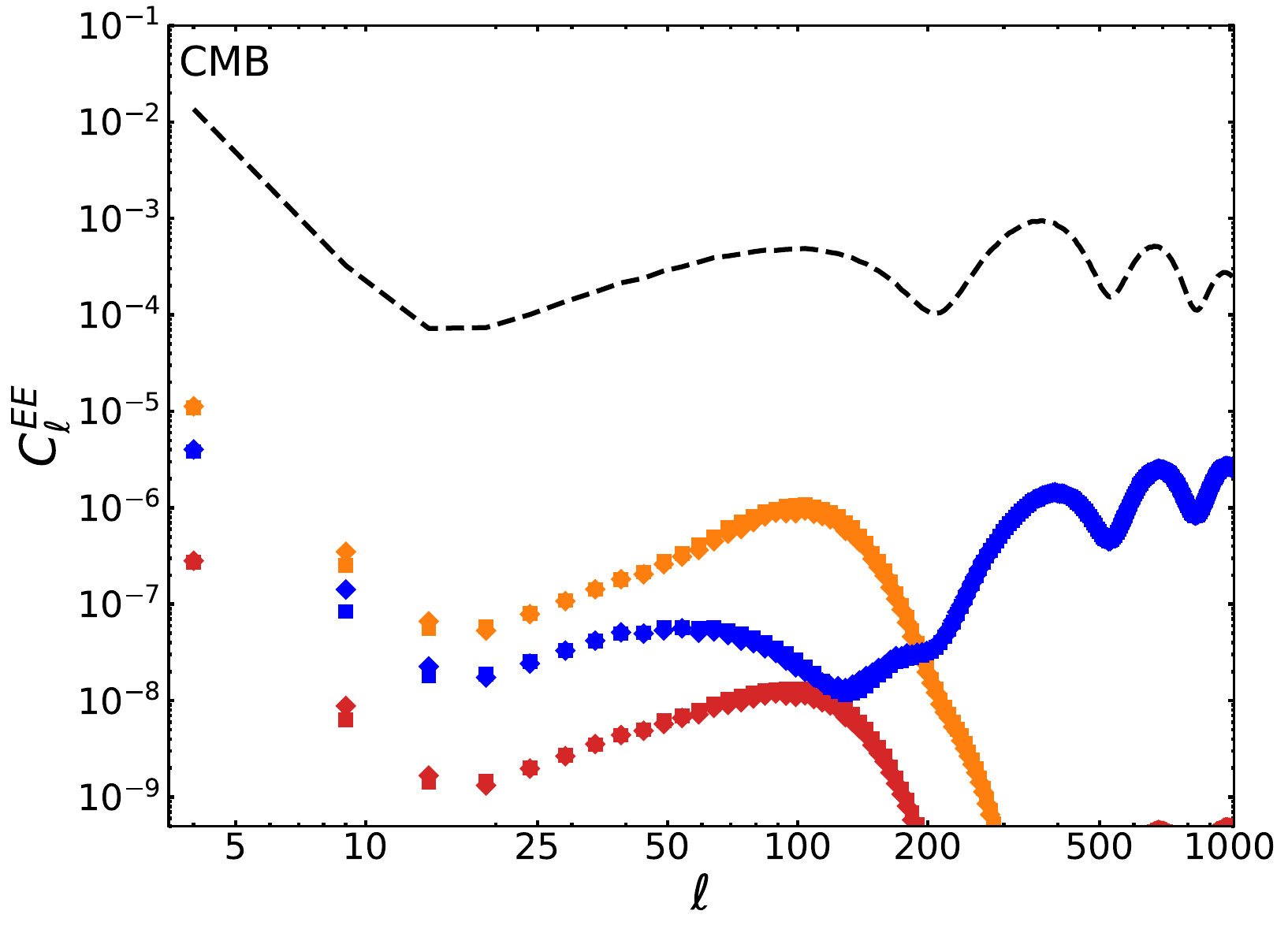}
\includegraphics[width=.32\textwidth]{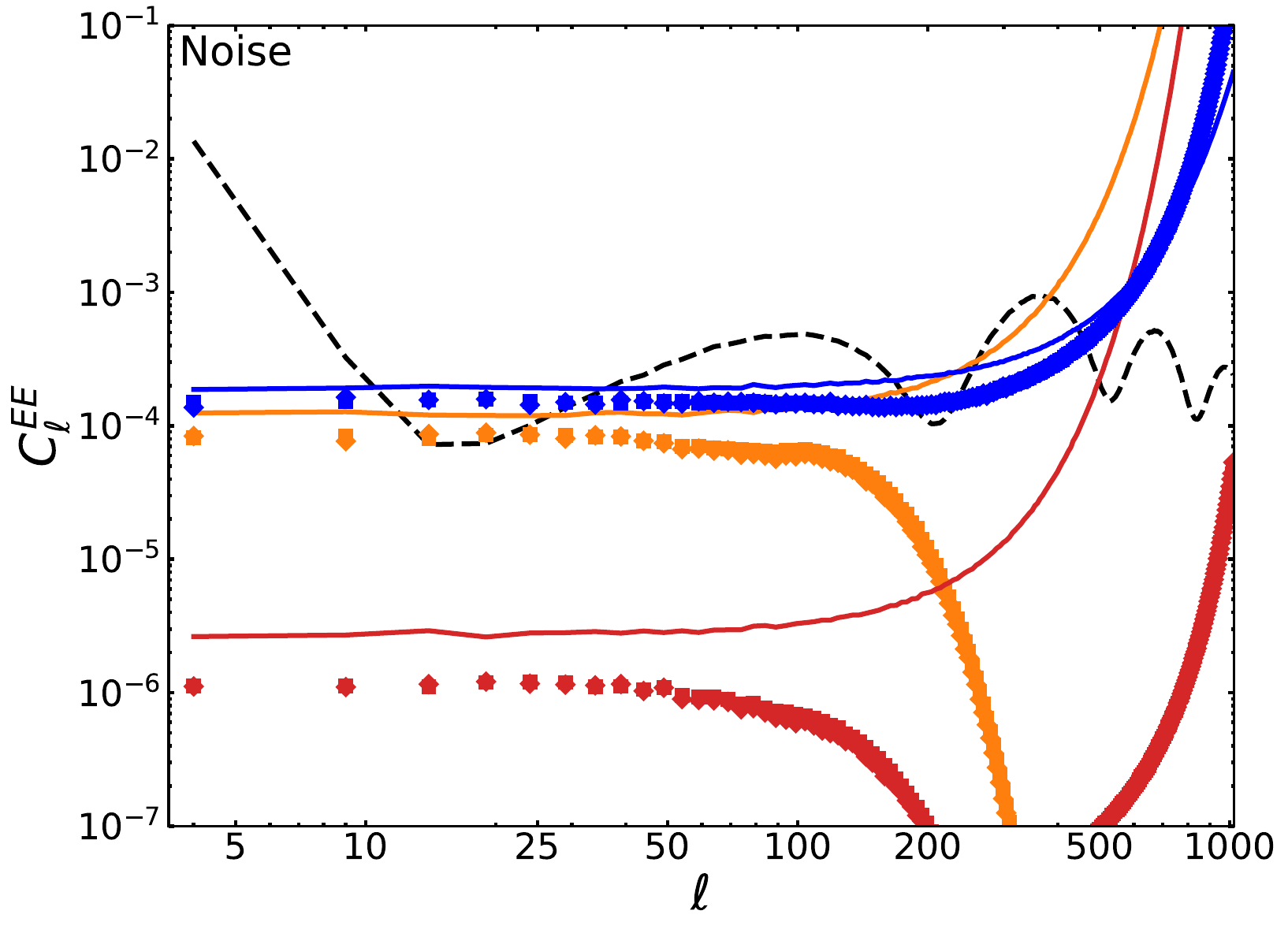}\\
\includegraphics[width=.32\textwidth]{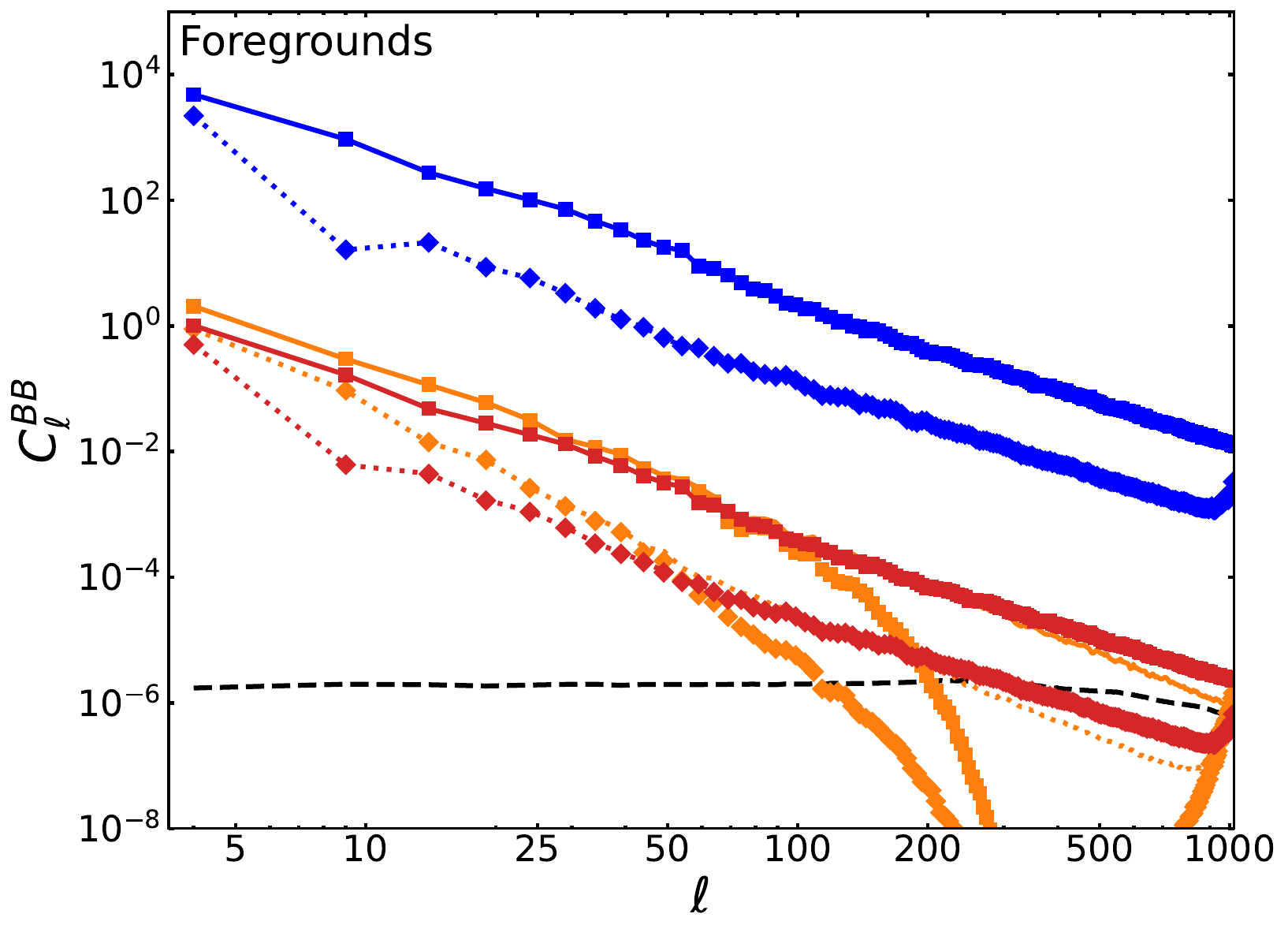} 
\includegraphics[width=.32\textwidth]{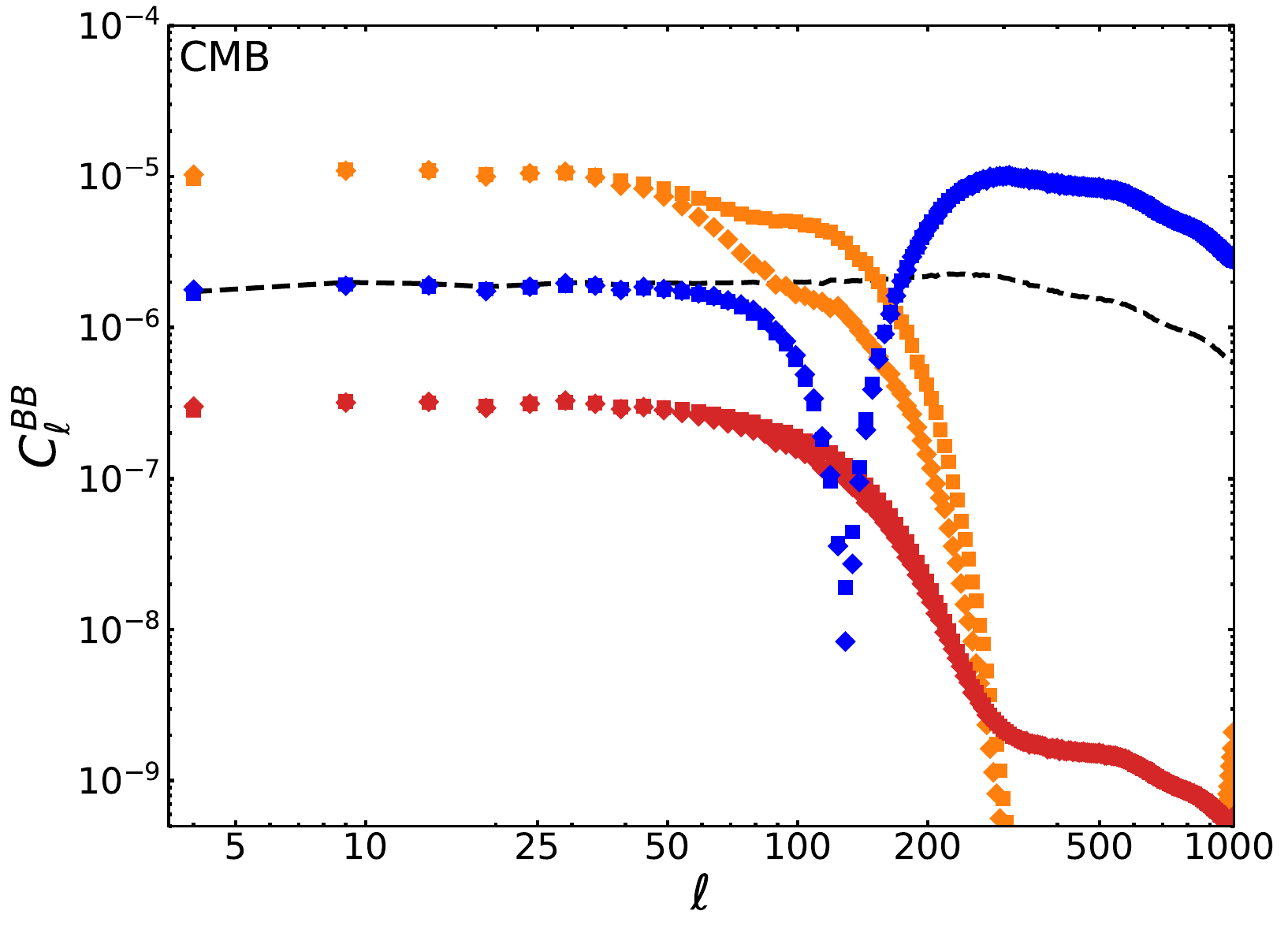} 
\includegraphics[width=.32\textwidth]{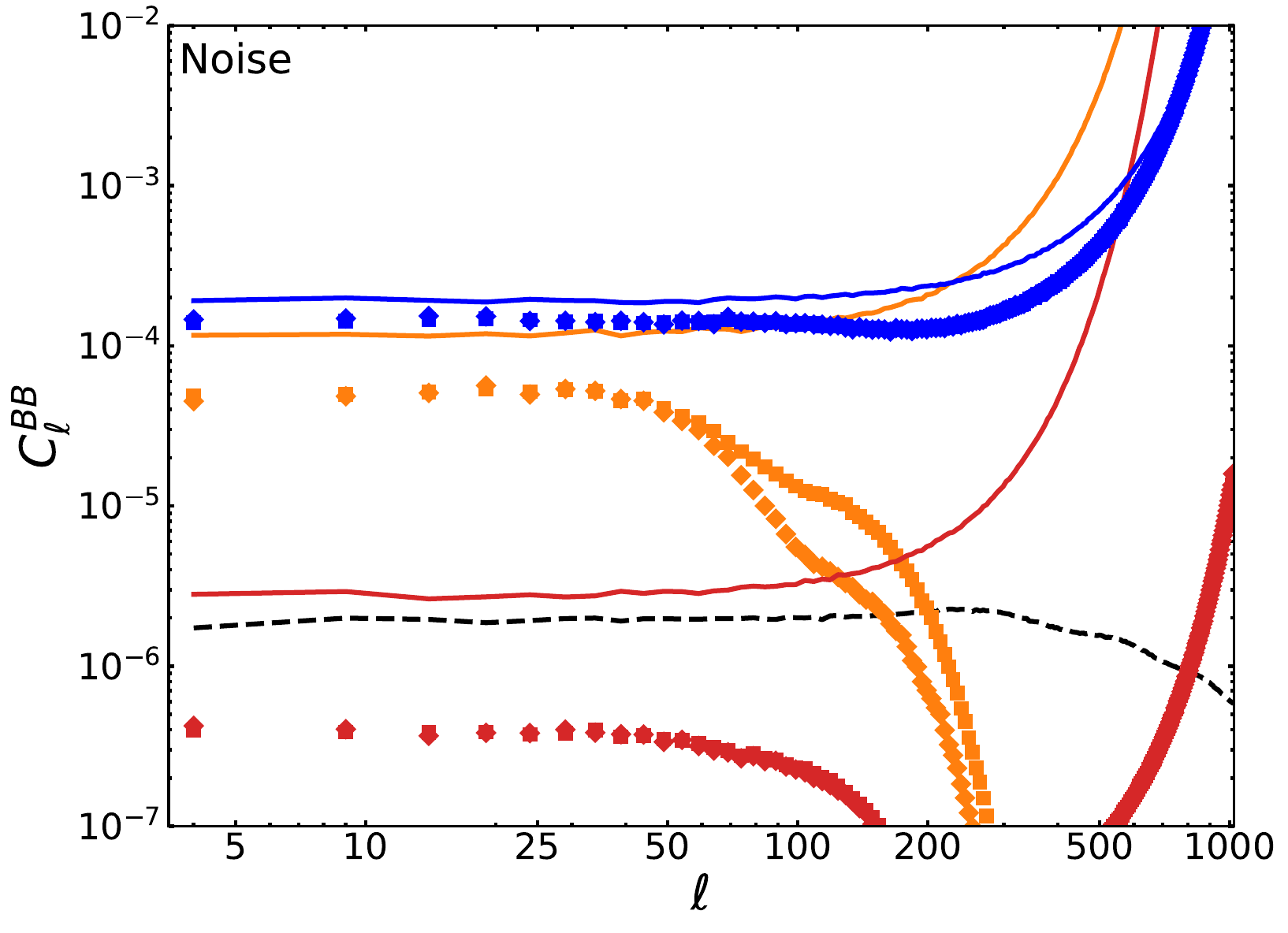} 
\caption{Angular power spectra of GILC input and output single components (total foregrounds: left; CMB: middle; noise: right) of temperature (top), $E$- (middle row) and $B$-mode (bottom) polarization. The results refer to average power spectra computed from an application to $10$ different realizations of LiteBIRD-like observations. Three different channels are reported: $40$ GHz (orange), $119$ GHz (red) and $402$ GHz (blue). Outputs are reported with markers while inputs with lines. Power spectra are computed either over the full sky (solid lines for the inputs and squares for the outputs) or adopting a Planck Galactic mask reatining $70\%$ of the sky (dotted lines for the inputs and diamonds for the outputs). The input theoretical angular power spectrum used for generating CMB simulations is shown with black dashed line. For the CMB panels, the input average angular power spectrum is not reported for simplicity but would collapse to the black dashed line.} 
\label{fig:gilc_spectra_LB}
\end{figure}
Figure~\ref{fig:gilc_spectra_LB} shows the corresponding angular power spectra for three representative frequency channels of the LiteBIRD-like configuration, namely $40$, $119$, and $402\,\mathrm{GHz}$. The plots display the angular power spectra of the individual GILC output components (total foreground, CMB, and noise), compared to those of the corresponding input components. The power spectra are computed both from full-sky maps and from masked maps, where the mask corresponds to the Planck intensity Galactic mask retaining a sky fraction of $f_{\rm sky}=70\%$.

Overall, we observe an optimal reconstruction of the foreground power spectrum at $119$ and $402\,\mathrm{GHz}$, both in temperature and polarization. At $40\,\mathrm{GHz}$, foreground recovery is limited to large angular scales due to the coarser angular resolution of the low-frequency channels, which prevents a reliable characterization of synchrotron emission on smaller scales. CMB contamination in the GILC output maps is significantly suppressed in both intensity and $E$-mode polarization. In contrast, for $B$ modes the level of residual CMB contamination depends on the frequency channel, with a more pronounced reduction in power in channels where the cosmological signal is comparatively dominant. Instrumental noise is attenuated in all polarization cases, whereas in temperature it is amplified as a consequence of the more aggressive deprojection of the CMB signal. Finally, the attenuation or amplification of both CMB and noise components shows only a weak dependence on the application of the sky mask. 
\begin{figure}
\centering
\includegraphics[width=.9\textwidth]{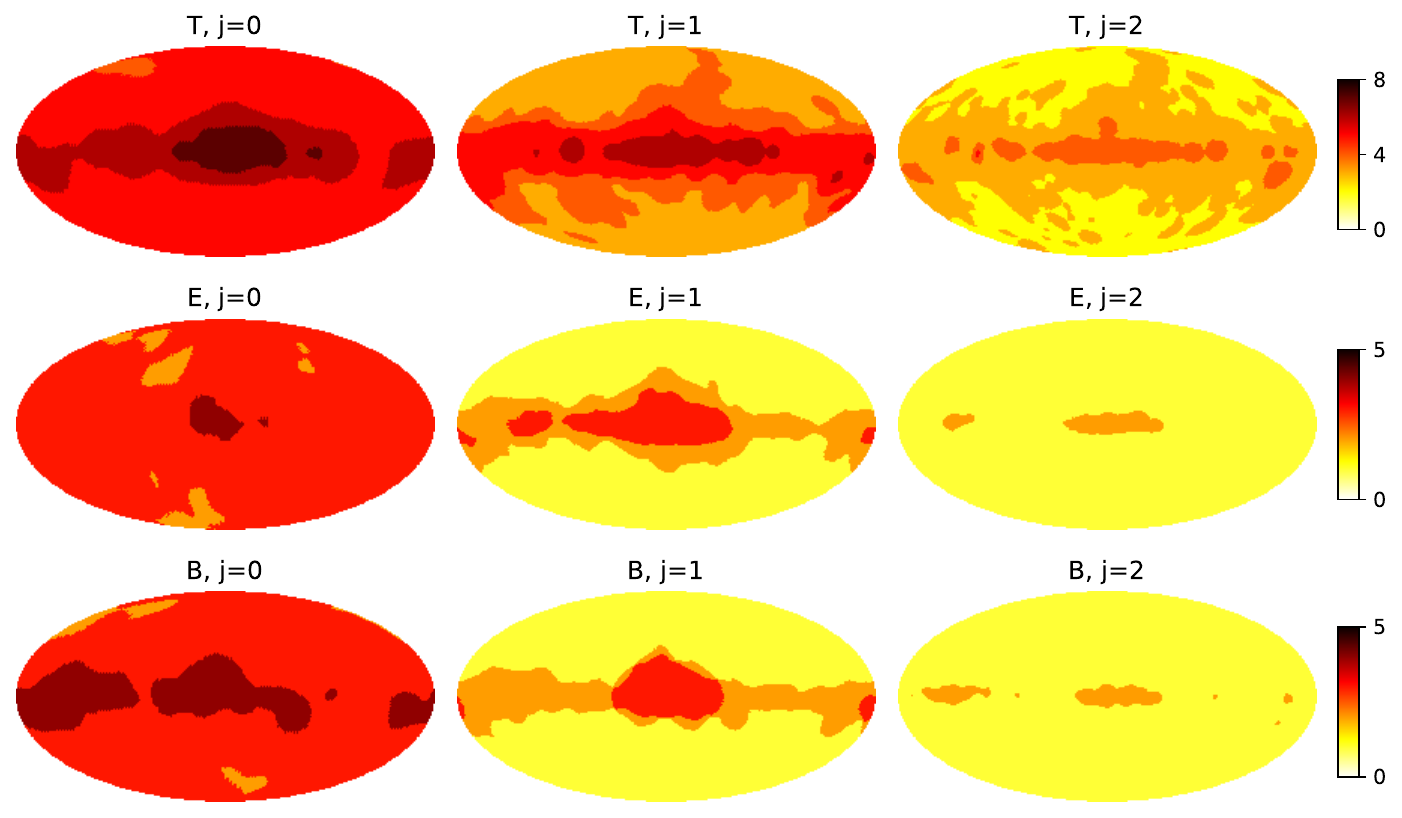}
\caption{Maps of the number of input covariance eigenmodes dominated by CMB foregrounds in temperature (top), $E$-mode (middle) and $B$-mode (bottom) polarization. Different columns refer to different needlet scales considered in the analysis. } 
\label{fig:diag_fgds_lb}
\end{figure}

The foreground diagnostic described in Section~\ref{sec:fgds_rec} has also been applied to and validated on LiteBIRD-like simulations. This diagnostic estimates the number of eigenmodes of the multifrequency input covariance matrix that are dominated by foreground emission or, more in general, by the non-nuisance components. In this case, we consider as nuisance components CMB and instrumental noise. As shown in \cite{ocMILC}, this information can also be exploited to improve the effectiveness of component-separation pipelines.
Results for temperature and for $E$- and $B$-mode polarization, computed for the first three needlet scales associated with the bands shown in Figure~\ref{fig:needlets}, are presented in Figure~\ref{fig:diag_fgds_lb}. As expected, we find that foregrounds exhibit a higher level of complexity in temperature than in polarization, with broadly similar trends observed for $E$ and $B$ modes. Another clear and anticipated behavior is the progressive reduction in the number of foreground-dominated modes as smaller angular scales (i.e., higher needlet scales $j$) are considered. This reflects the fact that foregrounds tend to dominate a larger fraction of the covariance at large angular scales, where their contribution is more significant relative to both the CMB signal and instrumental noise. As expected, we also observe a clear dependence on Galactic latitude, with higher foreground complexity primarily concentrated near the Galactic plane. 
We further anticipate that increased complexity extends to higher Galactic latitudes in polarization $E$- and $B$-mode maps. This is due to the non-local nature of the $Q/U \rightarrow E/B$ transformation, which redistributes polarized Galactic emission away from the plane.

The last pipeline to be validated within \texttt{BROOM} is the derivation of the spatial distribution of foreground residuals. This is obtained by combining the foreground emission estimated with GILC with the component-separation weights used for the reconstruction of the CMB signal, for example. As this procedure has already been validated in a recent study \cite{Carones_marg} using the \texttt{BROOM} routines, we refer the interested reader to that work for further details.

\subsection{Ground-based experiment case}
\label{sec:results_ground}
In this section, we present a representative subset of the results obtained from the validation of the \texttt{BROOM} pipeline for a ground-based CMB experiment. Specifically, we consider an SO-like instrumental setup, adopting the specifications of the SO Small Aperture Telescopes (SO-SATs) as reported in \cite{Wolz2024}. 
The SO-SATs are designed to achieve state-of-the-art reconstruction of the large-scale CMB polarization signal, combining unprecedented sensitivity with relatively modest angular resolution. Located at the Cerro Toco site in the Chilean Atacama Desert, the SO-SATs observe the microwave sky in six frequency channels spanning $27$ to $280\,\mathrm{GHz}$ \cite{SO_2019}. Science observations commenced in 2024.

\begin{figure}
\centering
\includegraphics[width=.6\textwidth]{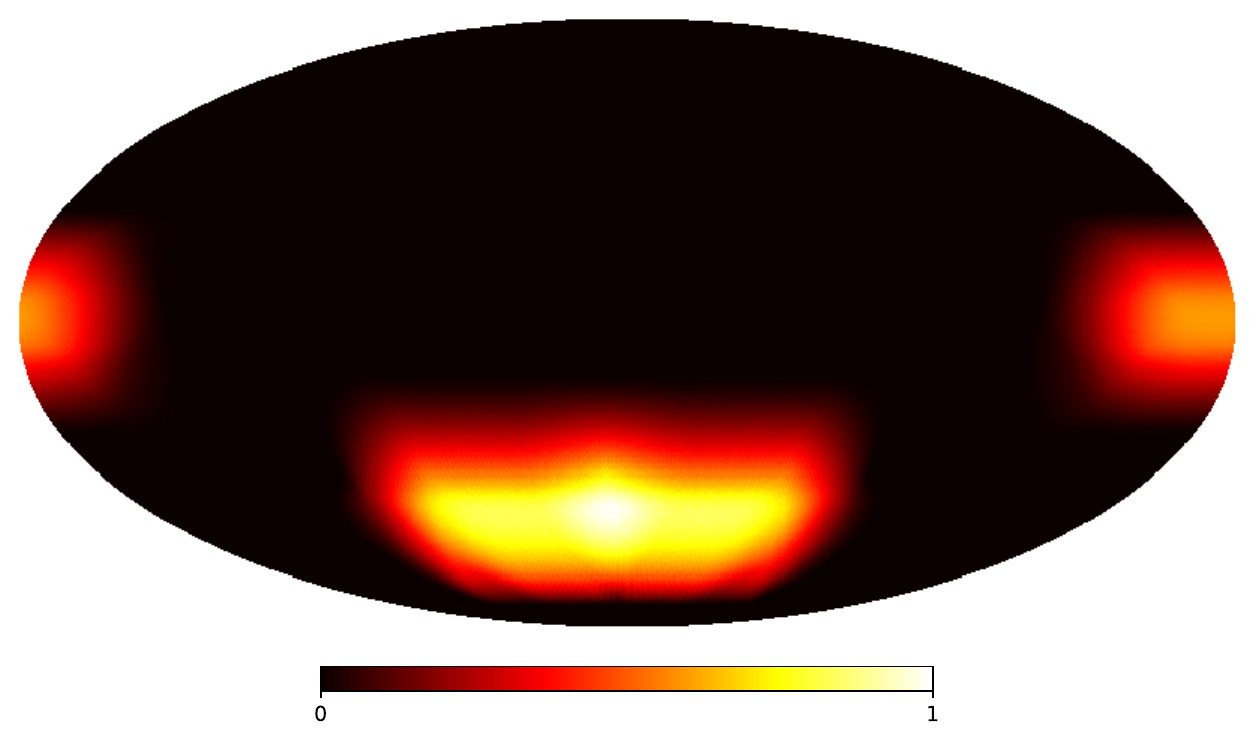}
\caption{Forecasted hit counts for the SO-SATs, taken from the analysis in \cite{Wolz2024}.}
\label{fig:so_hits}
\end{figure}

For ground-based observations, the assumption of isotropic noise is generally unrealistic due to the non-uniform scanning strategy of the telescopes. In such cases, the use of a hits-count map is preferable and can be readily implemented within \texttt{BROOM} through the instrument dictionary specifications (see Appendix~\ref{app:broom_sims} for details). In this validation, we adopt the hits-count map shown in Figure~\ref{fig:so_hits}, which matches the one used in the latest official SO-SAT forecast \cite{Wolz2024}. In this validation analysis, we assume the same hits-count map for all SO-SAT telescopes, but in principle sets of noise maps assuming a different hits-count map per frequency channel can be simulated within \texttt{BROOM}.

Although SO-SATs, like the LiteBIRD experiment, employ polarization modulators, the effective polarization noise spectrum in the reconstructed maps cannot be assumed to be white, even under idealized conditions. This is due to the filtering applied to the time-ordered data in standard analysis pipelines, which is designed to remove modes associated with, for example, atmospheric emission, detector thermal drifts, and scan-synchronous or HWP-synchronous signals. We refer the interested reader to \cite{Carlos_trasnfer_func} for a recent analysis of these effects in SO simulated data.
While such filtering simplifies the mapmaking process, it also leads to biased maps due to the simultaneous removal of signal modes. To account for this, transfer-function corrections are typically applied at the power-spectrum level. In practice, this enables an unbiased recovery of the input signal power, but results in an increase in the noise power, particularly on large angular scales.
As a useful approximation, the impact of this mode-loss correction can be modeled at the map level as a combination of coherent cosmological and astrophysical signals, together with a noise component characterized by a scale-dependent power spectrum that effectively captures the transfer-function correction induced by the filtering.
Specifically, the noise power spectrum is assumed to have a functional form as in Equation \ref{eq:cl_noise}
where $N_{w}$ denotes the white-noise level, while $\ell_{\mathrm{knee}}$ and $\alpha_{\mathrm{knee}}$ characterize the contribution arising from filtering corrections.
All these noise components can be incorporated in the noise-generation step within \texttt{BROOM}. The specific values adopted for $\sigma^{T/P}$, $\ell_{k}$ and $\alpha$ are taken from Table~3 of \cite{Wolz2024} considering the baseline optimistic instrumental scenario.

We note that a comprehensive implementation of ILC pipelines for frequency maps affected by frequency-dependent filtering, including the derivation of the corresponding ILC transfer functions and the assessment of the impact of such filtering on effectiveness of component separation, is currently ongoing.

\begin{figure}
\centering
\includegraphics[width=.85\textwidth]{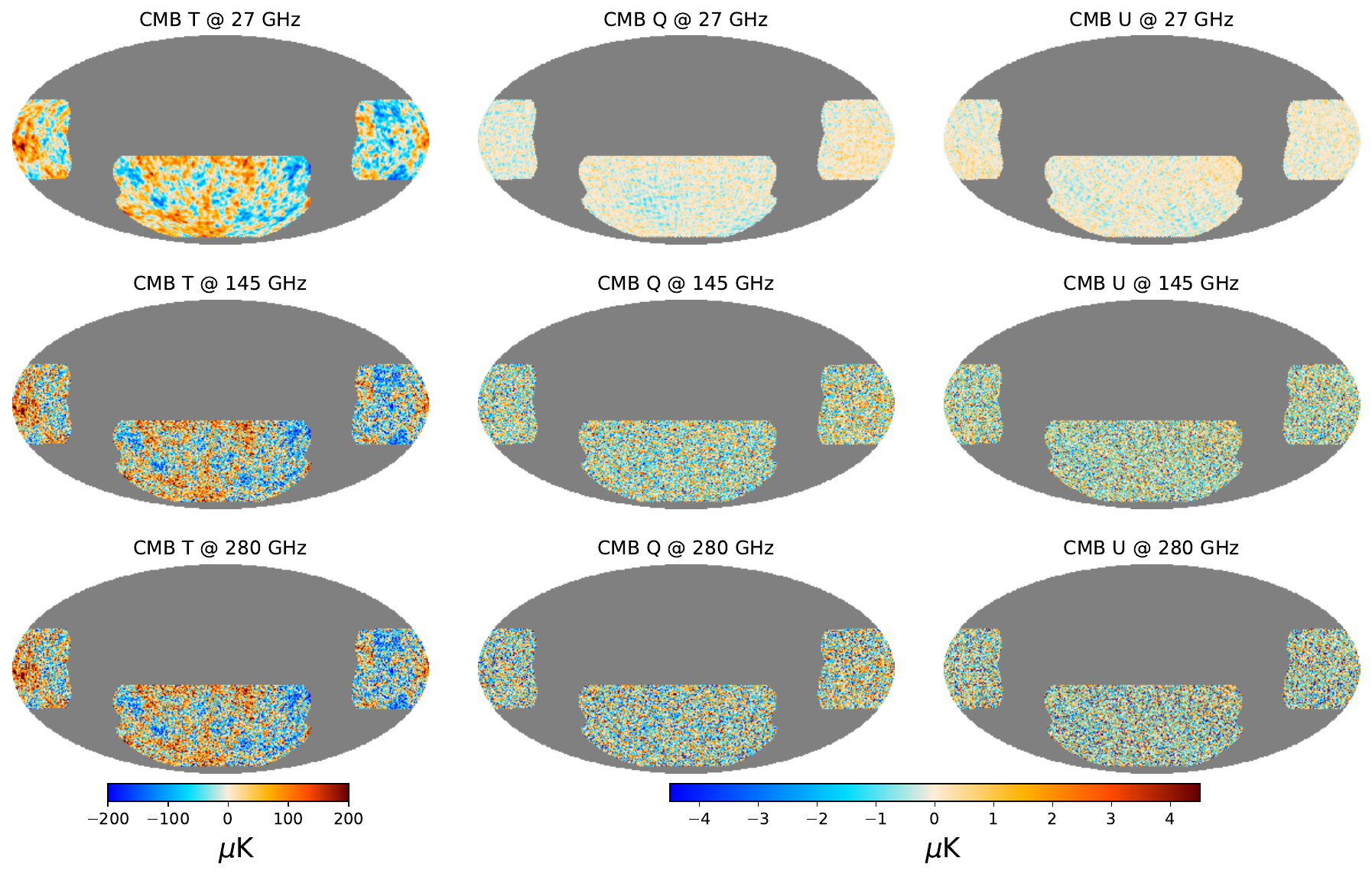}
\caption{Examples of the input CMB Stokes parameter maps: temperature $T$ (left panels), and polarization $Q$ (central panels) and $U$ (right panels). Three representative SO-SAT-like frequency channels are shown: a low-frequency channel ($27$\,GHz, top), a central channel ($145$\,GHz, middle), and a high-frequency channel ($280$\,GHz, bottom). The apparent differences in power across channels arise from the convolution of the same underlying CMB signal with the instrument beams, whose FWHM decreases with increasing frequency.
} 
\label{fig:so_cmb}
\end{figure}

\begin{figure}
\centering
\includegraphics[width=.85\textwidth]{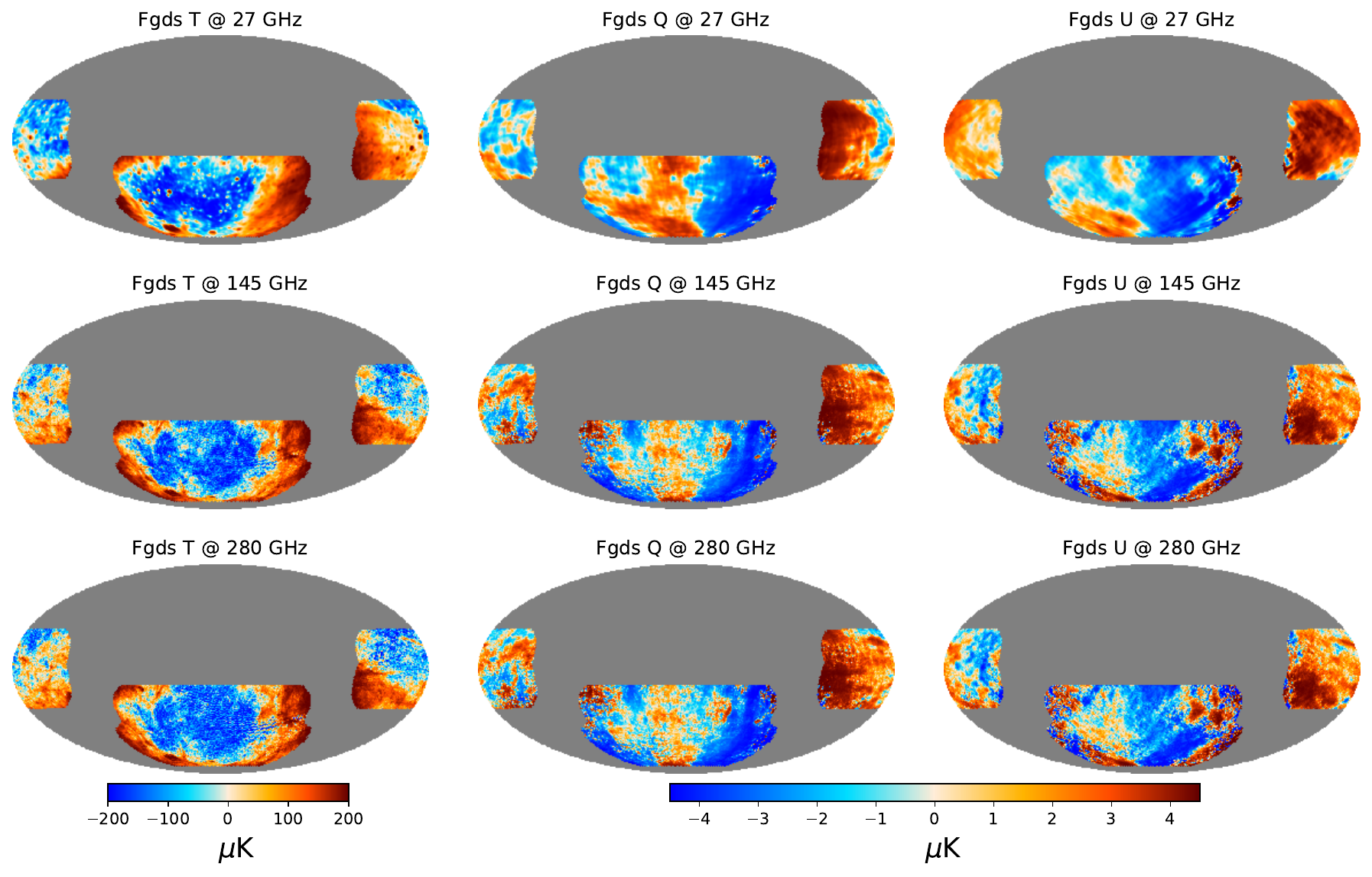}
\caption{Examples of the input Stokes parameter maps of the Galactic and extragalactic CMB foregrounds: temperature $T$ (left panels), and polarization $Q$ (central panels) and $U$ (right panels). Three representative SO-SAT-like frequency channels are shown: a low-frequency channel ($27$\,GHz, top), a central channel ($145$\,GHz, middle), and a high-frequency channel ($280$\,GHz, bottom). They are all represented assuming Equatorial coordinates.} 
\label{fig:so_fgds}
\end{figure}

\begin{figure}
\centering
\includegraphics[width=.85\textwidth]{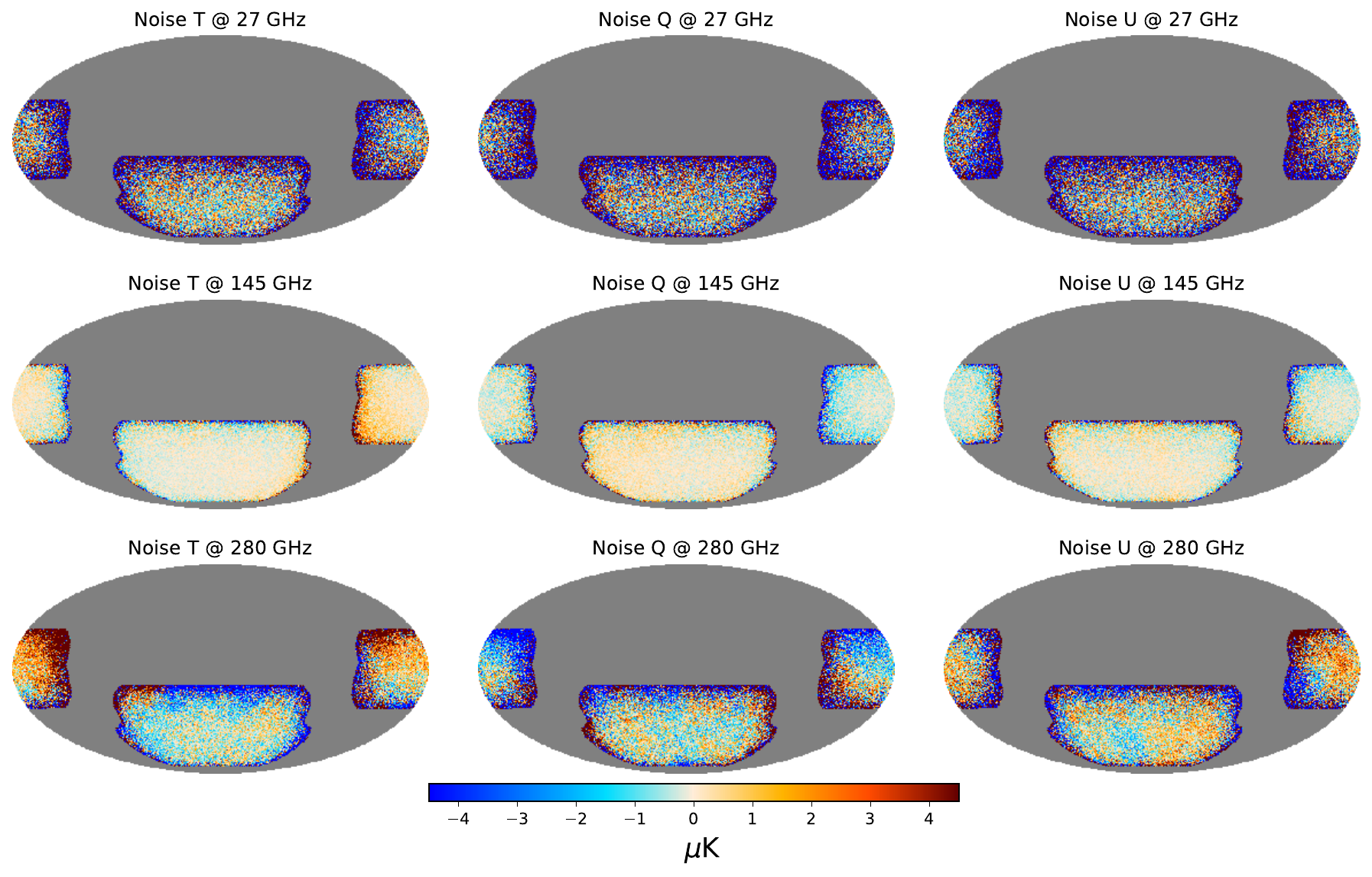}
\caption{Examples of the input noise Stokes parameter maps: temperature $T$ (left panels), and polarization $Q$ (central panels) and $U$ (right panels). Three representative SO-SAT-like frequency channels are shown: a low-frequency channel ($27$\,GHz, top), a central channel ($145$\,GHz, middle), and a high-frequency channel ($280$\,GHz, bottom). They are realizations of anisotropic and correlated noise made assuming the noise power spectrum of Equation \ref{eq:cl_noise} and the hits count map in Figure \ref{fig:so_hits}. Maps are in Equatorial coordinates.} 
\label{fig:so_noise}
\end{figure}

Examples of input maps—including CMB, total foregrounds, and noise—for three representative SO-SATs frequency channels ($27$, $145$, and $280,\mathrm{GHz}$) are shown in Figures~\ref{fig:so_cmb} (CMB), \ref{fig:so_fgds} (total foregrounds), and \ref{fig:so_noise} (noise). A visual inspection clearly reveals the anisotropic pattern of the noise. We also note that the sky-signal maps at different frequency channels have different intrinsic angular resolutions, as they are smoothed according to the corresponding instrumental beams. All maps presented in this validation of the ground-based configuration are shown in equatorial coordinates. The displayed signals are restricted to the forecasted SO-SAT observed region, as defined by the hit-count map shown in Figure~\ref{fig:so_hits}.

In this validation study, for simplicity we report results only for $E$- and $B$-mode polarization. As for satellite experiment case, we consider several component-separation pipelines, even though in reduced number. We first highlight the main procedural differences with respect to the satellite-based case. When analyzing simulations, signal maps such as the CMB and foregrounds are generated over the full sky; however, the data set effectively available for component separation is just over a limited sky region (shown in Figure \ref{fig:so_hits}). In such a framework, simulated data are then treated as partial-sky observations by specifying an observation mask—which sets all unobserved pixels to zero—via the \texttt{mask\_observation} parameter in \texttt{BROOM}. Given the highly anisotropic scanning strategy, component separation can account for the resulting uneven noise distribution by assigning larger weights to pixels, or more generally to sky regions, that are more frequently observed in the computation of the data covariance matrix (Equation~\ref{eq:data_cov}). This is achieved by specifying a weighting mask—corresponding here to the normalized hit-count map—through the \texttt{mask\_covariance} parameter in \texttt{BROOM}. We found this weighting procedure in the covariance computation to lead to improved component-separation performance.

As in the full-sky satellite case, polarization outputs can be provided either as $Q U$ or $E B$ maps, even for partial-sky data, as specified by the \texttt{field\_out} keyword. As anticipated at the end of Section~\ref{sec:basics}, $EB$ leakage in polarization component separation can be accounted for within the \texttt{BROOM} package. This can be achieved prior to the component-separation step, by applying one of the following map-based techniques: i) a standard purification procedure~\cite{Bunn2003,Smith2006,Smith2007}, as implemented in the \texttt{NaMaster}\footnote{\url{https://github.com/LSSTDESC/NaMaster/tree/master/pymaster}} package~\cite{pymaster}; ii) recycling technique~\cite{2019PhRvD.100b3538L} followed by iterative harmonic deprojection of residual ambiguous modes or diffusive inpainting~\cite{NILC_cutsky}; iii) diffusive inpainting followed by iterative harmonic deprojection of residual ambiguous modes. This approach enables the reconstruction of cleaned $E$- and $B$-mode maps with minimal residual leakage contamination over a partial sky, thereby allowing a reliable subsequent computation of the angular power spectra. Alternatively, the leakage can be propagated through the component-separation step without prior correction and subsequently addressed at the power-spectrum level. In this case, the outputs must be reconstructed in terms of $Q$ and $U$ maps to enable a proper correction using the \texttt{NaMaster} implementation for spin-2 fields. All these approaches are supported within \texttt{BROOM} through the appropriate configuration of the \texttt{leakage\_correction} parameter. Further details are provided in Appendix~\ref{app:compsep_broom}.

In this validation study, output maps are presented in terms of $E$- and $B$-mode components, while angular power spectra are computed from the corresponding $Q$ and $U$ maps. In this case, a final purification of the polarization harmonic coefficients is performed before estimating the angular power spectra. All angular power spectra are reported over the multipole range $\ell \in [30, 500]$, using a binning scheme with $\Delta\ell = 10$.

As in the satellite validation case, all outputs from the component-separation routines considered here have an intrinsic angular resolution corresponding to a Gaussian beam with $\mathrm{FWHM}=30^{\prime}$. Consequently, input maps at different frequencies are first brought to a common angular resolution. Adopting $\mathrm{FWHM}=30^{\prime}$ implies that the lowest-frequency channels must be convolved with an effective harmonic transfer function that enhances small-scale modes. While such an operation does not typically introduce severe artifacts in full-sky analyses, it becomes more problematic for partial-sky data, where harmonic coefficients are affected by mode mixing and potential ringing.
To mitigate these effects, when a \texttt{mask\_observations} is present, a regularization procedure is applied in cases of aggressive convolutions (i.e., when the effective transfer function exceeds a given threshold). In practice, the transfer function is tapered to zero at high multipoles. Although this approach does not provide a full correction of the beam window function, it ensures a stable and reliable reconstruction of modes up to well-defined angular scales for the lowest-frequency channels.

We finally note that methodologies based on minimizing the power of the polarization intensity,
$P^{2} = Q^{2} + U^{2}$, such as \texttt{P(R)ILC}, \texttt{cP(R)ILC}, and \texttt{GP(R)ILC}, may in principle appear to avoid the $E$--$B$ leakage issue, since they are not applied separately to $E$- and $B$-mode maps. However, in practice, these approaches still require the $QU$ maps to be brought to a common angular resolution prior to component separation. This operation generally involves a harmonic transformation, which can introduce mixing between polarization modes on a cut sky. As a consequence, $E$--$B$ leakage remains, in practice, an unavoidable issue in polarization analyses of partial-sky data.

The validation run for the ground-based experiment is performed using the \texttt{script\_paper} \texttt{\_groundbased.py} script provided in the \texttt{tutorials} directory of the GitHub repository. Additional representative examples are available in the \texttt{tutorial\_groundbased.ipynb} notebook, located in the same folder.

Further complementary insights into the application of ILC-based component-separation techniques to simulated SO-SAT data can also be found in \cite{2026_Mustafa}.

\subsubsection{Reconstruction of the CMB signal}
\label{subsec:cmb_SO}

\begin{figure}
\centering
\includegraphics[width=.9\textwidth]{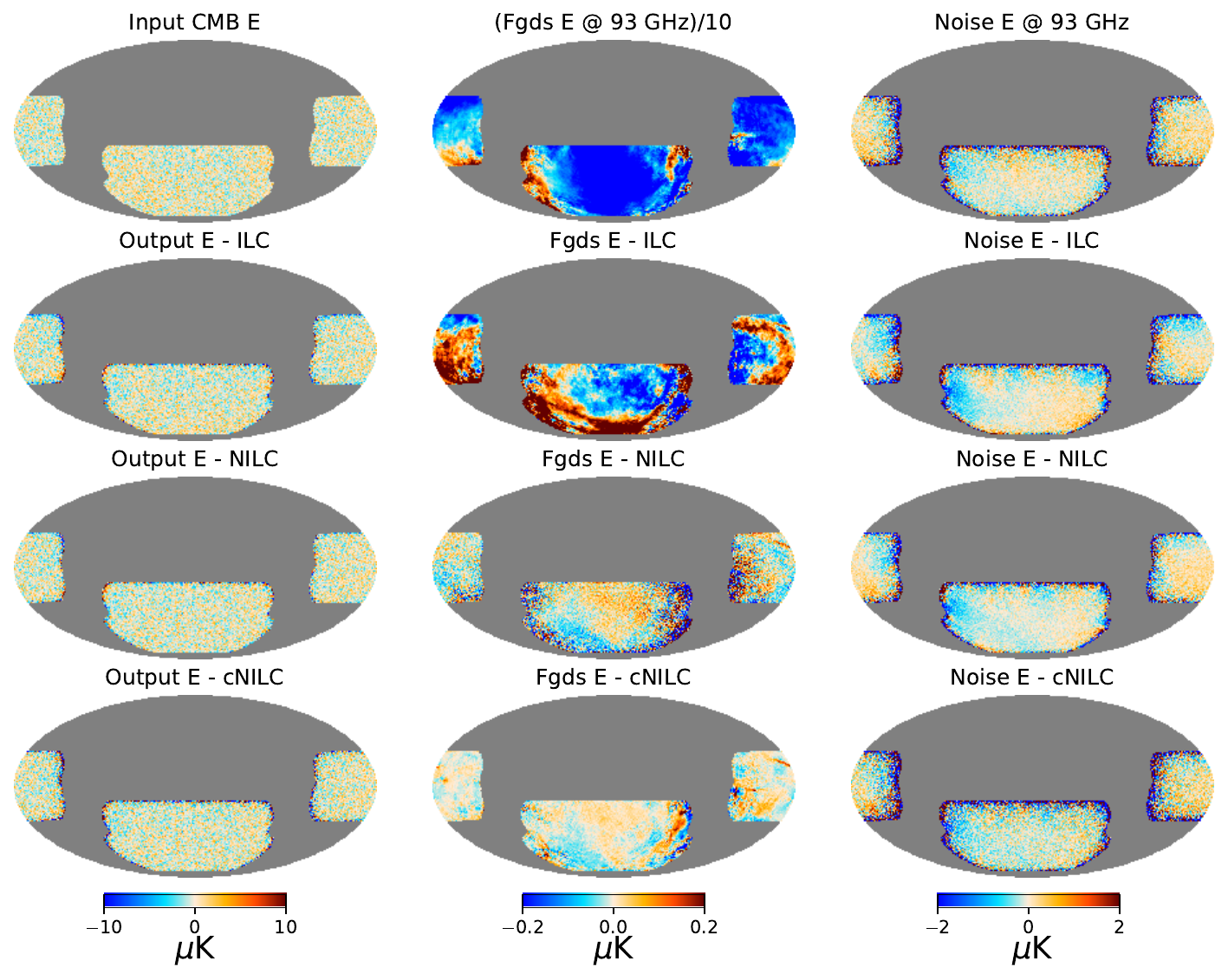}
\caption{$E$-mode maps from SO-SATs simulations. The first row illustrates the individual input components for a representative SO-SAT-like simulation at $93\,\mathrm{GHz}$: the CMB signal (left), foreground emission (middle, reduced for visualization purposes by a factor $10$), and instrumental noise. The other rows show the corresponding outputs from the ILC (second), NILC (third), cNILC (last) component separation. The three columns display, respectively, the total reconstructed signal, the foreground residuals, and the noise residuals. The maps in the last two columns are obtained by applying the component-separation weights to the input foreground-only and noise-only maps. All maps are shown in Equatorial coordinates and share a common angular resolution corresponding to a Gaussian beam with $\mathrm{FWHM}=30'$.} 
\label{fig:SO_ouputs_E}
\end{figure}
\begin{figure}
\centering
\includegraphics[width=.9\textwidth]{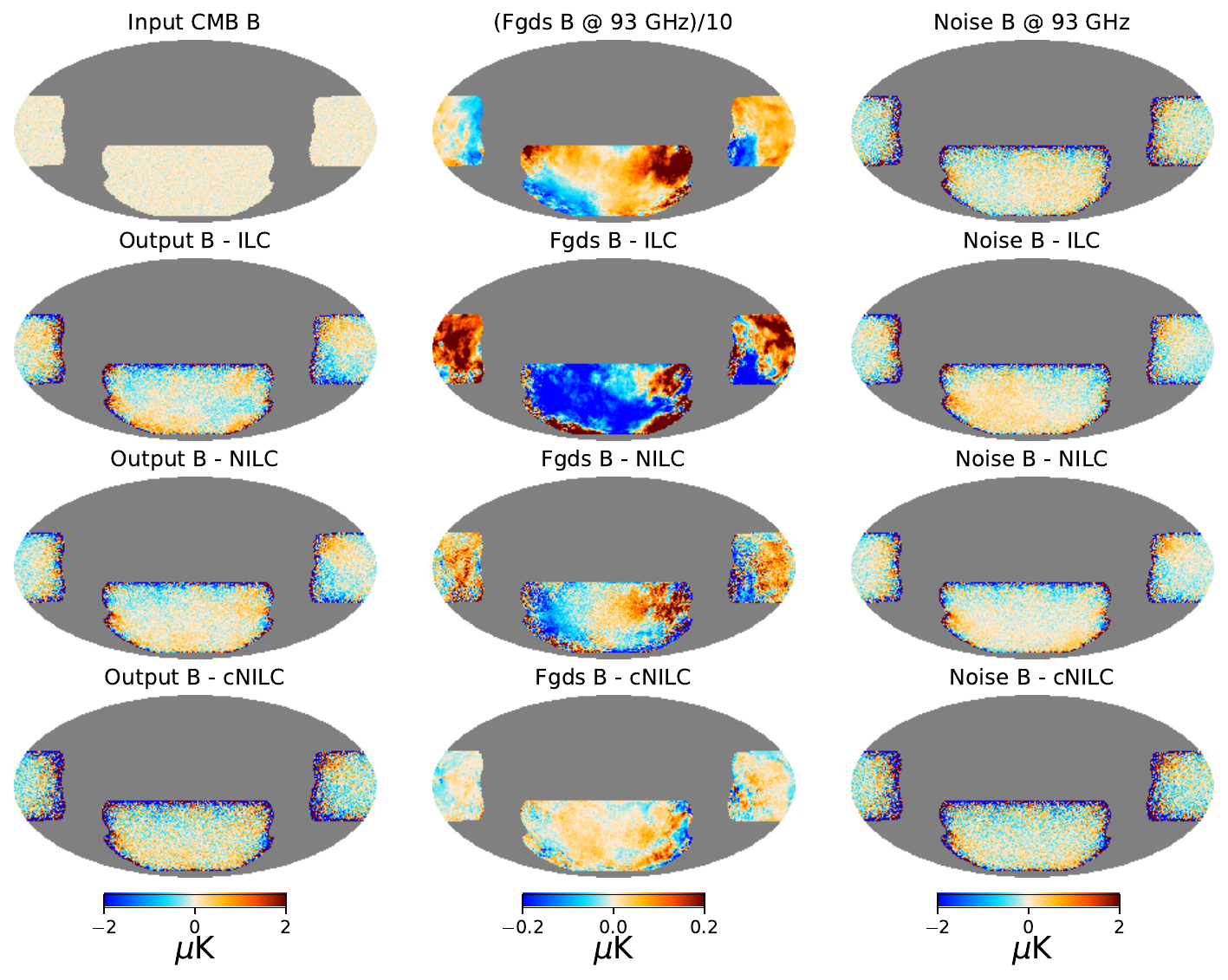}
\caption{$B$-mode maps. The first row illustrates the individual input $B$-mode components for a representative SO-SAT-like simulation at $93\,\mathrm{GHz}$: the CMB signal (left), foreground emission (middle, reduced for visualization purposes by a factor $10$), and instrumental noise. The other rows show the corresponding outputs from the ILC (second), NILC (third), cNILC (last) component separation. The three columns display, respectively, the total reconstructed signal, the foreground residuals, and the noise residuals. The maps in the last two columns are obtained by applying the component-separation weights to the input foreground and noise maps. All maps are shown in Equatorial coordinates and share a common angular resolution corresponding to a Gaussian beam with $\mathrm{FWHM}=30'$.} 
\label{fig:SO_ouputs_B}
\end{figure}

\begin{figure}
\centering
\includegraphics[width=.325\textwidth]{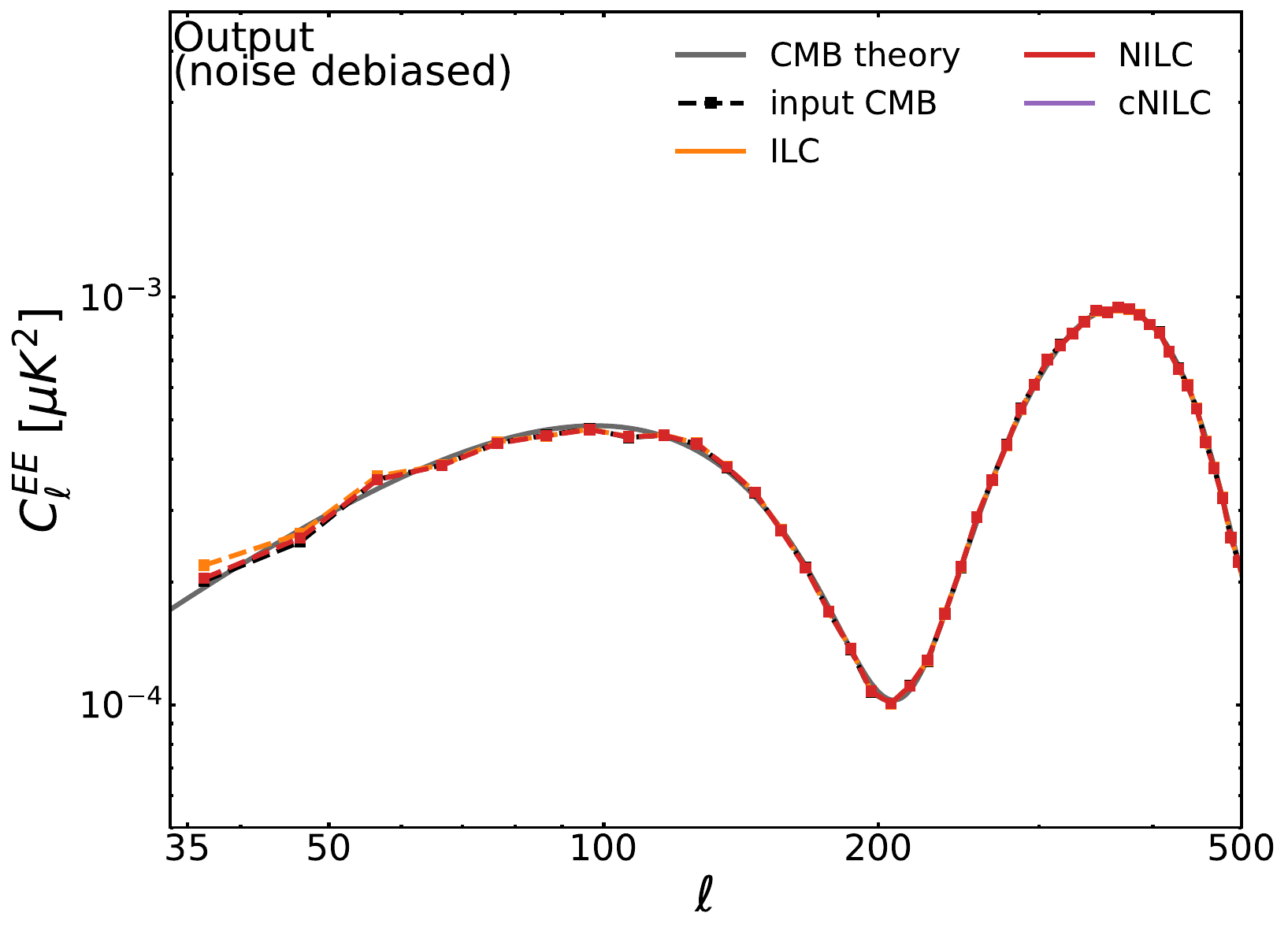}
\includegraphics[width=.325\textwidth]{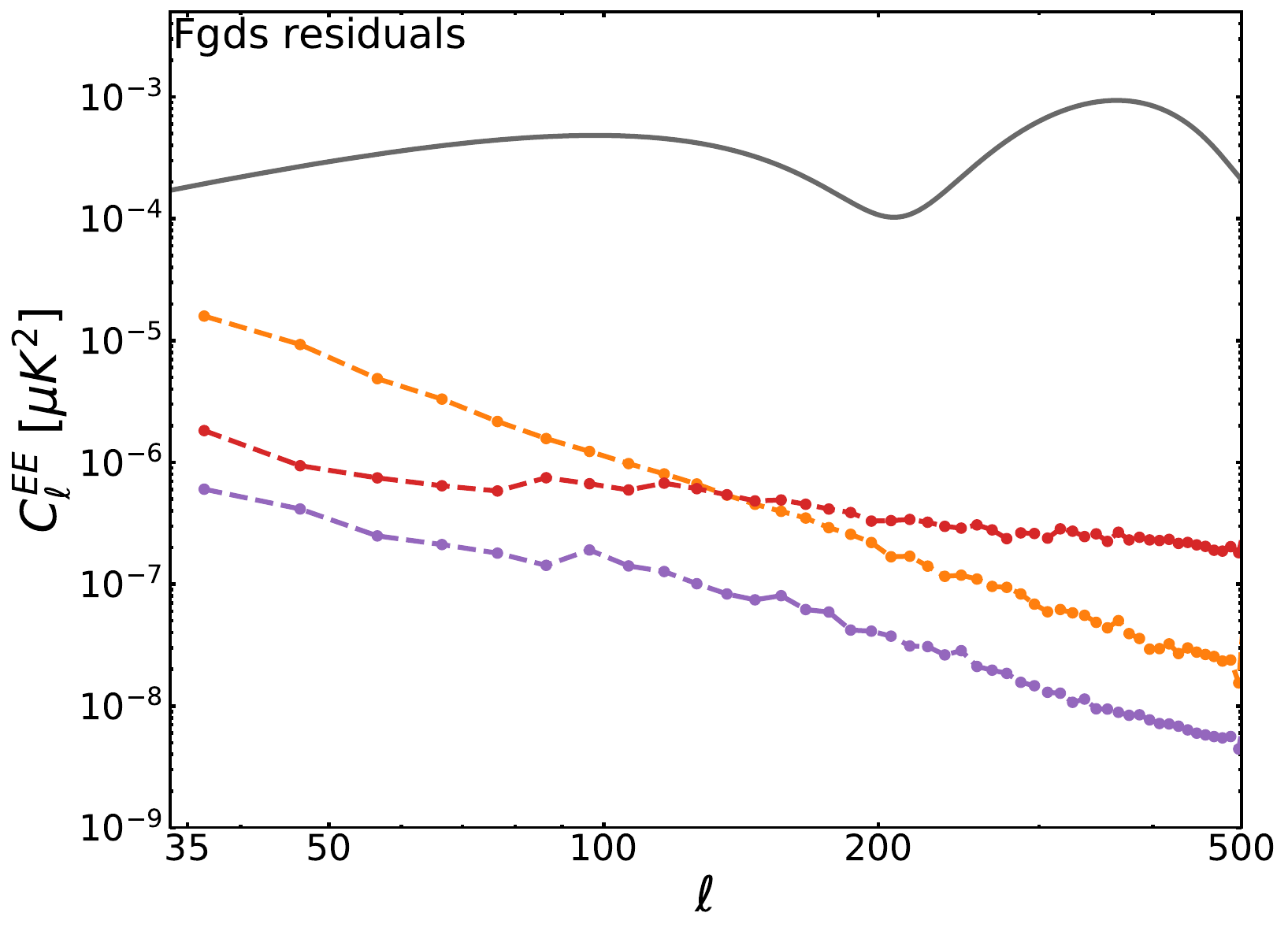}
\includegraphics[width=.325\textwidth]{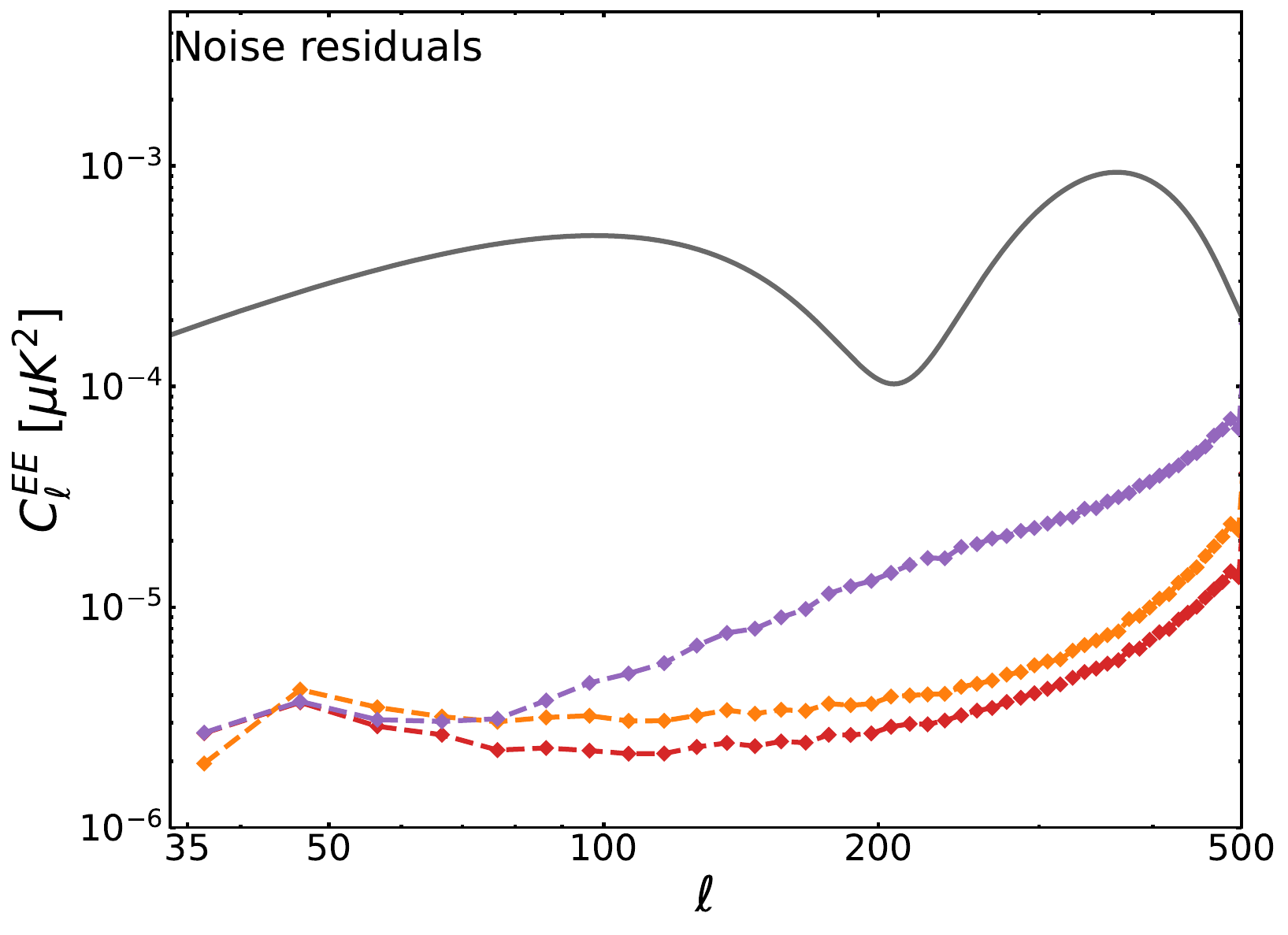}
\includegraphics[width=.325\textwidth]{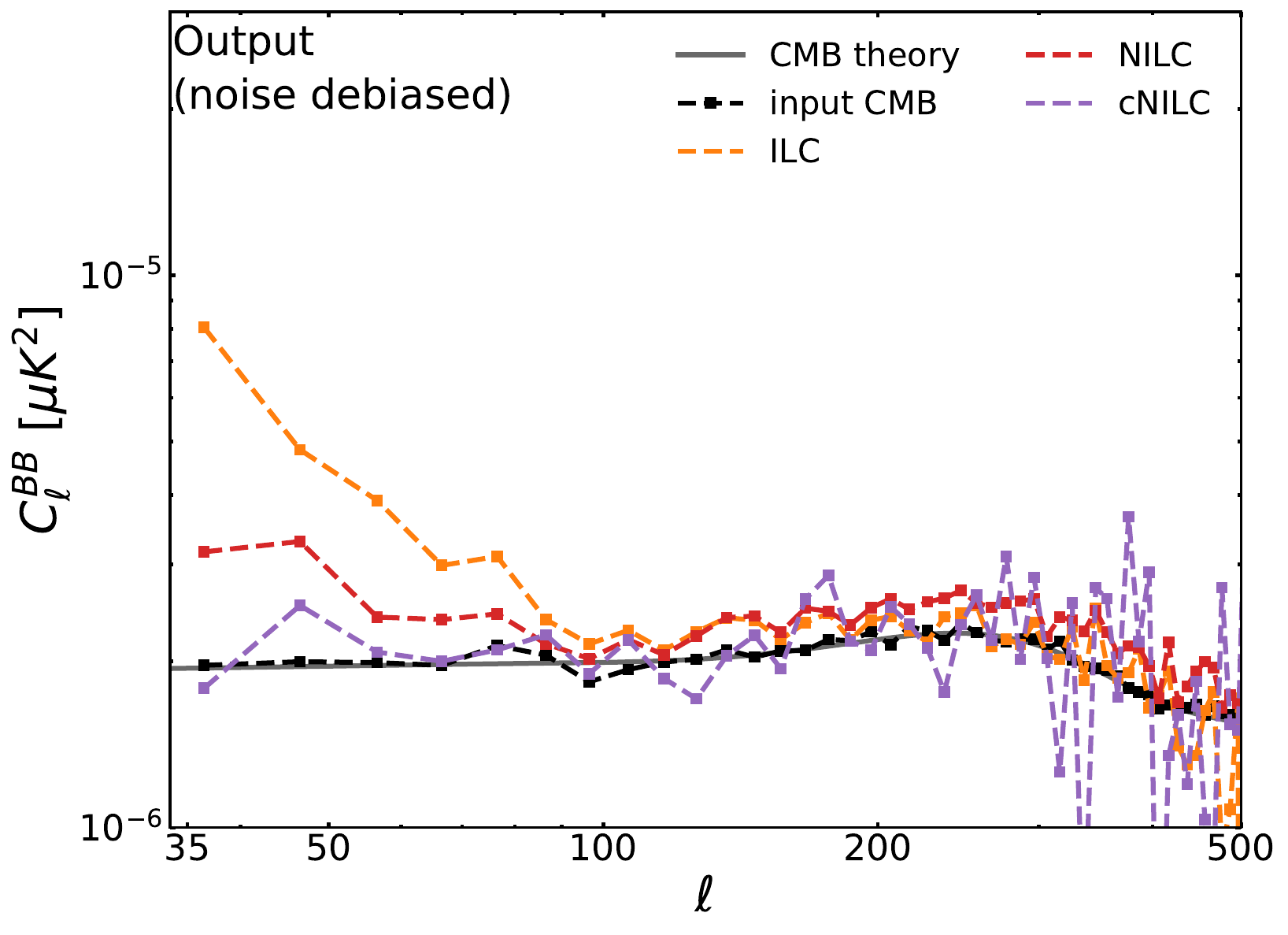}
\includegraphics[width=.325\textwidth]{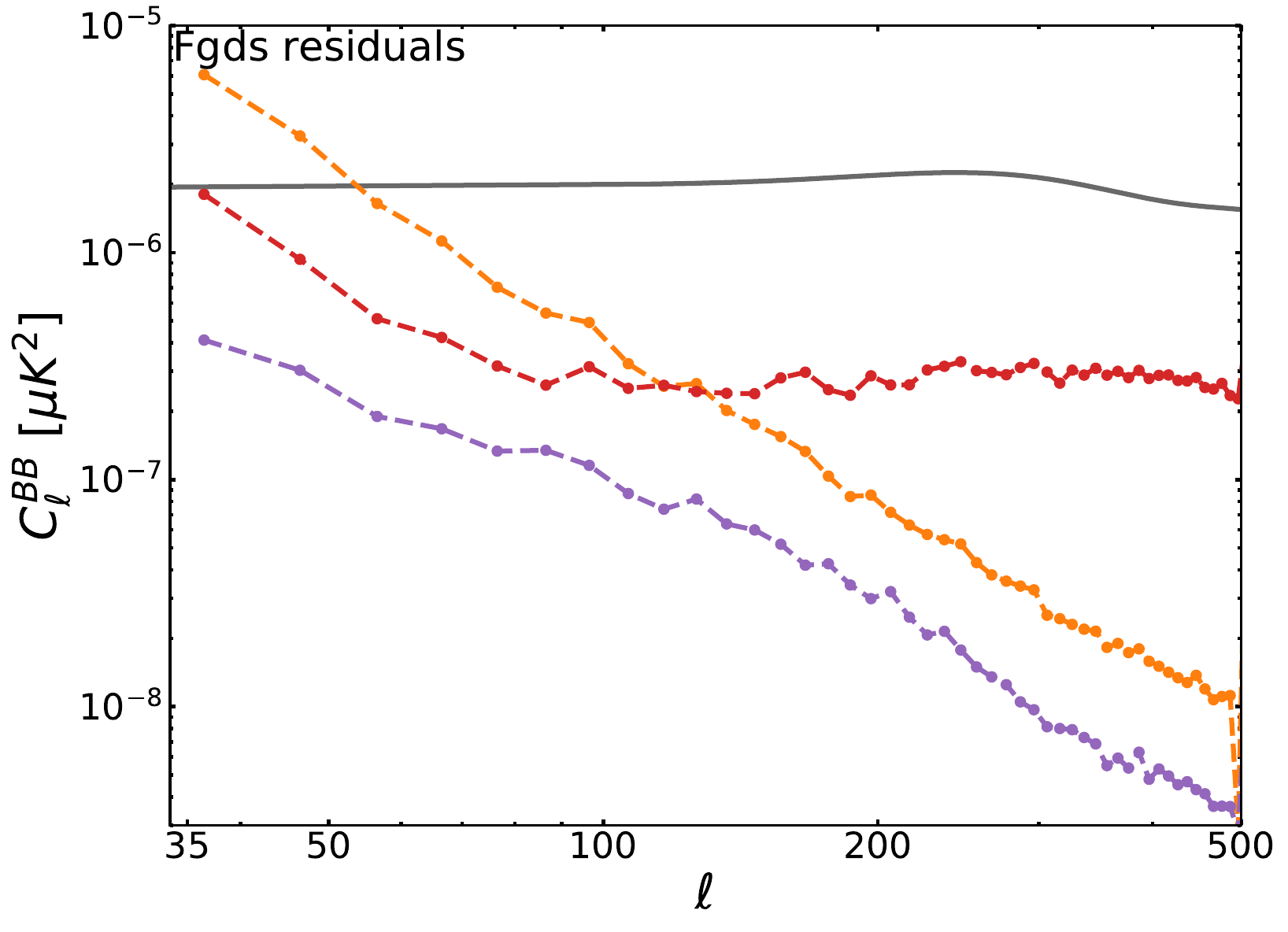}
\includegraphics[width=.325\textwidth]{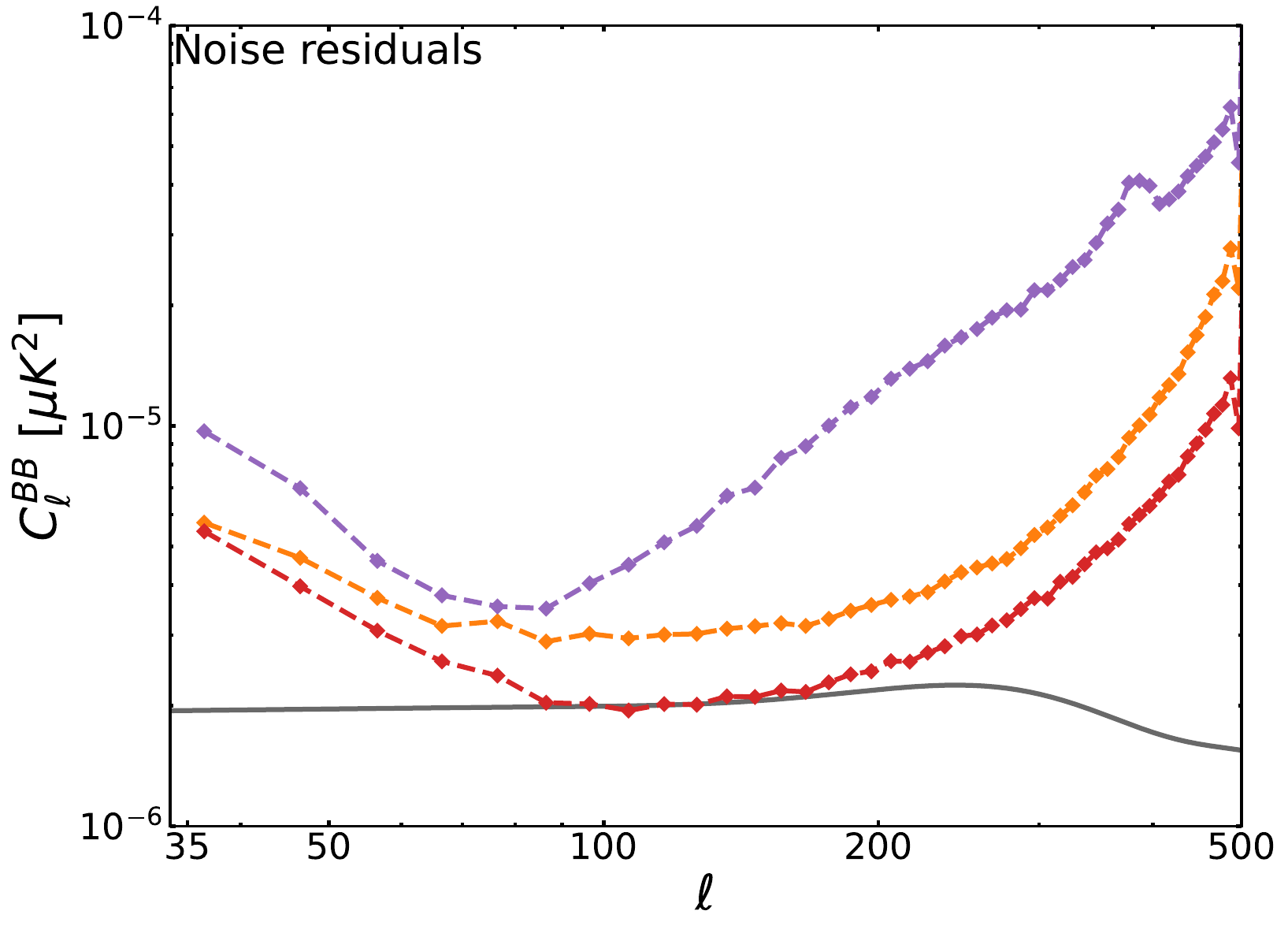}
\caption{Angular power spectra, averaged over $10$ distinct realizations computed on output products of component-separation applications to simulated SO-SAT-like data. Left, central and right panels report, respectively, power spectra of the output solution (denoised), foreground and noise residuals. The two rows report in order $C_{\ell}^{EE}$ and $C_{\ell}^{BB}$. The denoised output is obtained by computing cross spectra between two simulated solution splits with uncorrelated noise residuals. Different colors reflect results for different component separation pipelines considered for this comparison. Angular power spectra are computed adopting the normalized SO-SAT hit-count map shown in Figure \ref{fig:so_hits} as a mask. Average angular power spectrum of input CMB signal is shown in black in the left panels (in some cases it fully or partially overlaps with other curves). The theoretical input spectrum used to generate CMB realization is also shown in all panels with a grey solid line.} 
\label{fig:cls_SO}
\end{figure}
For the reconstruction of the polarized CMB signal from an SO--SAT--like experiment, we consider three alternative pipelines, already adopted in the satellite validation case: \texttt{ILC} (ILC in pixel space), \texttt{NILC} (ILC in needlet space), and \texttt{cNILC} (constrained ILC in needlet space with deprojection of the zeroth-order moments of both thermal dust and synchrotron emission). Unless otherwise stated, all component-separation outputs are obtained without applying any \emph{a priori} mitigation of $E$--$B$ leakage at the map level.

The resulting total solutions in $E$ and $B$ modes are shown in Figures~\ref{fig:SO_ouputs_E} and~\ref{fig:SO_ouputs_B}, respectively, together with maps of the corresponding foreground and noise residuals for simulation~0. These maps are obtained by linearly combining the foreground-only and noise-only input multifrequency data sets using the component-separation weights estimated by each minimum-variance routine.

In both the $E$- and $B$-mode cases, we observe a reduction of the foreground contamination by at least an order of magnitude with respect to the input level at the central frequency channel. For the standard ILC approach, as expected, residual large-scale fluctuations remain visible in the systematic residual maps. Overall, the deprojection of the zeroth-order moments of thermal dust and synchrotron emission leads to a further suppression of the bulk foreground contamination, at the cost of an increased reconstruction noise. 

We note that, while in the NILC and cNILC residual maps we observe features whose morphology closely reflects that of the input foreground maps at $93\,\mathrm{GHz}$ (which are dust dominated), the ILC foreground residual maps exhibit different structures that can be attributed to polarized synchrotron emission. This behaviour can be understood as a consequence of the variance-minimization procedure in pixel-space ILC, which is predominantly driven by the smallest angular scales available, namely those around the size the effective beam with $\mathrm{FWHM}=30^{\prime}$ to which the input data have been homogenized.
These small angular scales are not accessible at the lowest frequency channels in the simulated data set, which have a coarser intrinsic resolution. As a result, large-scale synchrotron emission from these channels leaks into the ILC residual maps. This effect is mitigated in needlet-based implementations, such as NILC and cNILC, which treat different angular scales independently. 

In the reconstructed $E$- and, in particular, $B$-mode maps, we observe residual high-amplitude noisy structures near the boundaries of the observed patch. These features are propagated through the pipeline from the input noise patterns associated with the scanning strategy. In the $B$-mode case, a comparison between the reconstructed solutions and the input target signal indicates that the reconstruction is not signal dominated at the pixel level.

Figure~\ref{fig:cls_SO} shows the angular power spectra computed from the reconstructed CMB maps, as well as from the corresponding foreground and noise residual $QU$ maps. The output power spectra are estimated as cross-spectra between two data splits, linearly combined using the component-separation weights derived from the corresponding component-separation run performed on the simulated full-mission data set. The reported spectra represent the average over the $10$ independent realizations considered in this validation.

Prior to power-spectrum estimation, these maps are weighted by a mask corresponding to the normalized hit-count map shown in Figure~\ref{fig:so_hits}, thereby down-weighting sky regions expected to be more affected by residual noise. 
The reconstruction of the $E$-mode power spectrum is highly accurate across the full multipole range considered, with residual power levels that remain significantly below the cosmological signal. In contrast, the $B$-mode power-spectrum recovery is affected by systematic biases on the largest angular scales, with residual contamination persisting for the NILC case even at smaller angular scales.  

By inspecting the recovered $E$- and $B$-mode angular power spectra of the residuals, we find that needlet-based minimum-variance implementations, as expected, achieve a stronger reduction of polarized foreground contamination on the largest angular scales. At higher multipoles, however, the residual foreground power increases relative to the pixel-based ILC approach. An opposite trend is observed for the reconstruction noise. The multipole at which the transition between the regimes where needlet-based and pixel-based variance minimization is most effective occurs depends on the overall sensitivity of the data set under analysis.
When moment deprojection is included in the weight constraints, an overall improvement in foreground cleaning is achieved, at the expense of a significantly increased noise level. This trade-off intrinsic to the cNILC implementation is particularly pronounced in the present case, owing to the limited number of frequency channels available for component separation.

Finally, as the last result of this validation study for the reconstruction of the polarized CMB signal from a simulated, representative ground-based experiment, we present the average angular power spectra obtained from NILC runs in different scenarios: (i) no map-based $E$--$B$ leakage correction is applied prior to component separation (the case considered so far); and (ii) various $E$--$B$ leakage-mitigation techniques are applied to the input maps before component separation.
The considered leakage-correction strategies are:
\begin{enumerate}
    \item a recycling technique followed by harmonic deprojection of residual ambiguous modes;
    \item a recycling technique followed by an additional correction obtained by subtracting a leakage-residual template constructed via diffusive inpainting;
    \item a diffusive inpainting correction followed by harmonic iterative deprojection of residual ambiguous modes.
\end{enumerate}
Additional details on these map-based strategies for correcting $EB$ leakage can be found in \cite{NILC_cutsky}.

\begin{figure}
\centering
\includegraphics[width=.325\textwidth]{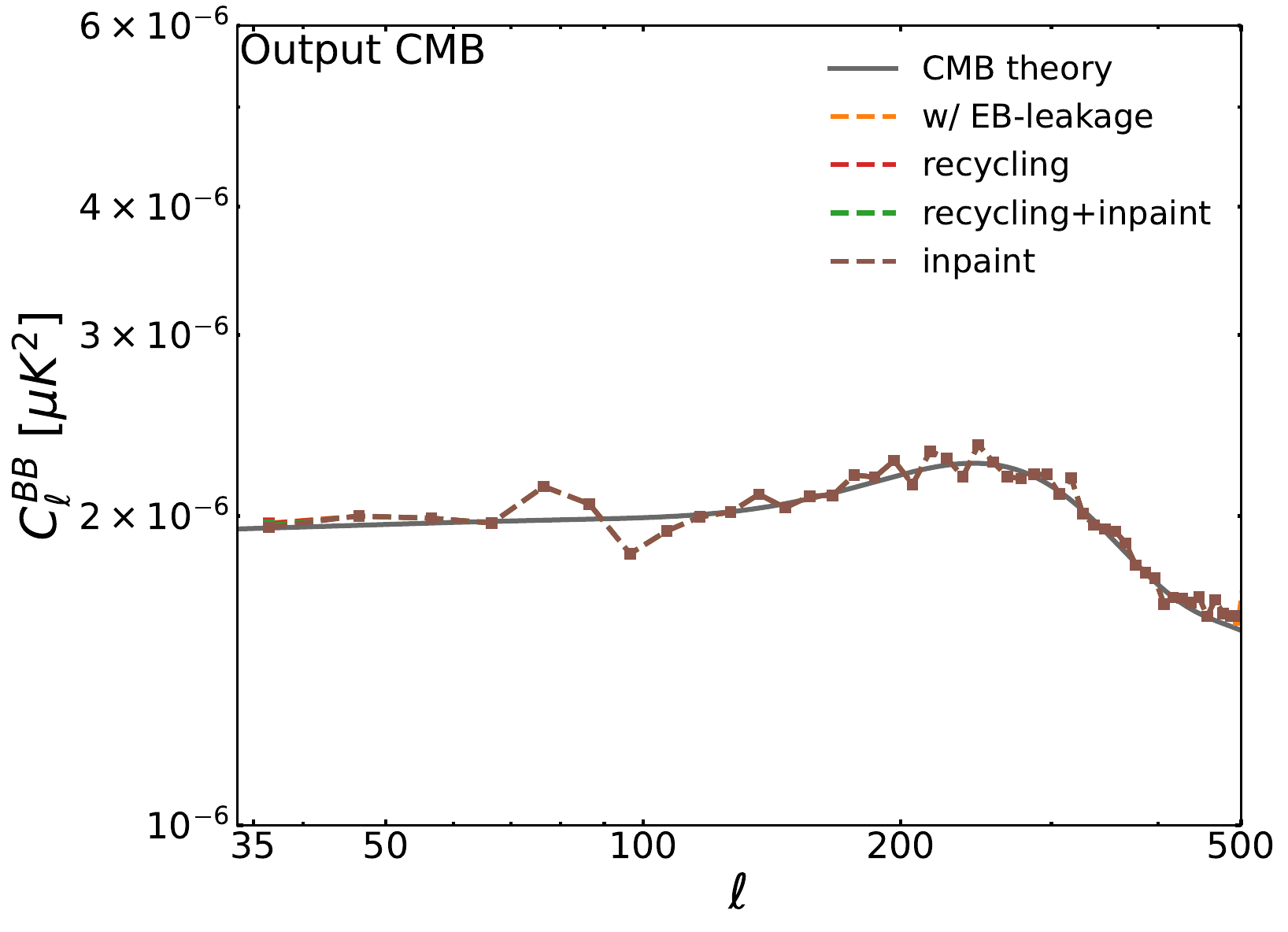}
\includegraphics[width=.325\textwidth]{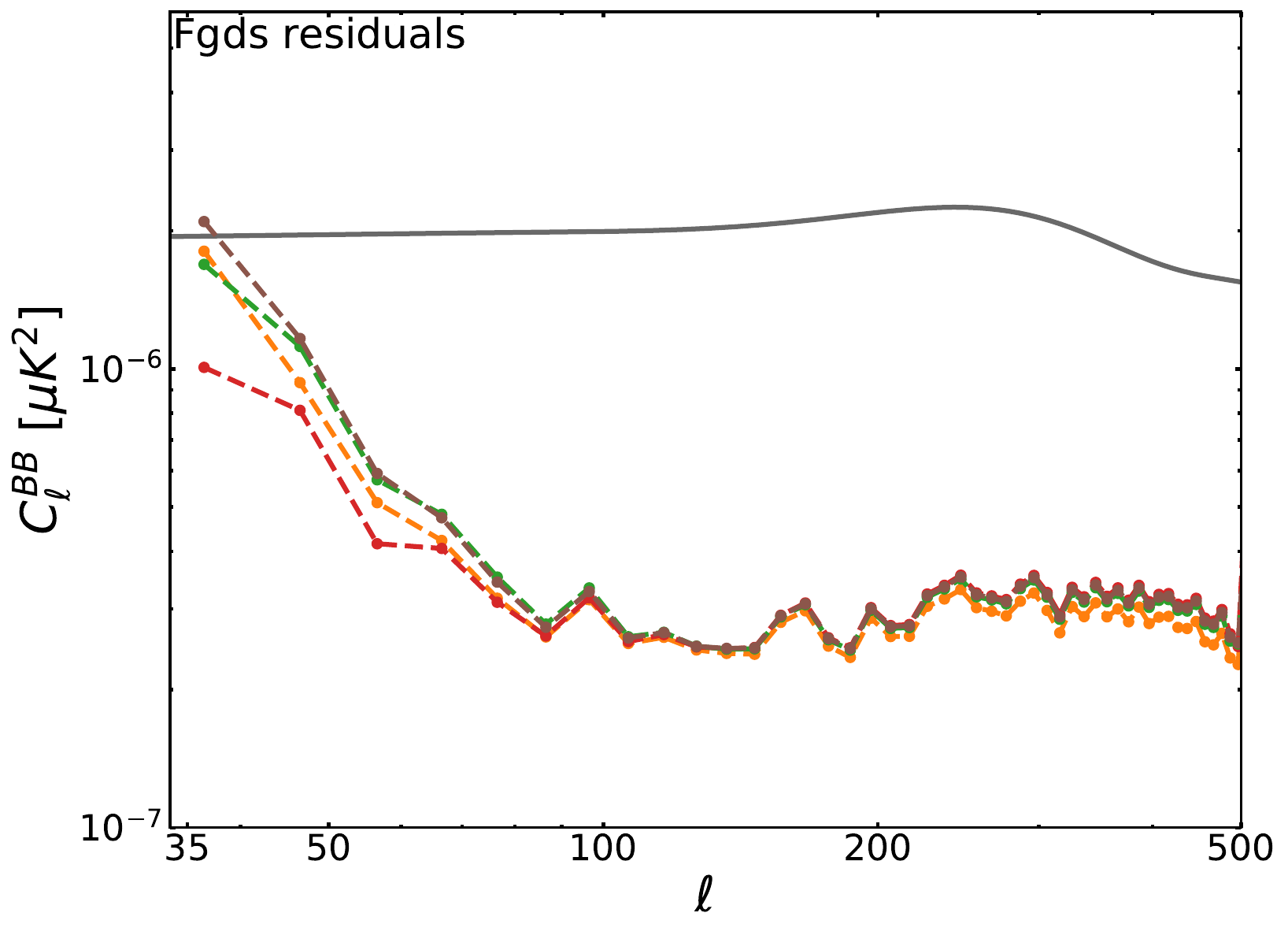}
\includegraphics[width=.325\textwidth]{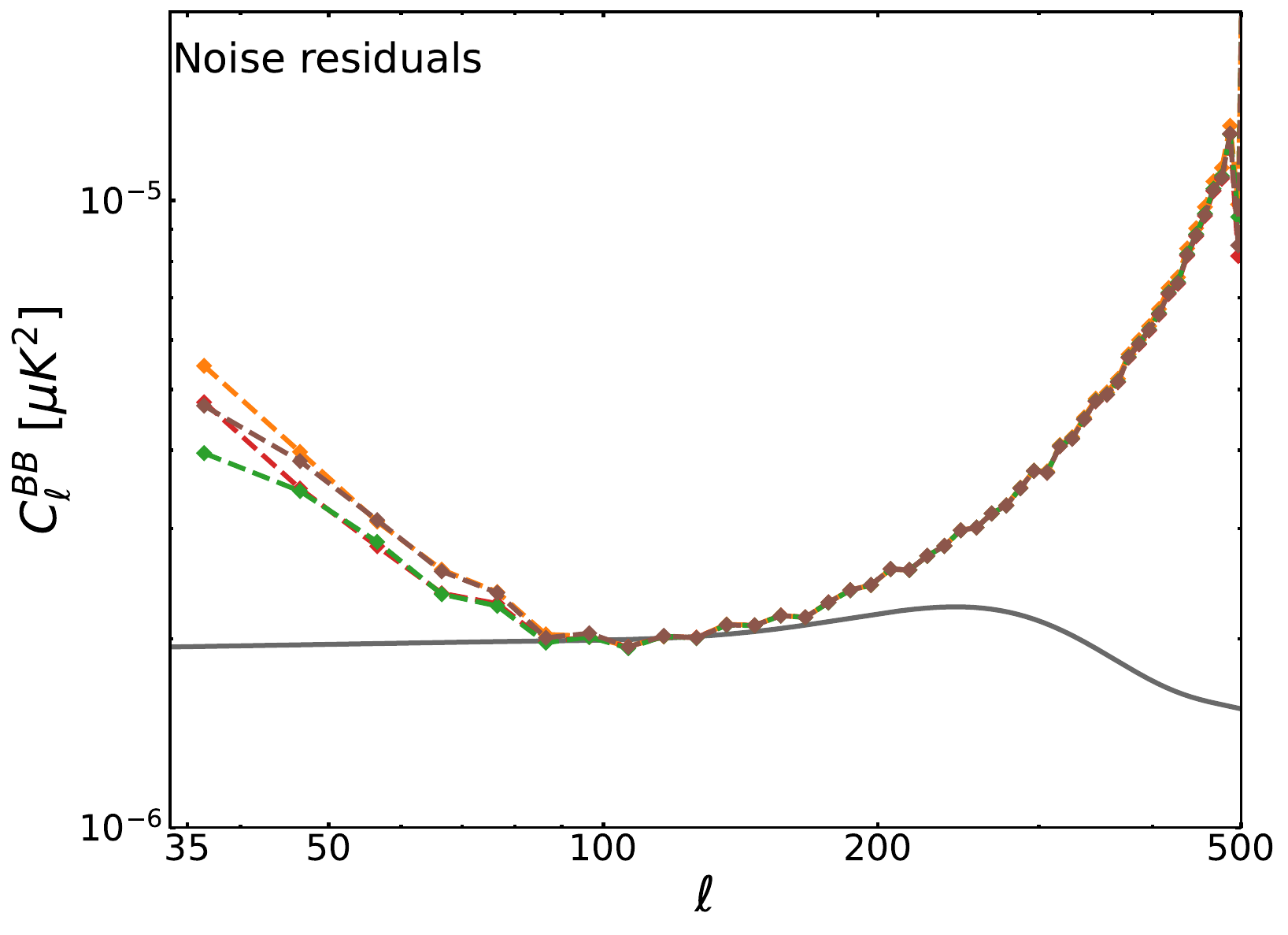}
\caption{$B$-mode angular power spectra, averaged over $10$ distinct realizations computed on output products from NILC component-separation application to simulated SO-SAT-like data. Left, middle and right panels report, respectively, power spectra of the propagated CMB component, foreground and noise residuals, obtained by propagating the corresponding multifrequency maps of each individual component through the component-separation pipeline. Different colors reflect results for different approaches to deal with $E$-$B$ leakage prior to component separation: i) orange: no correction; ii) red: recycling technique followed by three harmonic deprojections of residual ambiguous modes; iii) green: recycling technique followed by subtraction of a template of leakage residuals obtained with diffusive inpainting; iv) brown: diffusive inpainting and afterwards three harmonic deprojections of residual ambiguous modes. All angular power spectra are computed from $QU$ maps after final purification before power spectra estimation. Angular power spectra are computed adopting the normalized SO-SAT hit-count map shown in Figure \ref{fig:so_hits} as a mask. The theoretical input spectrum used to generate CMB realizations is shown in all panels with a grey solid line.} 
\label{fig:so_bmodes_leaks}
\end{figure}

A comparison of the resulting $B$-mode angular power spectra for these cases is shown in Figure~\ref{fig:so_bmodes_leaks}. The leftmost panel displays the angular power spectra of the CMB-only component propagated through all four cases, in order to highlight any potential residual bias in the recovered $B$-mode signal. All pipelines yield unbiased reconstructions, demonstrating that the adopted masking strategy—which strongly down-weights the boundaries of the observed patch—effectively mitigates $E$--$B$ leakage.
We note that these power spectra are computed from output $Q U$ maps that are further purified prior to power-spectrum estimation. However, consistent results are obtained when using output $B$-mode maps directly, even though additional purification before power-spectrum estimation is not possible in this case.
Finally, by comparing the residual foreground and noise power across angular scales, we do not observe significant differences among the considered approaches. This indicates that the choice of map-based $EB$-leakage correction applied to the input maps does not substantially affect the variance minimization implemented for the reconstruction of $B$ modes. A marginal improvement in foreground residual suppression is observed on the largest angular scales when adopting the recycling correction followed by iterative harmonic deprojection of residual ambiguous modes.

\begin{figure}
\centering
\includegraphics[width=.9\textwidth]{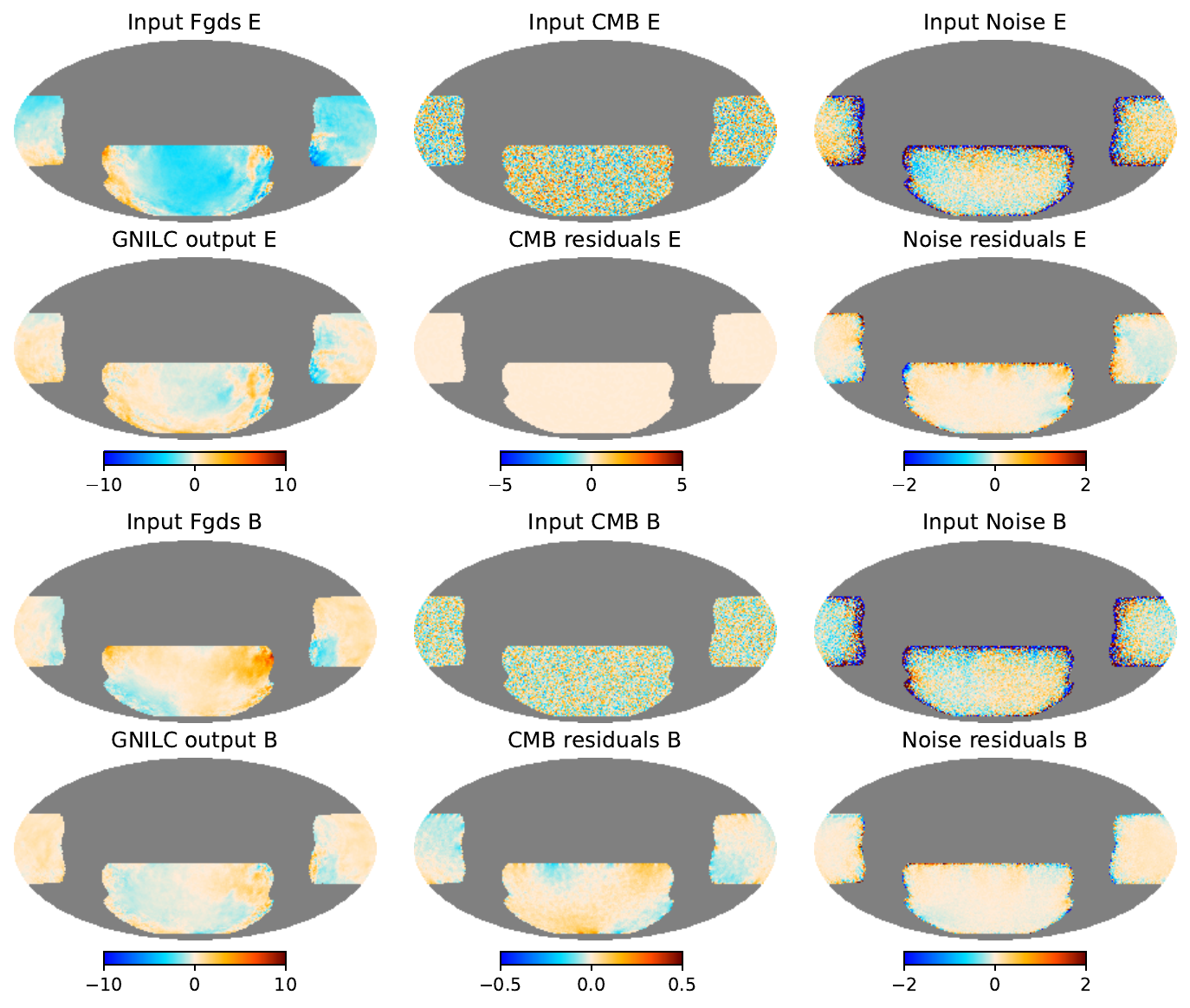}
\caption{Input and GILC-reconstructed maps for the $93$ GHz frequency channel of a simulated realization of the SO-SAT-like experiment. Each pair of rows corresponds, from top to bottom, to a different polarization scalar field: either $E$ or $B$. In the left column, odd rows show the input total foreground emission, while the corresponding even rows display the total foreground emission reconstructed by GILC. The middle and right columns present, respectively, the input and output CMB and noise contributions. All maps are smoothed with a Gaussian beam of FWHM$=30'$.} 
\label{fig:maps_gnilc_so}
\end{figure}
\begin{figure}
\centering
\includegraphics[width=.32\textwidth]{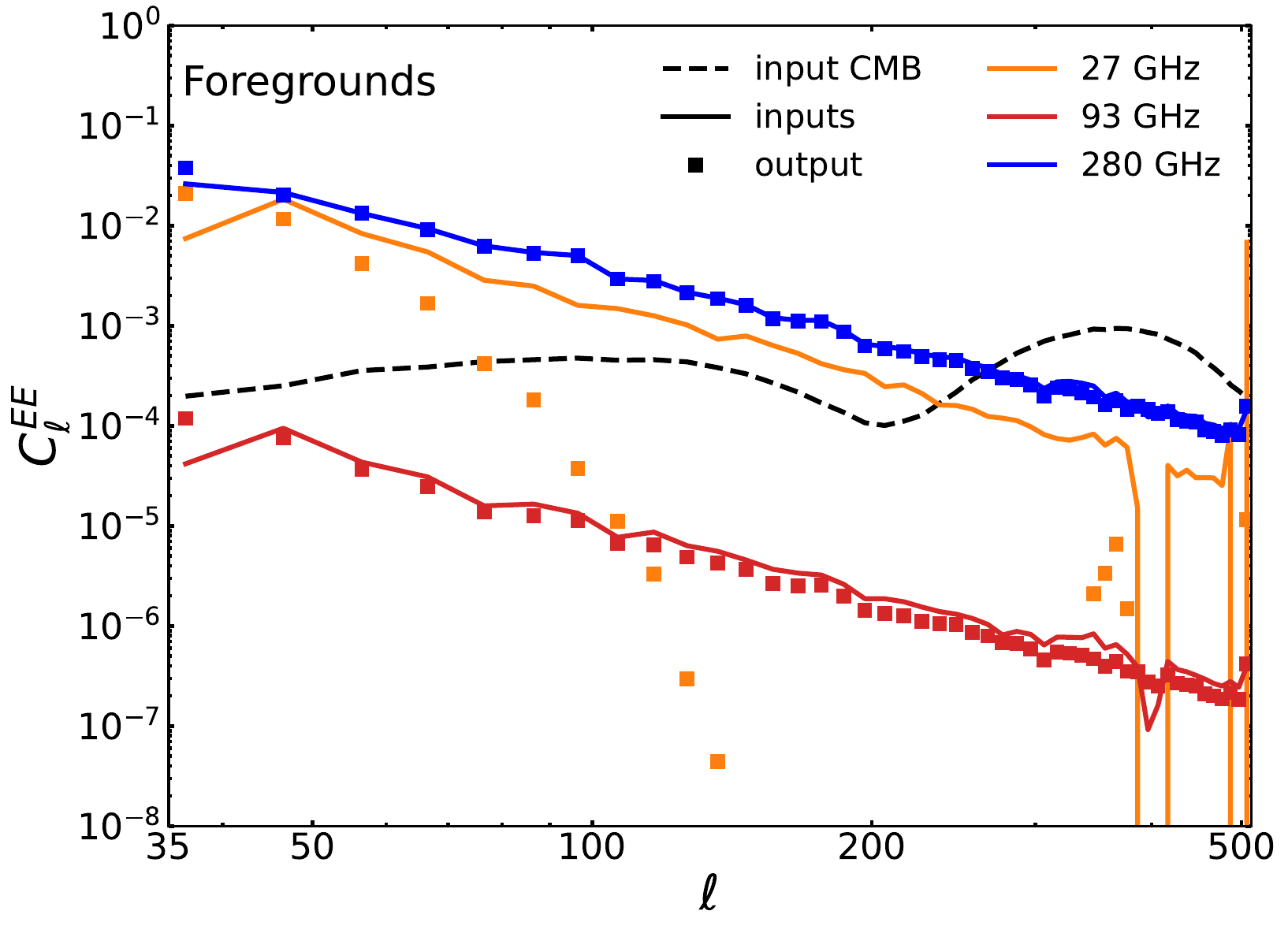} \includegraphics[width=.32\textwidth]{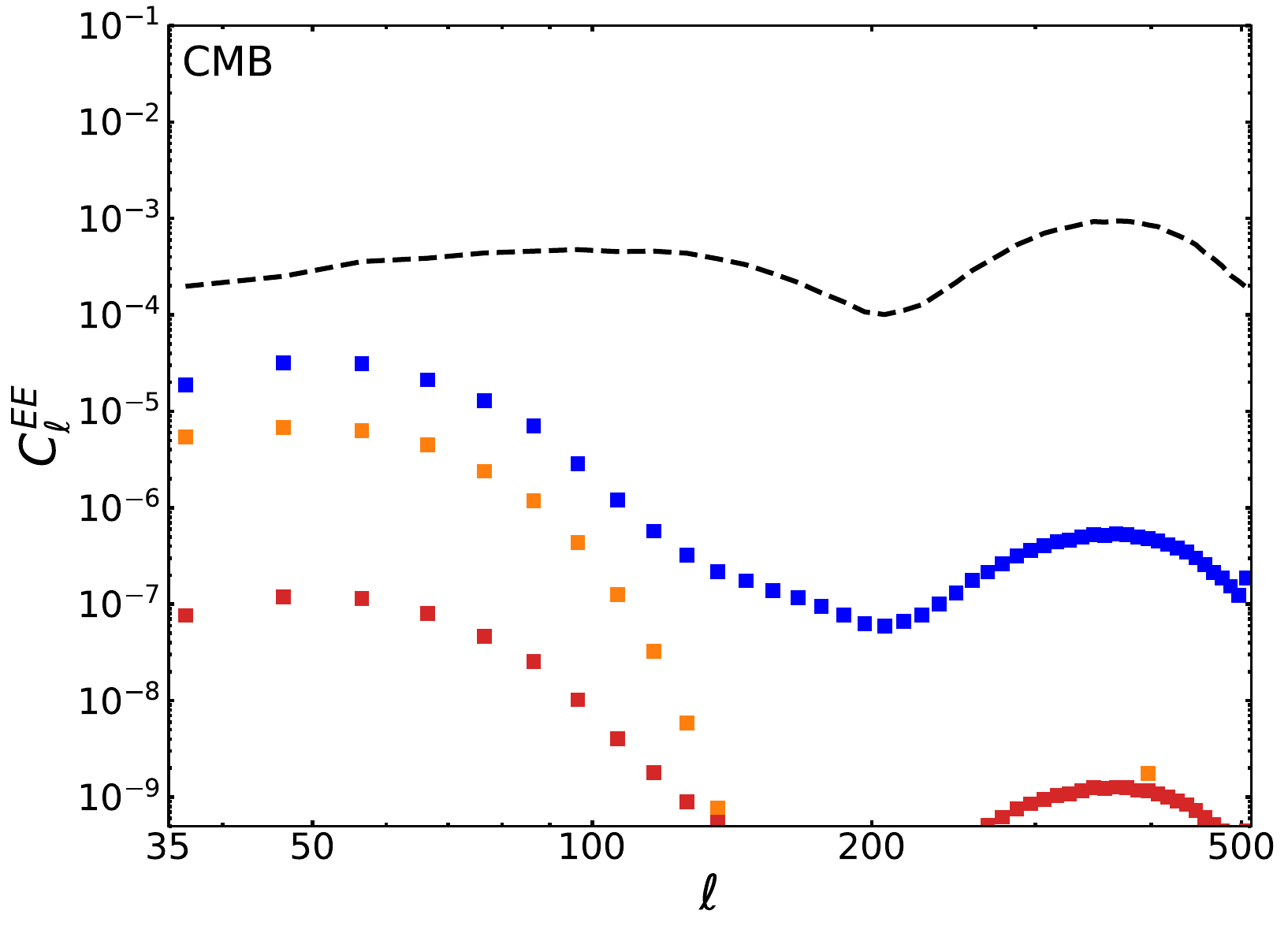}
\includegraphics[width=.32\textwidth]{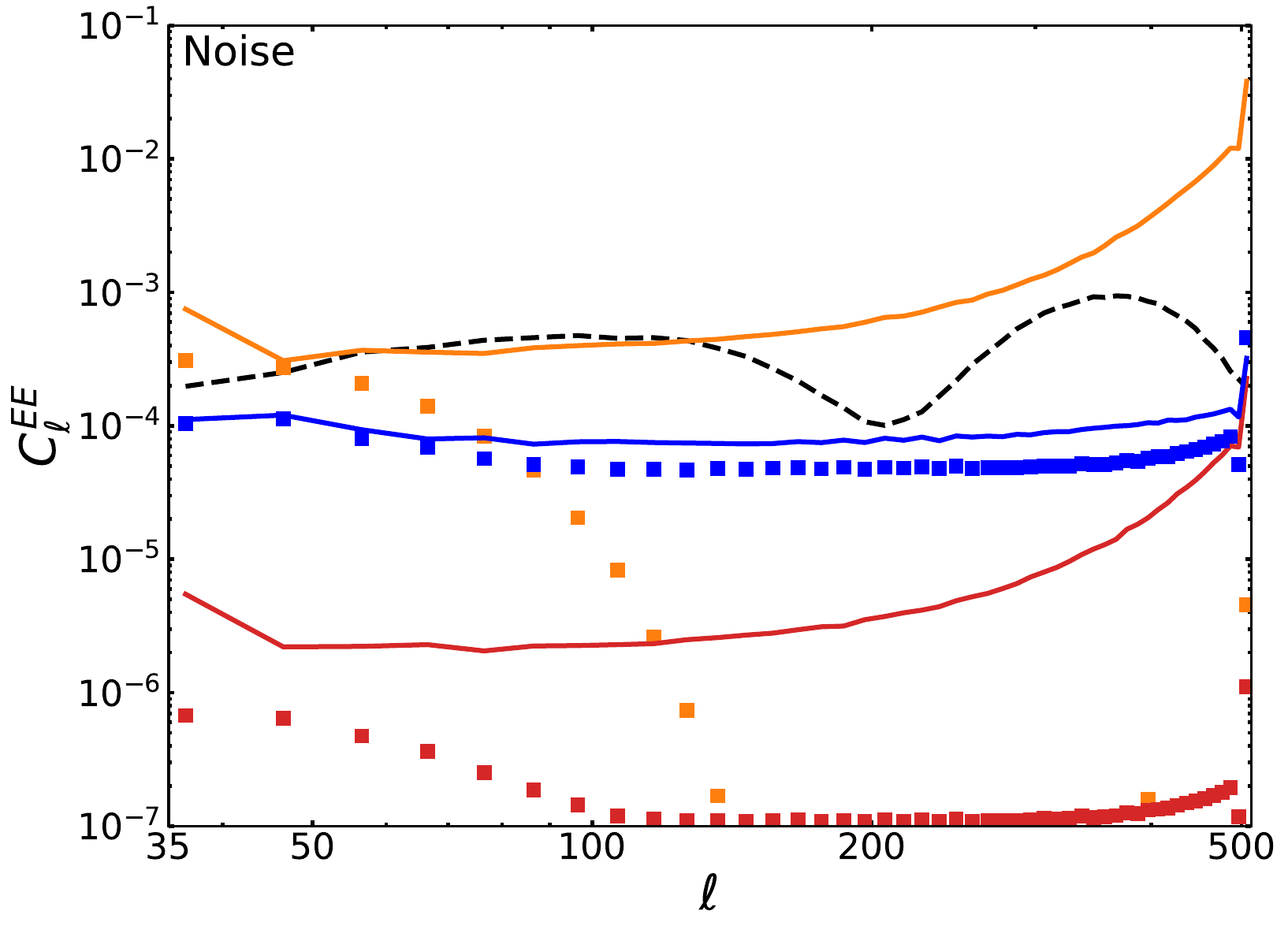}\\
\includegraphics[width=.32\textwidth]{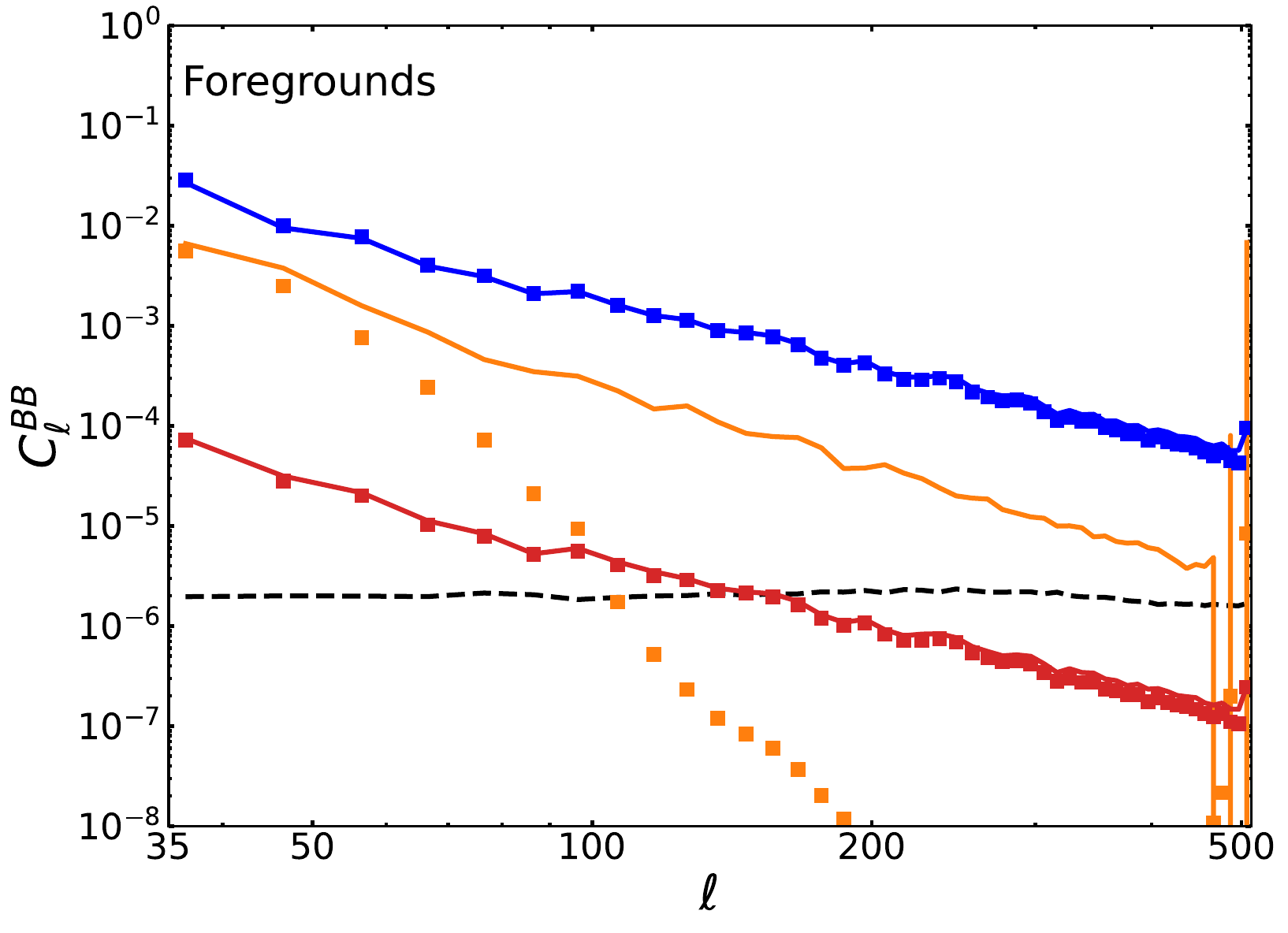} 
\includegraphics[width=.32\textwidth]{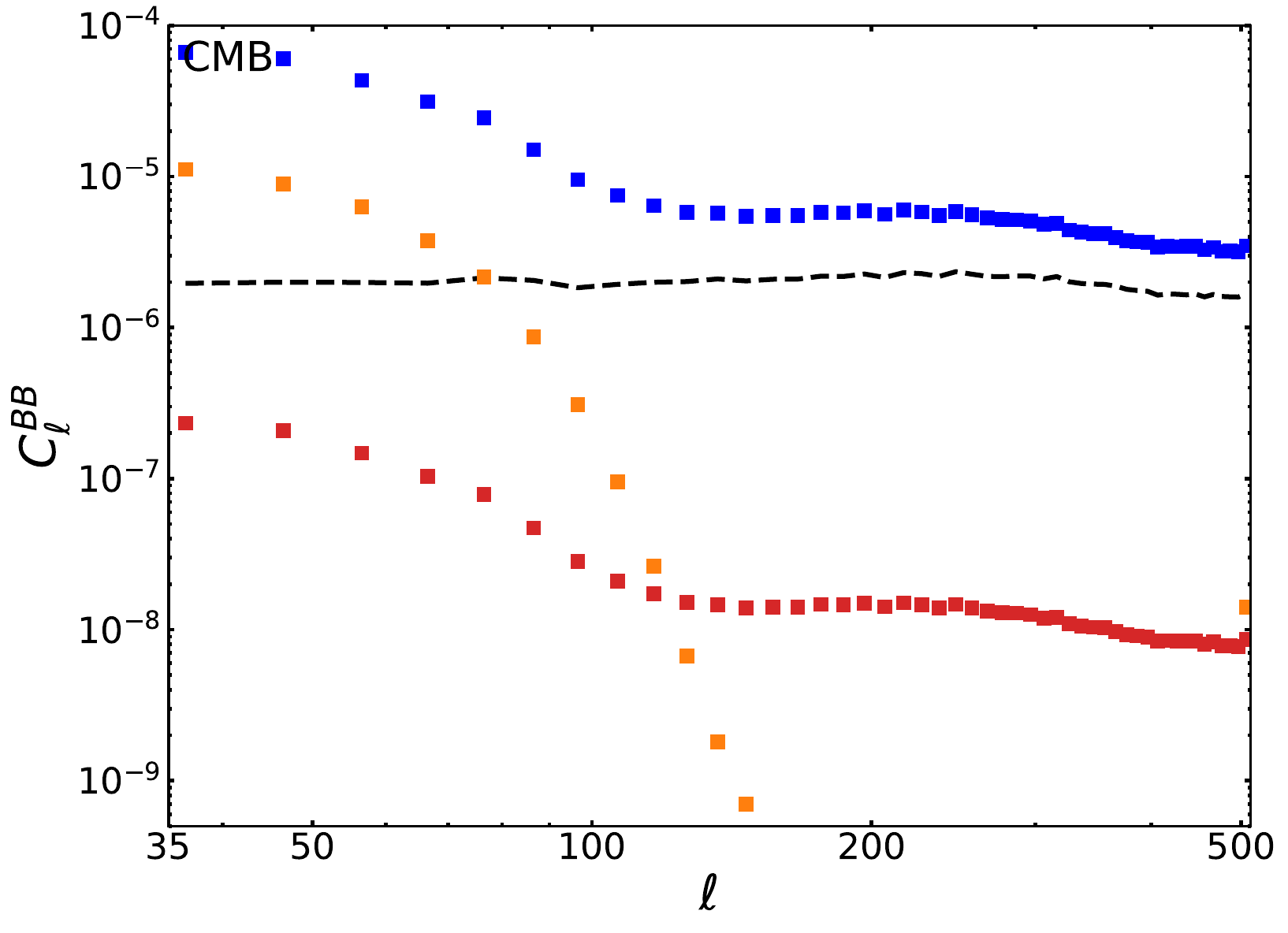} 
\includegraphics[width=.32\textwidth]{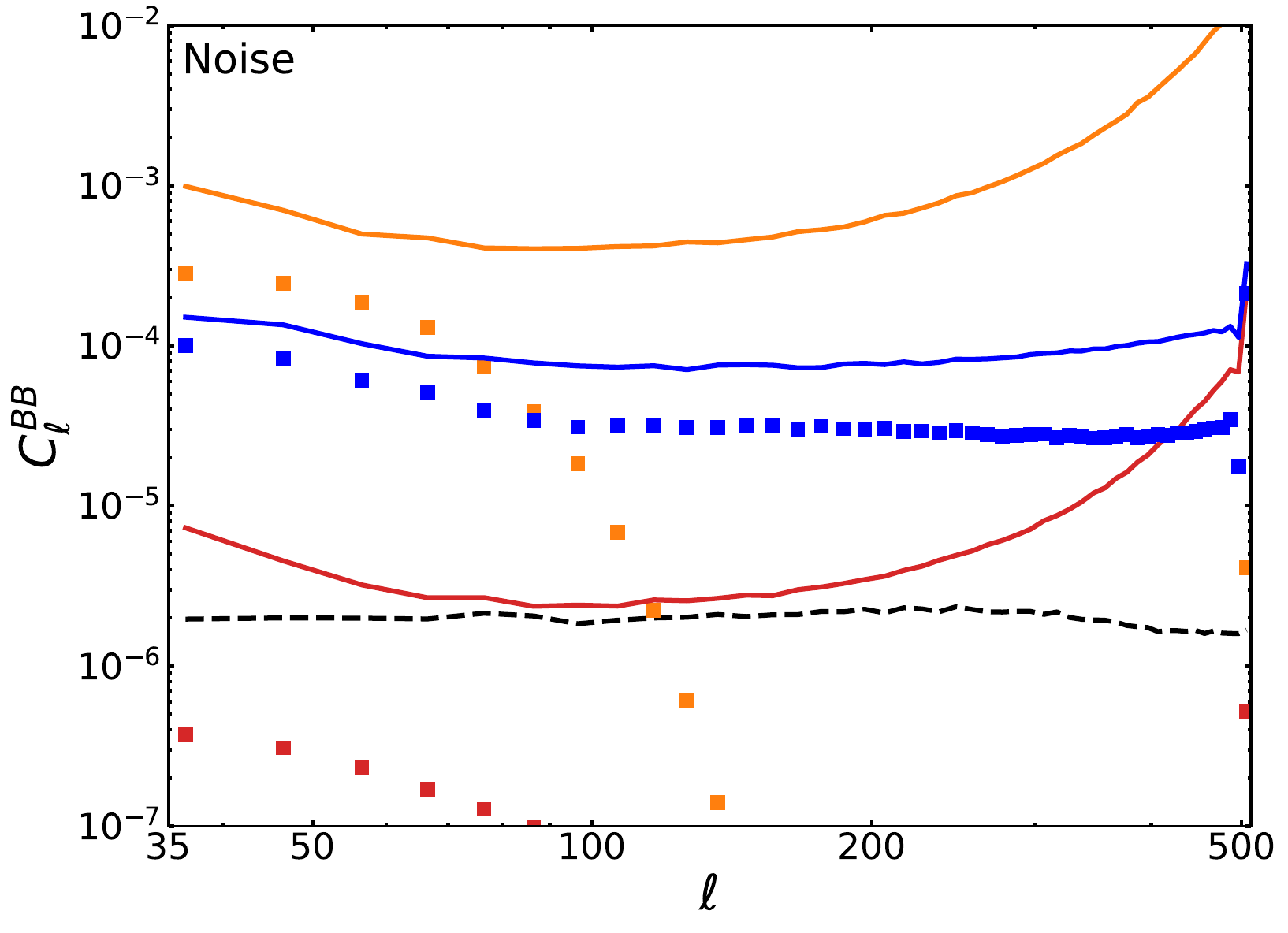} 
\caption{Angular power spectra of GNILC input and output single components (total foregrounds: left; CMB: middle; noise: right) of $E$- (top) and $B$-mode (bottom) polarization. The results refer to average power spectra computed from an application to $10$ different realizations of SO-SAT-like observations. Three different channels are reported: $27$ GHz (orange), $93$ GHz (red) and $280$ GHz (blue). Outputs are reported with markers while inputs with lines. All angular power spectra are computed from $QU$ maps after final purification before power spectra estimation. Angular power spectra are computed adopting the normalized SO-SAT hit-count map shown in Figure \ref{fig:so_hits} as a mask. The input theoretical angular power spectrum used for generating CMB simulations is shown with black dashed line. For the CMB panels, the input average angular power spectrum is not reported for simplicity but would collapse to the black dashed line.} 
\label{fig:gilc_spectra_SO}
\end{figure}
\subsubsection{Foreground reconstruction}
\label{sec:fgds_rec_SO}
As in the satellite-validation case, we repeat the analysis aimed at extracting cleaned templates of foreground emission through the model-independent approach implemented in \texttt{GILC}, now applied to a ground-based experiment configuration. As for the CMB reconstruction, we restrict the discussion to polarization for simplicity.

Map-level results are shown in Figure~\ref{fig:maps_gnilc_so} for the representative central frequency channel ($93\,\mathrm{GHz}$), in order to highlight the capability of the pipeline to recover foreground information in a channel traditionally considered CMB dominated. We observe that both CMB and noise contamination are blindly deprojected to a significant extent in both $E$ and $B$ modes, although some residual power remains detectable on the largest angular scales. At the same time, this deprojection induces a partial loss of signal in the reconstructed foreground map at this central frequency.

These findings are further supported by the corresponding angular power spectra shown in Figure~\ref{fig:gilc_spectra_SO}, where results for the $27$, $93$, and $280\,\mathrm{GHz}$ channels are reported. We recall that these spectra are computed from output $Q U$ maps, with $B$-mode purification applied prior to power-spectrum estimation. From Figure~\ref{fig:gilc_spectra_SO}, a broader picture emerges. For the $93\,\mathrm{GHz}$ channel, we observe a clear attenuation of CMB and noise contamination across all angular scales, together with an overall accurate reconstruction of the foreground power spectrum. However, consistent with the map-level inspection, a mild suppression of foreground power—particularly in $E$ modes—is visible relative to the input spectrum. This behaviour is expected, as the cosmological signal is stronger in $E$ modes, leading to a more challenging foreground separation.

In contrast, at the highest-frequency channel ($280\,\mathrm{GHz}$), no significant loss of foreground power is observed. In this case, CMB and noise are efficiently attenuated in $E$ modes, while the GILC weights slightly amplify the residual CMB contribution in $B$ modes. Finally, at $27\,\mathrm{GHz}$, the foreground signal is reliably reconstructed only on the largest angular scales, as the broader beam limits access to smaller-scale modes.

\section{Conclusions}
\label{sec:concl}

In this work, we presented \texttt{BROOM}, a comprehensive and publicly available python package for the model-independent analysis of microwave astronomical data. The framework implements a broad range of blind, minimum-variance component-separation techniques, including ILC, NILC, and their constrained and multi-clustering extensions, applicable to scalar fields and polarization intensity. We emphasize that these techniques are not limited to the reconstruction of primary CMB anisotropies, but can be applied to any signal with a priori known SED, such as SZ distortions or foreground spectral moments. In addition, it provides a full implementation of the Generalized ILC (GILC) methodology for blind foreground reconstruction and diagnostic, together with tools for nuisance-covariance estimation, residual propagation, and angular power-spectrum computation.

A key strength of \texttt{BROOM} lies in its unified and modular design. Within a single framework, users can simulate realistic microwave observations for arbitrary experimental configurations, perform component separation in either pixel or needlet space, account for partial-sky effects and $E$--$B$ leakage, propagate individual components through the separation weights, and compute angular power spectra of the resulting products under flexible masking strategies. This end-to-end capability makes \texttt{BROOM} suitable both for validation studies based on simulations and for the analysis of real data sets.

We have validated the package in two representative experimental scenarios: a full-sky satellite mission with LiteBIRD-like specifications and a partial-sky ground-based configuration inspired by the SO-SATs. In the satellite case, we demonstrated accurate reconstruction of CMB temperature and polarization anisotropies across multiple ILC-based pipelines, including constrained and multi-clustering extensions. In the ground-based case, we showed that the framework robustly handles anisotropic noise, masking, and $E$--$B$ leakage mitigation, while preserving an unbiased recovery of the CMB angular power spectra over most of the explored multipole range. As expected, the reconstruction of the $B$-mode signal remains more challenging and requires further refinement and optimization of the implemented techniques.

The validation results confirm several expected trends. Needlet-based implementations improve foreground suppression at large angular scales, while constrained approaches reduce residual contamination at the cost of increased reconstruction noise. Multi-clustering strategies enhance foreground mitigation in regimes of complex spectral variability. The GILC module successfully reconstructs foreground emission with reduced nuisance contamination and provides a robust, model-independent diagnostic of foreground complexity across angular scales and sky regions. Importantly, the propagation of foreground templates through the component-separation weights enables the construction of residual spectral templates that can be incorporated into cosmological likelihood analyses.

Overall, \texttt{BROOM} provides a flexible and extensible platform for blind microwave data analysis, with particular relevance for current and forthcoming CMB experiments targeting high-precision measurements of polarization anisotropies. Future developments will include optimized moment-selection strategies (e.g., ocMILC implementations), support for additional pixelization schemes, and performance optimizations for large-scale simulations. By making the package fully open-source and documented, we aim to facilitate reproducible research, cross-validation of analysis pipelines, and further methodological developments within the CMB community.

\acknowledgments

AC acknowledges partial support by the Italian Space Agency Project (ASI Grants No. 2020-9-HH.0 and 2016-24-H.1-2018), as well as the RadioForegroundsPlus Project HORIZON-CL4-2023-SPACE-01, GA 101135036 and through the Project SPACE-IT-UP by the Italian Space Agency and Ministry of University and Research, Contract Number 2024-5-E.0.
This work has also received support by the European Union’s Horizon 2020 research and innovation programme under
grant agreement no. 101007633 CMB-Inflate. This work is not an official Simons Observatory paper.

\appendix

\section{Configuration parameters}
\label{app:broom_pars}

In this Appendix, we introduce many of the parameters that can be set within \texttt{BROOM}. This introduction follows the same structure of Section~\ref{sec:basics}. In particular, we distinguish four parameter categories: (i) general parameters; (ii) parameters for generating realistic simulations of the observed microwave sky; (iii) parameters for component separation; and (iv) parameters for power-spectrum computation. The practical procedure for defining and providing them is described in Section~\ref{sec:broom_func}. 

\subsection{General settings}
General parameters to be defined are:
\begin{itemize}
    \item \texttt{nside} (integer): The \texttt{HEALPix} resolution parameter used to define the outputs of component separation;
    \item \texttt{nside\_in} (integer): This parameter specifies the \texttt{HEALPix} resolution used to generate simulation maps of any microwave component when running the \texttt{BROOM} \texttt{get\_input\_data} function. 
More generally, \texttt{BROOM} component-separation modules can accept as input either sets of maps or spherical-harmonic coefficients $a_{\ell m}$. If harmonic coefficients are provided, \texttt{nside\_in} defines the \texttt{HEALPix} resolution of the maps from which they were derived. If maps are provided directly, this parameter is inferred from the map headers. 
This information is used to apply the appropriate correction for the input pixel window function, when relevant. If not explicitly specified, \texttt{nside\_in} defaults to the value of \texttt{nside}.

    \item \texttt{lmin} and \texttt{lmin\_in} (integers): Minimum multipole retained in component separation and in the generation of simulations, respectively. In the latter case, it can mimic a rough large-scale time-ordered data filtering. In practice, this filtering is implemented through a harmonic filter $h_{\ell}$ defined as:
    \begin{equation}
        h_{\ell} = \begin{cases}
       0   & \text{if}\ \ell \leq \ell_{\rm min}-4 \\
	\cos\left( \frac{\ell_{\rm min}\, -\, \ell}{\ell_{\rm min}\, -\, 2}\frac{\pi}{2}\right)   & \text{if}\ \ell_{\rm min}-4 < \ell \leq \ell_{\rm min} \\
    1 & \ell > \ell_{\rm min}
   \end{cases}\,.
    \end{equation}
    Default value: $2$
\item \texttt{lmax} and \texttt{lmax\_in} (integers): The former is the maximum multipole used in component separation pipelines and angular power spectrum computation, while the latter in the generation of simulations. If not specified, the default values are $2$\texttt{nside} for \texttt{lmax} and $3$\texttt{nside\_in}$-1$ for \texttt{lmax\_in}.

\item \texttt{data\_type} (string): Specifies the format of the generated simulations to be returned and/or saved, as well as the format of the input data provided to the component-separation modules. Regardless of this setting, the outputs from component separation are always saved and/or returned as maps. Accepted values are \texttt{"maps"} or \texttt{"alms"}.

\item \texttt{nsim\_start} and \texttt{nsims} (integers): These parameters can be used when generating or performing component separation on multiple simulated realizations of the observed microwave sky, as well as when computing angular power spectra from the corresponding outputs. They define, respectively, the initial simulation index and the number of simulations to be considered. Their default values are $0$ and $1$.  
At the current stage, \texttt{BROOM} routines are not designed to run multiple simulations in parallel; however, the same routines can easily be applied in parallel across different simulations within a python script. As detailed in the next section, some \texttt{BROOM} functions are designed to apply a specific routine serially over the provided range of simulations. One example is the \texttt{get\_nuisance\_covariance}, where the number of realization specified with \texttt{nsims} is loaded from disk or simulated on the fly.  In general, all realization-dependent outputs are saved with filenames retaining the simulation index.

\item \texttt{units} (string): Units associated with the data. They are used both to properly return the generated simulations and to calibrate the required SEDs within any ILC pipeline (see, for instance, Equations~\ref{eq:ilc_w} and \ref{eq:ILC_QU_w}), as well as the moments to be deprojected within the cILC and cPILC methods (see Section~\ref{sec:cmb_rec}). The default unit is CMB temperature in $\mu\text{K}$ ($\mu\text{K}_{\text{CMB}}$). Methods aimed at reconstructing signals with fixed SEDs will always return outputs in $\mu\text{K}_{\text{CMB}}$, regardless of the chosen input units, while those reconstructing foreground emission will produce outputs in the same units as the input data. Accepted units include brightness and antenna temperature (with all corresponding multiples and sub-multiples), as well as intensity units such as $\text{Jy}/\text{sr}$ and its multiples and sub-multiples.

\item \texttt{coordinates} (string): Coordinate system associated with the data. Default: Galactic (\texttt{"G"}). Other options are \texttt{"E"} (ecliptic) and \texttt{"C"} (equatorial). This parameter is used to generate simulated foreground maps and to rotate pre-computed Galactic masks. It must match the coordinate system of any provided hit-counts or depth maps (see Section~\ref{app:broom_sims} for further details).

\item \texttt{verbose} (boolean): Verbosity. If set to \texttt{True}, the code prints information about the run. Default value: \texttt{False}.

\end{itemize}

\subsection{Processing and simulating input data} 
\label{app:broom_sims}
In the \texttt{BROOM} framework, inputs for component separation can either be newly simulated (and optionally saved) or loaded from existing files containing previously generated simulations or external data sets. The following parameters are used to generate simulations of the microwave sky for a given instrumental configuration or to process them prior to the application of a component-separation routine. We therefore note that some of these parameters are also required as inputs at later stages—such as component separation or angular power-spectrum computation—primarily to correctly reconstruct the file paths where the outputs of the component-separation modules are, optionally, to be saved or loaded. Depending on the specified \texttt{data\_type}, simulated data are returned either as pixelized $TQU$ maps with $N_{\text{side}} = \texttt{nside}$ or as $TEB$ harmonic coefficients with $\ell_{\text{max}}$ defined by the corresponding parameter \texttt{lmax\_in}.

Common parameters to be provided are:
\begin{itemize}
    \item \texttt{save\_inputs} (boolean): If \texttt{True}, the simulated sky components and co-added maps are saved in the corresponding provided paths (see details below). Default value is \texttt{False}.
    \item \texttt{pixel\_window\_in} (boolean): If \texttt{True}, simulated inputs are convolved with the pixel window function associated with \texttt{nside\_in} and provided data are corrected for it before component separation. Default: \texttt{False}.

    \item \texttt{bandpass\_integrate} (boolean): If \texttt{True}, simulated foreground and CMB maps are generated by integrating over the instrumental bandpasses, rather than assuming a $\delta$-function spectral response (see Equation \ref{eq:bandpass}). Default: \texttt{False}. The assumed bandpass shapes depend on the corresponding keyword value specified in the instrument dictionary (see details below).
    \item \texttt{data\_splits} (boolean): If set to \texttt{True}, two data splits (\texttt{split1} and \texttt{split2}) are generated or loaded for both the total co-added signal and the instrumental noise. When simulations are produced, the noise in the two splits is generated as statistically uncorrelated, with each split having a noise level increased by a factor of $\sqrt{2}$ relative to the full-mission map. Default: \texttt{False}.
    \item \texttt{only\_splits} (boolean): If set to \texttt{True}, the input data sets are generated or loaded exclusively as splits, without including the full-mission total co-added maps or instrumental noise components. This parameter is ignored if \texttt{data\_splits} is \texttt{False}. Default: \texttt{False}.

    \item \texttt{experiments\_file} (string): Path to the \texttt{yaml} file containing the instrumental settings used for simulation generation and pre-processing of input data. This parameter must specify the full path, including the file name, of the instrument configuration file. By default, it is set to the \texttt{experiments.yaml} file located in the \texttt{utils} folder of the installed \texttt{BROOM} package.
    \item \texttt{experiment} (string): Label of the experiment to be used. This label must correspond to a dictionary defined in the \texttt{experiments\_file}. The dictionary may include several items. The ones specifically adopted for the generation of one specific component will be presented in the corresponding section while here we report the general ones used as input for simulating a general component or pre-processing the input data:
    \begin{itemize}
        \item \texttt{frequency} (list): List of central frequencies (floating or integer numbers in $\text{GHz}$ units) corresponding to the frequency channels observed by the telescope.  
        \item \texttt{channel\_tags} (list): List of tags (strings) identifying the different frequency channels. These tags are used to optionally load beams, depth maps, bandpasses, or hits count maps files, and they also appear in the filenames of output foreground templates at the different frequencies as returned by GILC or GPILC (see Section~\ref{sec:compsep}).  
        If not provided, tags are automatically generated from \texttt{frequency} (e.g.\ \texttt{"40.0GHz"} for a $40$\,GHz channel). If multiple channels share the same central frequency, alphabetical letters are appended to these automated labels.
        \item \texttt{beams} (string): Specifies the nature of the instrumental beams. Accepted values are \texttt{"gaussian"} (Gaussian symmetric beams), \texttt{"file\_l"} (symmetric but non-Gaussian beams), and \texttt{"file\_lm"} (asymmetric beams). Default: \texttt{"gaussian"}.  
        If \texttt{"gaussian"} is selected, the FWHM of the beams at each frequency channel must be specified as detailed below. Otherwise, the code will search for a different beam \texttt{FITS} file for each frequency channel, with paths provided via the relevant parameter. Information on the beam characteristics is required to convolve the simulated sky-signal maps at the different frequencies and to bring input data to a common resolution (as in Equation~\ref{eq:beams_1}) prior to component separation, if needed.
         \item \texttt{fwhm} (list): List of float numbers specifying the FWHM of the gaussian beam associated to the different frequency channels. It should have same dimension of \texttt{frequency} and be expressed in units of $\text{arcmin}$. This parameter is used if \texttt{beams} is set to \texttt{"gaussian"}.
         \item \texttt{path\_beams} (string): if \texttt{beams} is not \texttt{"gaussian"}, this keyword is used to identify the full path associated to the beams to be used to smooth or deconvolve sky signals at different frequency channels. The beam for a specific frequency channel with associated label \texttt{channel\_tag} (from the \texttt{channels\_tags} list) will have full path corresponding to: \texttt{path\_beams}~+~\texttt{"\_\{channel\_tag\}.npy"}.
         \item \texttt{path\_bandpasses} (string): Path to the directory containing the bandpass files to be used. Bandpasses are assumed to refer to intensity units. This parameter must be specified if non–top-hat bandpass shapes are required (otherwise, see the \texttt{bandwidth} parameter below). It is used only when \texttt{bandpass\_integrate} is set to \texttt{True}. For each frequency channel, the code expects a file with full path to bandpass file given by \texttt{path\_bandpasses}~+~\texttt{"\_\{channel\_tag\}.npy"}, where \texttt{channel\_tag} corresponds to an element of the \texttt{channel\_tags} list. Each file must contain a 2D array specifying the frequency (in~GHz) and the corresponding transmission.
         \item \texttt{bandwidth} (list): List of relative bandwidths for top-hat bandpasses (e.g., \texttt{0.3} for a $30\%$ bandwidth). Even in this case, the top-hat shape is assumed to be defined in intensity units. This parameter is used when \texttt{path\_bandpasses} is not provided and \texttt{bandpass\_integrate} is set to \texttt{True}.
    \end{itemize}
\end{itemize}

All \texttt{BROOM} routines return, save, or load sets of multifrequency maps (for a specific component or the co-added signal) with shape $N_{\nu} \times N_{\text{fields}} \times N_{\text{pix/lm}}$, where $N_{\nu}$, $N_{\text{fields}}$, and $N_{\text{pix/lm}}$ denote, respectively, the number of observed frequency channels (matching the length of the \texttt{frequency} list), the number of simulated or stored fields, and the number of pixels or harmonic coefficients.

\paragraph{Foreground sky.} To simulate (load from file) a multi-frequency set of foreground maps, the user must set the parameter \texttt{generate\_input\_foregrounds} to \texttt{True} (\texttt{False}). The default value is \texttt{True}. In addition to the full co-added foreground signal, individual foreground components can also be stored (or loaded) if the parameter \texttt{return\_fgd\_components} is set to \texttt{True} (default: \texttt{False}). 

The \texttt{PySM} model for each foreground component must be specified, either for generating the simulation or for loading from existing files. This is done by setting \texttt{foreground\_models} as a python list containing the \texttt{PySM} acronyms of the selected models. The order of the components does not matter as it will be internally ordered according to a pre-defined scheme. The available microwave components in \texttt{BROOM} are listed in Section~\ref{sec:sims}.  

Foreground maps are saved to (or loaded from) \texttt{.npy} files in the directory specified by \texttt{fgds\_path}. The file name format is \texttt{fgds\_path}~+~\texttt{"\_\{model\}.npy"}, where \texttt{model} corresponds to the concatenation of all acronyms in \texttt{foreground\_models} for the full co-added maps, or to a single acronym for individual component maps.

If \texttt{generate\_input\_foregrounds} is \texttt{False} and \texttt{fgds\_path} is provided, foreground maps will be loaded from disk, while if \texttt{fgds\_path} is not specified, no foreground maps are either generated or loaded. An analogous statement holds for the other components and the values of their corresponding keywords.

\paragraph{CMB signal.} The generation or loading of CMB-only multifrequency data is controlled by the parameter \texttt{generate\_input\_cmb} (default: \texttt{True}). The corresponding maps are saved to, or loaded from, the directory specified by \texttt{cmb\_path}, with filenames formatted from the coaddition of \texttt{cmb\_path} and \texttt{"\_\{nsim\}.npy"}. They are loaded and not simulated if \texttt{generate\_input\_cmb} is \texttt{False} and \texttt{cmb\_path} is provided. The optional \texttt{nsim} element denotes the simulation index, which can be specified in the relevant \texttt{BROOM} routines when generating or loading multiple realizations.  

The random seed used for generating CMB realizations can be controlled with the parameter \texttt{seed\_cmb}, which sets the seed value for the random generator. When multiple simulations are requested, the seed is automatically incremented by the simulation index starting from \texttt{seed\_cmb}. By default, \texttt{seed\_cmb} is set to a random value.

Generating a Gaussian CMB realization requires a fiducial (theoretical) angular power spectrum provided as a \texttt{FITS} file, whose full path is specified via \texttt{cls\_cmb\_path}. By default, \texttt{cls\_cmb\_path} points to a theoretical angular power spectrum computed with the \texttt{CAMB} Boltzmann code\footnote{\url{https://github.com/cmbant/camb}}~\cite{camb}, assuming the $\Lambda$CDM best-fit cosmological parameters from the \textit{Planck} 2018 analysis~\cite{Planck_cosmopars}. This file is stored in the \texttt{utils} folder of the installed \texttt{BROOM} package and named \texttt{Cls\_Planck2018\_lensed\_r0.fits}.  

The ordering of the power spectra in the provided file can be specified using the parameter \texttt{cls\_cmb\_new\_ordered}. If set to \texttt{True} (default), the expected order is $[\text{TT}, \text{EE}, \text{BB}, \text{TE}, \text{EB}, \text{TB}]$; if set to \texttt{False}, the order is $[\text{TT}, \text{TE}, \text{TB}, \text{EE}, \text{EB}, \text{BB}]$.

\paragraph{Instrumental noise.} Instrumental noise can be either simulated or loaded by setting the parameter \texttt{generate\_input\_noise} to \texttt{True} (default) or \texttt{False}, respectively. If loading from file, the code needs the \texttt{noise\_path} keyword and searches for multifrequency noise data stored as \texttt{.npy} files in the directory specified by \texttt{noise\_path}. As for the CMB data, the expected filename is given by the concatenation of \texttt{noise\_path} and \texttt{"\_\{nsim\}.npy"}, where the optional \texttt{nsim} index identifies the simulation realization.  

The random seed for noise generation can be controlled via the parameter \texttt{seed\_noise}. The seed is automatically incremented across fields, frequency channels, and simulation indices to ensure independent noise realizations.

Specifically, instrumental noise in \texttt{BROOM} is generated by drawing realizations from the fiducial power spectrum defined in Equation~\ref{eq:cl_noise}. The parameters of this spectrum can be specified by setting the following items within the \texttt{experiment} dictionary located in the \texttt{experiments\_file}:

\begin{itemize}
    \item \texttt{path\_depth\_maps} (string): Path to \texttt{HEALPix} maps of per-pixel map-depth ($\sigma^{T/P}$ in Equation~\ref{eq:cl_noise}). These maps allow for the simultaneous propagation of both the instrument's overall sensitivity and its observational scanning strategy. The code searches for a \texttt{FITS} file named \texttt{path\_depth\_maps}~+~\texttt{"\_\{channel\_tag\}.fits"} for each frequency channel. The maps must be provided in units of $\mu\text{K}\cdot\text{arcmin}$.  
    If the file contains two fields, the first is assumed to correspond to $\sigma^{T}$ and the second to $\sigma^{P}$; if only one field is present, it is assumed to represent polarization, and $\sigma^{T} = \sigma^{P}/\sqrt{2}$. In practice, the noise simulation routines first set $\sigma^{T} = \sigma^{P} = \Omega_{\text{pix}}/a2r$ in Equation~\ref{eq:cl_noise} (where $\Omega_{\text{pix}}=4\pi/N_{\text{pix}}$ is the pixel area and $a2r$ is the converting factor from arcminutes to radians), draw a realization from the corresponding fiducial spectrum, and subsequently rescale this realization using the loaded depth maps.  
    If \texttt{path\_depth\_maps} is not provided, the routines will instead use the \texttt{depth\_I} and \texttt{depth\_P} specifications.
    \item \texttt{depth\_I} (list): List of noise map-depths (one per frequency) in units of $\mu \text{K}\cdot\text{arcmin}$ for intensity maps ($\sigma^{T}$). If not provided, they will be set to map-depths of polarization maps (\texttt{depth\_P}) divided by $\sqrt{2}$.
    \item \texttt{depth\_P} (list): List of noise map-depths (one per frequency) in units of $\mu \text{K}\cdot\text{arcmin}$ for polarization maps ($\sigma^{P}$). If not provided, they will be set to map-depths of intensity maps (\texttt{depth\_I}) multiplied by $\sqrt{2}$. If \texttt{path\_depth\_maps}, \texttt{depth\_I} and \texttt{depth\_P} are all not provided, the noise generation code will return an error.
    \item \texttt{path\_hits\_maps} (string): If \texttt{path\_depth\_maps} is not provided but the user still wishes to simulate the effect of a scanning strategy, this parameter can be defined. It specifies the path to \texttt{HEALPix} maps of hit counts. If \texttt{path\_hits\_maps} corresponds to a full file path (ending with \texttt{.fits}), the same hit-count map is assumed for all frequency channels; otherwise, one file per channel is expected, named as the concatenation of \texttt{path\_hits\_maps} and \texttt{"\_\{channel\_tag\}.fits"}.  
    In practice, the noise realization drawn from the spectrum in Equation~\ref{eq:cl_noise} is weighted by a per-pixel factor of $1/\sqrt{\text{hits}}$, where $\text{hits}$ are the elements of the hit-count map normalized to its maximum value.
    \item \texttt{ell\_knee} and \texttt{alpha\_knee} (lists): Optional lists specifying the values of $\ell_{k}$ and $\alpha$ in Equation~\ref{eq:cl_noise}. These parameters define the $1/f$ noise contribution. If either of them is not provided, purely white noise is simulated.  

\texttt{ell\_knee} can take the following forms:
\begin{itemize}
    \item A single list of multipole values, in which case the $1/f$ noise is included only in temperature;
    \item A list containing two sub-lists, where the first corresponds to temperature and the second to polarization.
\end{itemize}
Similarly, \texttt{alpha\_knee} can be:
\begin{itemize}
    \item A single list of spectral slopes. In this case, $1/f$ noise is injected only in temperature (if \texttt{ell\_knee} is a single list) or in both temperature and polarization using the same slope values (if \texttt{ell\_knee} includes two lists);
    \item A list of two sub-lists, where the first corresponds to temperature and the second to polarization.
\end{itemize}
    \end{itemize}

\paragraph{Full co-added data.} Full co-added simulated data, can be generated from the individual simulated (or loaded) components by setting \texttt{generate\_input\_data} to \texttt{True} (default). If set to \texttt{False}, the full-sky maps are loaded from the directory specified by \texttt{data\_path}, with filenames formatted as \texttt{data\_path}~+~\texttt{"\_\{nsim\}.npy"} for each simulation realization. If \texttt{generate\_input\_data} is set to \texttt{False} and \texttt{data\_path} is not specified, no total maps are generated.

When simulated, the multi-frequency sky signal maps (foregrounds and CMB) are smoothed with the corresponding instrumental beams to account for the angular resolution of the experiment. If loaded from files, they are assumed to be either smoothed with distinct beams for each frequency channel or already brought to a common angular resolution corresponding to a Gaussian beam with a fixed FWHM. Further details on how to handle both cases prior to component separation are provided in Section~\ref{app:compsep_broom}.  

Similarly, all maps are convolved (if simulated) or assumed to be convolved (if loaded) with the pixel window function associated with \texttt{nside\_in}, provided that \texttt{pixel\_window\_in} is set to \texttt{True}.

When either simulated or loaded, both the individual components and the co-added maps are stored and returned as different attributes of a \textsc{SimpleNamespace} object, a lightweight built-in python class available in the \texttt{types} module. It provides a convenient container whose attributes can be dynamically defined and accessed using dot notation—similar to class members—while internally behaving like a mutable dictionary. Specifically, the total, foreground, CMB, and noise data are stored in the \texttt{total}, \texttt{fgds}, \texttt{cmb}, and \texttt{noise} attributes, respectively. Corresponding available attributes for single foreground components are \texttt{dust}, \texttt{synch}, \texttt{ame}, \texttt{co}, \texttt{freefree}, \texttt{cib}, \texttt{tsz}, \texttt{ksz}, \texttt{radio\_galaxies}. If \texttt{data\_splits} is set to \texttt{True}, additional attributes are created, namely \texttt{total\_split1} and \texttt{total\_split2}, as well as \texttt{noise\_split1} and \texttt{noise\_split2}.

When simulated, all components for each frequency channel feature three distinct entries, corresponding either to $TQU$ maps when \texttt{data\_type} is set to \texttt{"maps"} or to $TEB$ coefficients when \texttt{data\_type} is set to \texttt{"alms"}. No such constraints are imposed on loaded data. Therefore, when \texttt{generate\_input\_data} is set to \texttt{True}, the code will combine different components, and the user must ensure the coherence of all component fields when all or part of them are loaded from disk using \texttt{BROOM} routines.

\subsection{Component separation}
\label{app:compsep_broom}
The component-separation routine in \texttt{BROOM} takes as input a \textsc{SimpleNamespace} object, which may contain the following attributes: \texttt{total}, \texttt{fgds}, \texttt{cmb}, \texttt{noise}, \texttt{total\_split1}, \texttt{total\_split2}, \texttt{noise\_split1}, \texttt{noise\_split2}, as well as additional attributes corresponding to individual foreground components.
Each attribute consists of a set of multifrequency maps or spherical-harmonic coefficients with shape $N_{\nu} \times N_{\text{fields}} \times N_{\text{pix/lm}}$, where $N_{\nu}$ denotes the number of frequency channels, $N_{\text{fields}}$ the number of input fields, and $N_{\text{pix/lm}}$ the number of pixels or harmonic coefficients. The value of $N_{\text{fields}}$ must be consistent with the specification provided through the \texttt{field\_in} parameter, introduced below.

Some parameters, common to all component-separation executions, can be specified as described below:

\begin{itemize}
    \item \texttt{field\_in} (string): Input fields provided as data to be processed by component separation routines. It can be in \texttt{["T","E","B","QU","TQU","EB","TEB"]} if inputs are \texttt{HEALPix} maps (\texttt{data\_type}$=$\texttt{"maps"}) and in \texttt{["T","E","B","EB","TEB"]} for harmonic coefficients (\texttt{data\_type}$=$\texttt{"alms"}).
    \item \texttt{field\_out} (string): Fields to be returned and/or saved from the component-separation routines. This parameter can differ from \texttt{field\_in}, both in the type of fields (e.g., \texttt{field\_in = "TQU"} and \texttt{field\_out = "TEB"}) and in the number of fields (e.g., \texttt{field\_in = "TQU"} and \texttt{field\_out = "B"}, where the $T$ and $E$ modes are discarded during component separation).  
Overall, \texttt{field\_out} can take any of the following values:  
\texttt{["T", "E", "B", "QU", "QU\_E", "QU\_B", "TQU", "EB", "TEB"]},  
where \texttt{"QU\_E"} and \texttt{"QU\_B"} correspond to Stokes parameter maps containing only $E$-mode and $B$-mode signals, respectively. For instance, even if the user applies an ILC method (which operates independently on $E$ and $B$ modes), \texttt{field\_out} can still be set to \texttt{"QU"}; conversely, for a PILC application on $QU$ maps (see Section \ref{sec:compsep}), it can be set to \texttt{"EB"}.
\item \texttt{bring\_to\_common\_resolution} (boolean): If set to \texttt{True} (default), the routine assumes that the provided sky signals at different frequencies are smoothed with their respective instrumental beams. In this case, before performing component separation, the inputs are brought to a common angular resolution as described in Equation~\ref{eq:beams_1}. If set to \texttt{False}, the input data are assumed to already share a common resolution and are propagated through the component-separation pipeline without additional smoothing or deconvolution.

\item \texttt{fwhm\_out} (float): Sets the common angular resolution at which the component separation is performed, as well as the resolution of the returned and/or saved outputs. The default value is $2.5 \times d_{p}$, where $d_{p}$ denotes the pixel size corresponding to the \texttt{nside} parameter. This parameter is also used to correct the angular power spectra computed from the component-separation outputs.

\item \texttt{pixel\_window\_out} (boolean): If set to \texttt{True}, the output products are convolved with the pixel window function corresponding to the chosen \texttt{nside} parameter. Consequently, the angular power spectra computed from these outputs will be corrected for this effect. Default: \texttt{False}.

\item \texttt{mask\_observations} (string or array): Full path to a \texttt{FITS} file or a \emph{NumPy} array containing a \texttt{HEALPix} mask that defines the observed and unobserved regions of the sky for the considered instrument. At the current stage, this must be a single file—thus not accounting for different observed regions across frequency channels—although it can represent a combination of them.  
The loaded mask may be apodized; this feature is required only for the map-based $EB$ leakage correction based on standard field purification (see Section~\ref{sec:compsep}). The reconstructed fields from component separation will therefore be apodized as well. This effect is properly accounted for when computing angular power spectra, as the corresponding routines automatically retrieve and use the features of the provided mask. If not specified, full-sky observations are assumed. In the framework of component separation, it is used when computing harmonic coefficients from provided maps if \texttt{data\_type}$=$\texttt{"maps"}.

\item \texttt{mask\_covariance} (string): Full path to a \texttt{FITS} file or a \emph{NumPy} array containing a \texttt{HEALPix} mask that defines the sky region over which component separation is performed. Using this mask, the user may exclude regions that are otherwise retained by \texttt{mask\_observations}. The mask may be apodized; in that case the code assigns the corresponding pixel \emph{weights} when computing the covariance—useful, for example, to downweight sparsely observed, intrinsically noisier regions and thus better control astrophysical emissions. If not provided, this is set equal to \texttt{mask\_observations}.

\item \texttt{leakage\_correction} (string): If specified and polarization maps are propagated through the component-separation pipeline, the code applies an $E$--$B$ leakage correction at the map level before running the component-separation routines. This can be useful in pipelines targeting $B$-mode data, to prevent additional variance from leaked $E$-mode signal.
The option can be set as \texttt{"\{field\}\_\{method\}"}, where:
\begin{itemize}
  \item \textbf{\texttt{field}} can be either \texttt{E}, \texttt{B} or \texttt{EB}, indicating that the leakage correction is applied to $E$-, $B$-mode maps or both, respectively.
  \item \textbf{\texttt{method}} specifies the correction technique and can take the following values:
  \begin{enumerate}
    \item \texttt{purify}: The standard map-based purification implemented in \texttt{NaMaster} (see \cite{pymaster} and references therein);
    \item \texttt{recycling} or \texttt{recycling\_iterationsN}: Applies the recycling technique, optionally including a subsequent iterative harmonic deprojection of ambiguous modes with \texttt{N} iterations in the latter case (see \cite{NILC_cutsky} and references therein).
    \item \texttt{recycling\_inpainting}: Applies the recycling technique, followed by the subtraction of a leakage-residual template estimated via diffusive inpainting.

\item \texttt{inpainting} or \texttt{inpainting\_iterationsN}: Subtracts a leakage template obtained through diffusive inpainting. If the \texttt{inpainting\_iterationsN} option is used, the procedure is additionally followed by an iterative harmonic deprojection of residual ambiguous modes, where the number of iterations is specified by \texttt{N}.

  \end{enumerate}
\end{itemize}
This keyword is only used if \texttt{mask\_observations} is provided. By default, no leakage correction is applied.

\item \texttt{save\_compsep\_products} (boolean): If set to \texttt{True}, the output products from the component-separation, residual-estimation, or component-propagation processes are saved in predefined subdirectories with standardized filenames under the root directory specified by \texttt{path\_outputs} (see below). Default: \texttt{False}.

\item \texttt{path\_outputs} (string): This specifies the root directory where all outputs from component separation are saved if \texttt{save\_compsep\_products} is set to \texttt{True}.

\item \texttt{return\_compsep\_products} (boolean): If set to \texttt{True} (default), the output products from the component-separation, residual-estimation, or component-propagation processes are returned by the corresponding routines and can therefore be reused for any other purpose within the same python script.
\end{itemize}

\noindent The component-separation methodologies to be applied must be specified through the configuration parameter \texttt{compsep}. The \texttt{compsep} parameter is a dictionary containing one or more subdictionaries, each defining a specific component-separation setup, including all required hyperparameters for the run to be applied to the input data.
Common hyperparameters that must or may be specified include:
\begin{itemize}
    \item \texttt{domain} (string): domain in which the method is applied. It can take the values \texttt{"pixel"} (the technique is applied directly in pixel space, i.e., a full-scale analysis) or \texttt{"needlet"} (the technique is applied independently across different needlet bands).

    \item \texttt{method} (string): component-separation technique to be used. It can be one among the following methodologies for the reconstruction of signals with known SED: 
    \begin{itemize}
        \item \texttt{"ilc"}: standard minimum-variance technique to be applied on scalar fields ($T,\ E,\ B$). It comprises ILC and NILC (depending on the provided \texttt{domain}).
        \item \texttt{"cilc"}: cILC to be applied on scalar fields ($T,\ E,\ B$). ILC is augmented with additional constraints which partially or fully deproject nuisance components with known SEDs (such as SZ effect or foreground moments in the CMB reconstruction). 
        \item \texttt{"mcilc"}: Multi-Clustering ILC, at the current stage, just implemented for application to $B$-mode data.
        \item \texttt{"pilc"} (\texttt{"prilc"}) or \texttt{"cpilc"} (\texttt{"cprilc"}): Same as \texttt{ilc} and \texttt{cilc}, respectively, but with variance minimization performed on the polarization intensity $P$. These methods should therefore be used only when total-intensity outputs are not requested.  \texttt{"pilc"} and \texttt{"cpilc"} consider complex component-separation weights, while \texttt{"prilc"} and \texttt{"cprilc"} represent two subcases with just real weights.
        \item \texttt{"mc\_ilc"}, \texttt{"mc\_cilc"}, \texttt{"c\_ilc"}, \texttt{"c\_pilc"}, or \texttt{"c\_prilc"}: hybrid approaches designed for applications in the needlet domain. In these labels, the prefix before the underscore identifies the component-separation technique applied to the needlet bands specified by the \texttt{special\_nls} parameter, while the suffix after the underscore indicates the technique used in all remaining needlet bands.
    For example, \texttt{"mc\_cilc"} applies MC-ILC to the needlet bands listed in \texttt{special\_nls} and cILC to all other bands, whereas \texttt{"c\_ilc"} applies cILC to the specified bands and standard ILC elsewhere.

    \end{itemize}
    It can also be a technique which allows the reconstruction or diagnostic of foreground emission (see Section \ref{sec:fgds_rec} for methodological details):
    \begin{itemize}
        \item \texttt{"gilc"}: GILC technique for blind reconstruction of foreground scalar fields at the input frequency channels specified in \texttt{channels\_out}.
    \item \texttt{"gpilc"}: same as \texttt{"gilc"}, but with application to polarization only and minimization of the variance of the polarized intensity $P$.
    \item \texttt{"fgd\_diagnostic"}: diagnostic of foreground complexity for scalar fields.
    \item \texttt{"fgd\_P\_diagnostic"}: diagnostic of foreground complexity in polarized intensity.
    \end{itemize}
\item \texttt{component\_out} (string): Required for all techniques except those related to foreground reconstruction and diagnostics. This parameter specifies the component with known SED to be reconstructed using the selected ILC methodology. It can be any of 
\texttt{["cmb", "tsz", "0d", "0s", "1bd", "1bs", "1Td", "2bd", "2bs", "2Td", "2bdTd", "2Tdbd"]}. 
Tags ending with \texttt{d} correspond to MBB moments, while those ending with \texttt{s} correspond to power-law moments. In these two cases, the corresponding reference frequencies and pivot spectral parameters must be specified: 
\texttt{nu\_ref\_out\_d}, \texttt{beta\_d\_out}, and \texttt{T\_d\_out} for MBB components; 
\texttt{nu\_ref\_out\_s} and \texttt{beta\_s\_out} for power-law components.
Default: \texttt{"cmb"}.

\item \texttt{ilc\_bias} (float): For all techniques except MC-ILC, this parameter controls the size of the spatial domains over which the empirical data covariance (see Equation~\ref{eq:data_cov}) is estimated. When set to $0$ (the default value), the covariance is computed by averaging over the entire observed sky region.
In practice, this parameter sets the targeted level of residual ILC bias in the reconstructed signal: larger values correspond to smaller covariance domains and, consequently, to a larger expected ILC bias. For example, setting \texttt{ilc\_bias} to $0.001$ implies that a residual ILC bias at the level of one per mille is tolerated in the reconstructed power of the signal of interest. This choice automatically determines the size of the domains over which the covariance elements are estimated, including across different needlet bands.

\item \texttt{needlet\_config} (dictionary): A dictionary defining the needlet configuration to be used when \texttt{domain} is set to \texttt{"needlet"}. It must include the parameter \texttt{needlet\_windows}, which can take one of the following values: \texttt{"mexican"}, \texttt{"standard"}, or \texttt{"cosine"} (see Section~\ref{sec:needlets} for details). Additional hyperparameters must be specified depending on the selected window type.
If \texttt{needlet\_windows} is set to \texttt{"mexican"} or \texttt{"standard"}, the required hyperparameters are \texttt{width}, which defines the width of each needlet band, and \texttt{merging\_needlets}, an optional list specifying intervals of needlet bands to be merged according to Equation~\ref{eq:b_merge}. 
If \texttt{needlet\_windows} is set to \texttt{"cosine"}, the required parameter is \texttt{ell\_peaks}, a list of multipoles at which successive cosine needlets peak.

\item \texttt{save\_weights} (boolean): If set to \texttt{True}, the component-separation weights are saved to a dedicated path specific to each component-separation run. This is needed if one wants to then propagate other components through the same component-separation run.

\item \texttt{save\_needlets} (boolean): If set to \texttt{True} and \texttt{domain} is \texttt{"needlet"}, the adopted set of needlet bands is saved in the directory associated with the corresponding component-separation run. This file is required when propagating G(P)ILC outputs (e.g., for foreground-residual estimation) or any other component of interest through needlet-based component-separation pipelines.

\item \texttt{adapt\_nside} (boolean): If \texttt{True} and \texttt{domain} is \texttt{"needlet"}, the \texttt{HEALPix} resolution of the input multifrequency needlet maps—regenerated from needlet-filtered harmonic coefficients—is adjusted to the maximum multipole sampled by each needlet band. This reduces the computation time because the first bands typically sample only large angular scales.

\item \texttt{cov\_noise\_debias} (list or float): If provided, this parameter should be a float when \texttt{domain} is \texttt{"pixel"}, and a list when \texttt{domain} is \texttt{"needlet"} (in principle containing as many elements as the number of needlet bands). If different from 0 (or if any element is non-zero in the needlet-based component separation), a debiasing of the data covariance with respect to the noise contribution is performed, thereby mitigating its impact on the final variance minimization.
The noise covariance is loaded from disk if \texttt{load\_noise\_covariance} is \texttt{True} in the same subdictionary; otherwise, it is estimated from the \texttt{noise} attribute of the input \textsc{SimpleNamespace} object, if available.
In the needlet case, if the list contains fewer elements than the number of adopted needlet bands, the default value ($0.$) is assumed for the remaining bands.

\end{itemize}

An additional key parameter required to run cILC pipelines is \texttt{constraints}, which is a dictionary containing the following hyperparameters:

\begin{itemize}

\item \texttt{components} (list): List of string tags identifying one or more components to be (fully or partially) deprojected within the cILC formalism (see Equation~\ref{eq:cMILC_weights}). It can include 
\texttt{["0d", "0s", "1bd", "1bs", "1Td", "2bd", "2bs", "2Td", "2bdTd", "2Tdbd", "cmb", "tsz"]}. 
The entries \texttt{"cmb"} and \texttt{"tsz"} correspond to the CMB and tSZ components, which can be deprojected when \texttt{component\_out} does not correspond to them. 
All other entries correspond to the zeroth-, first-, and second-order moments of MBB (indicated by \texttt{d}) and power-law (indicated by \texttt{s}) SEDs. The presence of \texttt{bd}, \texttt{Td}, or \texttt{bs} indicates derivatives with respect to the MBB spectral index, MBB temperature, or power-law spectral index, respectively. 
This parameter can also be specified as a list of lists if different sets of components are to be deprojected in different needlet bands.

\item \texttt{deprojection} (list): Specifies the degree of deprojection for each component listed in \texttt{components}. It can be either a list or a list of lists (if different values are requested for different needlet bands). If an element is equal to $0.$, the corresponding component is fully deprojected; if equal to $0.5$, it is partially deprojected at the 50\% level, and so on.

\item \texttt{beta\_d}, \texttt{T\_d}, \texttt{beta\_s} (float or list): Respectively, the MBB spectral index, MBB temperature, and power-law spectral index assumed for generating the SEDs of the moments when such moments are included in \texttt{components}. These parameters can be provided as lists if different values are desired for different needlet bands. If given as single floating-point numbers, the same value is assumed across all needlet bands.

\end{itemize}

The following parameters are instead only used for GILC, GPILC and foreground diagnostic runs:

\begin{itemize}
    
\item \texttt{nuisance} (list or string): Specifies the string tags of the components to be treated as nuisance in the GILC, GPILC, or foreground diagnostic runs. This parameter is used whether the nuisance covariance is loaded from disk or computed on the fly from the attributes provided in the \textsc{SimpleNamespace} objects that store the input data and/or simulations. Default: \texttt{["noise", "cmb"]}.

\item \texttt{load\_nuisance\_covariance} (boolean): If \texttt{True}, loads the precomputed nuisance covariance matrix (which must include all components listed in \texttt{nuisance}) from disk. The user must therefore first run the corresponding \texttt{get\_nuisance\_covariance} routine (see below).

\item \texttt{depro\_cmb} (list or float): Used only for G(P)ILC runs. If provided, this parameter should be a float when \texttt{domain} is \texttt{"pixel"}, and a list when \texttt{domain} is \texttt{"needlet"} (in principle containing as many elements as the number of needlet bands). If different from \texttt{None} (or if any element is not \texttt{None} in the needlet-based component separation case), the weights are calibrated so that the CMB signal is preserved in the pixel domain or in the corresponding needlet band by the specified amount. For example, if set to $0.$, the CMB signal is fully deprojected; if set to $1.$, it is fully preserved. If \texttt{None} (default value), no specific constraint is imposed on the G(P)ILC weights. If the list contains fewer elements than the number of needlet bands, the default value (\texttt{None}) is assumed for the remaining bands.

\item \texttt{m\_bias} (list or integer): Used only for G(P)ILC runs. If provided, this parameter should be an integer when \texttt{domain} is \texttt{"pixel"}, and a list when \texttt{domain} is \texttt{"needlet"} (in principle containing as many elements as the number of needlet bands). 
If different from $0$ (or if any element is non-zero in the needlet-based component-separation case), the G(P)ILC weights are calibrated to preserve \texttt{m\_bias} (or \texttt{m\_bias[j]} in the $j$-th needlet band) eigenmodes of the data covariance matrix in Equation~\ref{eq:whitened_covar}. The sign of the integer determines whether more or fewer eigenmodes are preserved.
If set to $0$ (default value), no additional constraint is imposed on the G(P)ILC weights. If the list contains fewer elements than the number of needlet bands, the default value ($0$) is assumed for the remaining bands.

\end{itemize}

Many additional parameters, with varying degrees of relevance, can be specified within each entry of the \texttt{compsep} keyword. We refer interested readers and users to the available tutorial and the pedagogical configuration file in the \texttt{BROOM} GitHub page for further insights.

As already explained in Section~\ref{sec:broom_func}, it is possible to derive an estimate of the foreground residuals contaminating the reconstruction of a target signal with a known SED. As detailed in Section~\ref{sec:compsep}, this procedure requires a set of cleaned foreground maps, one for each frequency channel, obtained through the application of G(P)ILC to the data set. These foreground templates are then propagated using the same component-separation weights employed for the reconstruction of the signal of interest.
Within the \texttt{BROOM} framework, this residual estimation is performed using the \texttt{estimate\_residuals} function. The use of this functionality requires defining the \texttt{compsep\_residuals} list of dictionaries within the parameters dictionary. Among the key entries to be specified in each \texttt{compsep\_residuals} dictionary are \texttt{gilc\_path} and \texttt{compsep\_path}, which respectively indicate the location of the G(P)ILC maps to be used and the directory containing the component-separation outputs from which the corresponding weights are loaded. The full path to these data is obtained by concatenating \texttt{gilc\_path} and \texttt{compsep\_path} with \texttt{path\_outputs}. The keyword \texttt{gilc\_path} is required only if the optional argument \texttt{gilc\_outputs} of the \texttt{estimate\_residuals} function is not provided. If \texttt{gilc\_outputs} (a \textsc{SimpleNamespace} object) is supplied, the code automatically identifies the products originating from G(P)ILC runs and uses them for the residual estimation. 
If the corresponding component-separation run was performed in the needlet domain, the code also searches the directory obtained by concatenating \texttt{path\_outputs} and \texttt{compsep\_path} to retrieve the saved needlet bands.

Another important \texttt{compsep\_residuals} parameter is \texttt{gilc\_components}, which allows the user to specify which components (i.e. sets of maps at different observed frequency channels) stored in the G(P)ILC directory or in the \texttt{gilc\_outputs} \textsc{SimpleNamespace} object should be propagated through the weighted combination. The \texttt{compsep\_residuals} list may include multiple dictionaries, each corresponding to a different configuration, for example when distinct sets of G(P)ILC templates or component-separation weights are to be employed. The propagation of the full set of G(P)ILC foreground templates through the desired component separation pipeline can be saved into a folder named \texttt{fgres\_templates} within the same component-separation directory. All the specific configuration parameters, like the maximum multipole, the FWHM of the products, $N_{\text{side}}$ are by default inherited by the main configuration parameters but can be specified specifically for each run by the user.

In the same spirit, any set of multifrequency maps can be propagated through a component-separation run using the \texttt{BROOM} \texttt{combine\_with\_weights} function. In this case, the sets of components to be propagated must be provided to the function via a \textsc{SimpleNamespace} object, where each attribute corresponds to a different component (i.e. a collection of maps across the observed frequency channels).

The key configuration entry is the \texttt{compsep\_propagate} list of dictionaries. Each entry must specify the \texttt{compsep\_path} (as for \texttt{estimate\_residuals}), pointing to the component-separation output directory from which the weights are loaded (and, for needlet-based runs, the corresponding needlet bands). The component-separation run used for propagation can correspond either to a standard constrained reconstruction of a signal with known SED, or to a G(P)ILC solution.

In some specific cases, the user may need to pre-compute a nuisance covariance matrix from tailored data simulations. This is required, for instance, in realistic applications of G(P)ILC, for foreground diagnostics, or to debias the data covariance in the context of minimum-variance reconstructions. Within \texttt{BROOM}, this functionality is provided by the \texttt{get\_nuisance\_covariance} function, which estimates the nuisance covariance from an ensemble of simulations—whose size is set by the \texttt{nsims} parameter—and saves the resulting covariance to disk.
This function searches for a list of dictionaries in the configuration object under the \texttt{nuisance\_covariance} key. Each dictionary must specify a set of parameters, including several that are common to component-separation runs, such as \texttt{domain}, \texttt{needlet\_config} (for needlet-based implementations), and \texttt{ilc\_bias}. Additional parameters are specific to this routine; the main ones are summarized below:
\begin{itemize}
    \item \texttt{type} (\texttt{str}): specifies the type of nuisance covariance to be computed. Allowed values are \texttt{"scalar"}, \texttt{"P\_scalar"}, and \texttt{"Pr\_scalar"}, corresponding respectively to scalar fields (as specified in \texttt{field\_out}), the full polarization field (see Equation~\ref{eq:cov_QU}), or the polarization intensity only (i.e. the $C^{+}$ term in Equation~\ref{eq:cov_QU}).
    \item \texttt{nuisance} (\texttt{list}): list of components to be simulated or loaded from disk and treated as nuisance. This list may include \texttt{"cmb"}, \texttt{"noise"}, and any identifier corresponding to a \texttt{PySM} foreground model.
    \item \texttt{nuisance\_path} (\texttt{str}): path to the directory from which nuisance component simulations are loaded. If not provided, the required simulations are generated on the fly.
\end{itemize}

\subsection{Angular power spectrum computation}
\label{app:broom_spectra}
Below we summarize the main parameters relevant for the computation of angular power spectra of products generated within \texttt{BROOM}.
\begin{itemize}
\item \texttt{spectra\_comp} (string): Specifies the routine used to compute angular power spectra. It can be set to \texttt{"anafast"} or \texttt{"namaster"}, corresponding respectively to the \texttt{HEALPix} \texttt{anafast} routine or to the \texttt{MASTER} estimator implemented via the \texttt{NaMaster} package.

\item \texttt{ell\_min\_bpws} (list): List of minimum multipoles defining each bandpower. If not provided, the code will use the value specified by \texttt{delta\_ell} to construct uniform bins.

\item \texttt{delta\_ell} (integer): Sets the width of the bandpowers (i.e., the bin size). The default value is $1$. If \texttt{ell\_min\_bpws} is specified, this parameter is ignored.

\item \texttt{return\_Dell} (boolean): If set to \texttt{True}, the spectrum-computation routines return and/or save the quantity $D_{\ell} = \frac{\ell(\ell+1)}{2\pi} C_{\ell}$ instead of $C_{\ell}$.

\item \texttt{field\_cls\_out} (string or list): Specifies which angular power spectra to compute. This argument may be given either as a single string, for one spectrum, or as a list of strings, for multiple spectra. Allowed values include any auto- or cross-spectrum in \([TT,\ EE,\ BB,\ TE,\ EB,\ TB]\).

\item \texttt{return\_spectra} (boolean): If set to \texttt{True}, the spectrum-computation routine returns the angular power spectra as a \textsc{SimpleNamespace} object.

\item \texttt{save\_spectra} (boolean): If set to \texttt{True}, the computed spectra are saved in predefined subdirectories (sharing the same root path as the corresponding input products) using standardized filenames.

\item \texttt{save\_mask} (boolean): If a masking strategy is applied that does not rely on loading a pre-existing FITS mask and this parameter is set to \texttt{True}, the generated mask is saved to disk in a predefined directory with a standardized filename.

\item \texttt{compute\_spectra} (dictionary): This is the main configuration dictionary used to compute angular power spectra from any product generated within \texttt{BROOM}. As described in Section~\ref{sec:broom_func}, the spectrum-computation routines can take as input a \textsc{SimpleNamespace} object, and some or all of its attributes may be used to estimate the corresponding angular power spectra.
The \texttt{compute\_spectra} dictionary may contain one or more subdictionaries. If multiple subdictionaries are provided, they should typically correspond to the number of distinct runs that generated the products stored in the \textsc{SimpleNamespace} object. Alternatively, if \texttt{outputs} is not supplied directly to the spectrum-computation routine, the user may define as many subdictionaries as desired, each corresponding to a specific run or masking configuration. The main parameters to be specified in each subdictionary are:
\begin{itemize}
    \item \texttt{path\_method} (string): A complementary string that, when added to \texttt{path\_outputs}, defines the full directory path where the products of component separation, residual estimation, or component propagation are stored. This keyword is not required if a \textsc{SimpleNamespace} object is provided directly to the \texttt{\_compute\_spectra\_} routine via the \texttt{outputs} argument.

    \item \texttt{components\_for\_cls} (list): A list of strings identifying the tags of the components (from component separation, residual estimation, or component propagation) for which angular power spectra are to be computed. This keyword can be used both when a \textsc{SimpleNamespace} object is provided to the \texttt{\_compute\_spectra\_} routine and when products are loaded from disk through the \texttt{path\_method} parameter. In the latter case, it is mandatory; in the former, it is optional. Examples of valid tag lists can be found in the example configuration files and tutorials available on the \texttt{BROOM} GitHub repository. We note that, when computing spectra for products from G(P)ILC runs, the tag corresponding to each component must include the suffix \texttt{"\_\{channel\_tag\}"}, where \texttt{channel\_tag} identifies a specific frequency channel as defined in the \texttt{channel\_tags} entry of the instrument dictionary.

   \item \texttt{mask\_type} (string): Specifies the masking strategy to be applied. The available options are:

\begin{itemize}
    \item \texttt{"GAL*"}: Applies a Galactic mask from the Planck 2018 analysis with the specified retained sky fraction (e.g., \texttt{"GAL40"} retains $40\%$ of the sky).

    \item \texttt{"from\_fits"}: Loads a mask from the file specified by the \texttt{mask\_path} keyword (see below).

    \item \texttt{"GAL*+fgres"}: Loads the Planck Galactic mask with the specified sky fraction and further masks regions based on a foreground-residual map until the final sky fraction specified by the \texttt{fsky} keyword is reached.

    \item \texttt{"GAL*+fgtemp"} or \texttt{"GAL*+fgtemp\^{}3"}: Loads the Planck Galactic mask with the specified sky fraction and further masks regions using a foreground-residual template map (or its third power), until the final sky fraction defined by the \texttt{fsky} keyword is achieved. The path to the foreground-residual template must be provided through the \texttt{fgres\_temp\_for\_masking} parameter.

    \item \texttt{"fgres"}, \texttt{"fgtemp"}, or \texttt{"fgtemp\^{}3"}: Generates a mask by thresholding the foreground-residual map or template (or its third power), respectively, such that the final sky fraction matches the value specified by the \texttt{fsky} keyword.
\end{itemize}

If \texttt{mask\_type} is not specified but \texttt{mask\_covariance} is provided, the latter is used as the mask. If \texttt{mask\_covariance} is not provided, the code uses \texttt{mask\_observations}. If neither is specified, the angular power spectra are computed over the full sky.

    \item \texttt{field\_out} (string): Specifies the fields associated to the product files to be loaded from disk for power spectra computation when \texttt{outputs} is not provided to the \texttt{\_compute\_spectra\_} routine. If not explicitly set, it defaults to the common \texttt{field\_out} parameter defined in the configuration file or class.

    \item \texttt{mask\_path} (string): Full path to the mask file to be loaded when \texttt{mask\_type} is set to \texttt{"from\_fits"}. 
If multiple masks are stored in the FITS file and both temperature and polarization spectra are to be computed, the code uses the first field for temperature and the second field for polarization power spectra derived from $Q$ and $U$ maps. 
If the input products are provided in terms of $E$ and $B$ modes, different masks can be loaded from the same FITS file for the two components. In this case, the masks are read from fields $1$ and $2$ when temperature spectra are also computed, or from fields $0$ and $1$ if only polarization spectra are required.

   \item \texttt{fsky} (float): Desired final sky fraction retained after masking, applicable when \texttt{mask\_type} is one of \texttt{"GAL*+fgres"}, \texttt{"GAL*+fgtemp"}, \texttt{"GAL*+fgtemp\^{}3"}, \texttt{"fgres"}, \texttt{"fgtemp"}, or \texttt{"fgtemp\^{}3"}. It must be a real number between $0$ and $1$.

    \item \texttt{apodize\_mask} (string): Optional apodization scheme applied to the mask. Available options are \texttt{"gaussian"} (simple Gaussian smoothing), \texttt{"gaussian\_nmt"} (Gaussian smoothing as implemented in \texttt{NaMaster}), \texttt{"C1"}, and \texttt{"C2"}.

    \item \texttt{smooth\_mask} (float): Apodization scale of the mask, in degrees.

\item \texttt{nmt\_purify\_B} (boolean): Whether to purify the $B$-mode harmonic coefficients before computing the power spectra. This option can be used when \texttt{spectra\_comp} is set to \texttt{"namaster"} and $B$-mode spectra are computed from $QU$ products.

\item \texttt{nmt\_purify\_E} (boolean): Whether to purify the $E$-mode harmonic coefficients before computing the power spectra. This option can be used when \texttt{spectra\_comp} is set to \texttt{"namaster"} and $E$-mode spectra are computed from $QU$ products.

\item \texttt{smooth\_tracer} (float): Smoothing scale (in degrees) applied to the tracer (foreground residual map or template) used to generate the mask. This parameter is used when \texttt{mask\_type} is one of \texttt{"GAL*+fgres"}, \texttt{"GAL*+fgtemp"}, \texttt{"GAL*+fgtemp\^{}3"}, \texttt{"fgres"}, \texttt{"fgtemp"}, or \texttt{"fgtemp\^{}3"}.

\end{itemize}

\end{itemize}



\bibliographystyle{JHEP}
\bibliography{biblio}

\end{document}